\documentclass[fleqn,10pt]{wlscirep}
\usepackage[utf8]{inputenc}
\usepackage[T1]{fontenc}
\usepackage[labelformat=empty, position=top]{subcaption}
\usepackage[export]{adjustbox}

\title{Inferring firm-level supply chain networks with realistic systemic risk from industry sector-level data}

\author[1,*]{Massimiliano Fessina}
\author[2,3]{Giulio Cimini}
\author[1,4]{Tiziano Squartini}
\author[5,6,7]{Pablo Astudillo-Estévez}
\author[8,6,9]{Stefan Thurner}
\author[1,4,10]{Diego Garlaschelli}

\affil[1]{IMT School for Advanced Studies, 55100 Lucca (Italy)}
\affil[2]{Physics Department and INFN, University of Rome `Tor Vergata', 00133 Rome (Italy)}
\affil[3]{Enrico Fermi Research Center (CREF), 00184 Rome (Italy)}
\affil[4]{INdAM-GNAMPA Istituto Nazionale di Alta Matematica, 00185 Rome (Italy)}
\affil[5]{Colegio de Economía, Universidad San Francisco de Quito, 170901 Quito (Ecuador)}
\affil[6]{Complexity Science Hub, A-1080 Vienna (Austria)}
\affil[7]{Institute for New Economic Thinking, University of Oxford, OX1 3UQ Oxford (UK)}
\affil[8]{Section for Science of Complex Systems, Medical University of Vienna, A-1090 Vienna (Austria)}
\affil[9]{Santa Fe Institute, NM 87501 Santa Fe (USA)}
\affil[10]{Lorentz Institute for Theoretical Physics, University of Leiden, 2333 CA Leiden (Netherlands)}


\begin{abstract}
Production networks constitute the backbone of every economic system. They are inherently fragile as several recent crises clearly highlighted. Estimating the system-wide consequences of local disruptions (systemic risk) requires detailed information on the supply chain networks (SCN) at the firm-level, as systemic risk is associated with specific mesoscopic patterns. However, such information is usually not available and realistic estimates must be inferred from available sector-level data such as input-output tables and firm-level aggregate output data. Here we explore the ability of several maximum-entropy algorithms to infer realizations of SCNs characterized by a realistic level of systemic risk. We are in the unique position to test them against the actual Ecuadorian production network at the firm-level. Concretely, we compare various properties, including the Economic Systemic Risk Index, of the Ecuadorian production network with those from four inference models. We find that the most realistic systemic risk content at the firm-level is retrieved by the model that incorporates information about firm-specific input disaggregated by sector, indicating the importance of correctly accounting for firms' heterogeneous input profiles across sectors. Our results clearly demonstrate the minimal amount of empirical information at the sector level that is necessary to statistically generate synthetic SCNs that encode realistic firm-specific systemic risk.
\end{abstract}

\begin{document}

\flushbottom

\maketitle

\thispagestyle{empty}

\section*{Introduction}

Supply chains arise since firms purchase goods from other firms as inputs for their own production. Although this intricate set of relationships constitutes the backbone of any productive system, it is also the cause of their inherent fragility as disruptions can propagate along the channels defining interfirm networks, threatening the whole economy~\cite{choi2006supply,Craighead:2007aa}. An example is provided by the earthquake striking Fukushima in 2011, whose economic impact went far beyond the area directly affected by the disaster~\cite{inoue2019firm,guan2020global,Aldrighetti:2021aa,carvalho2021jap,Ivanov:2021aa,chowdhury2021covid}; another example is represented by the Covid-19 pandemic, so far-reaching to impede some countries from being capable of supplying the population with basic goods~\cite{guan2020global,chowdhury2021covid,pichler2020production,pichler2022simultaneous,Ivanov:2021aa}. Advising governments and policymakers with optimal strategies to contain the economic effects of supply chains disruptions has, thus, become an issue of paramount importance.

The propagation of economic shocks has been traditionally studied by employing industry-level input/output tables~\cite{Miller:2009aa,leontief1986input}. Only recently firm-level production networks have been considered to assess under which conditions economic shocks spreading across individual firms may generate macroeconomic fluctuations~\cite{acemoglu2012net,carvalho2019production,Gabaix:2011aa,diem2024estimating}: the first analyses of this kind appeared in the literature about the supply chain management, where the extent of the failures caused by a propagating shock - a phenomenon known as \emph{ripple effect}~\cite{dolgui2018ripple,ivanov2014ripple,ivanov2017simulation} - was proxied by the portion of the system undergoing a disruption. More recent tools to explicitly account for the role played by the structure of production networks in the propagation of shocks~\cite{choi2001supply} include the \emph{Nexus Supplier Index} (NSI), evaluating the riskiness of `hidden' suppliers (i.e. those belonging to the innermost layers of a supply chain)~\cite{yan2015theory,shao2018data} and the newly proposed \emph{Economic Systemic Risk Index} (ESRI)~\cite{diem2022quantifying}, proxying the riskiness of a firm with the output reduction experienced by the whole production network as a consequence of its failure.

As in the financial case, privacy issues prevent large-scale firm-level datasets from being publicly available~\cite{bacilieri2022firm}. The few proprietary datasets with global coverage (e.g. Capital IQ~\cite{Chakraborty:2020aa,chakraborty2024inequality}, Compustat~\cite{atalay2011network,Cohen:2008aa,Barrot:2016aa} and Factset~\cite{Konig:2022aa}, just to cite a few) are obliged to report the existence of a relationship between a company and any customer that is responsible for at least the $10\%$ of its revenues: hence, they solely cover the largest firms listed in the US stock exchanges; besides, they do not disclose any information about link weights~\cite{bacilieri2022firm}. Datasets that are less `geographically biased' exist; such a desirable feature, however, comes at the cost of being limited to specific industrial sectors: this is the case of MarkLines, gathering information about the automotive sector through surveys to car manufacturers~\cite{brintrup2015nested,fessina2024pattern}. At the national scale, only a handful of countries collect firm-level relationships, via either value-added tax (VAT) data (e.g. Belgium~\cite{dyne2015belgian}, Ecuador~\cite{mungo2023reconstructing}, Hungary~\cite{diem2022quantifying}, Spain~\cite{peydro2020production}) or payment data mediated by the major banks (e.g. Brazil~\cite{silva2020brasil}, Japan~\cite{tamura2012estimation,Fujiwara2021regional}, The Netherlands~\cite{ialongo2022reconstructing}). For what concerns Japan, a second dataset, compiled through surveys to Japanese firms about their major customers and suppliers~\cite{saito2007larger,ohnishi2009hubs,ohnishi2010network,fujiwara2010large}, is provided by the Tokyo Shoko Research company. In summary, national datasets are usually accurate, providing information on link weights as well: since, however, they contain highly sensitive information, can be accessed only under strict non-disclosure agreements.

The difficulties in accessing empirical data have motivated researchers to devise algorithms for inferring the unknown, structural details of production networks from incomplete information~\cite{mungo2024reconstructing}. More specifically, the \emph{link prediction} problem has been tackled by applying machine learning techniques, in particular classification algorithms, trained to infer the probability that a buyer-supplier pair of companies is connected from partial knowledge about the network topology~\cite{brintrup2018predicting,kosasih2021machine}, the financial status, industrial sector and geographical localisation of firms~\cite{mungo2023reconstructing} as well as from Internet news~\cite{brockmann2022supply}. For what concerns the \emph{network reconstruction} problem, instead, very diverse approaches exist. In~\cite{mungo2023revealing}, correlations between financial time series of companies are used to infer buyer-supplier relationships; in~\cite{reisch2022monitoring}, a whole  national production network is inferred from mobile phone communications data. Approaches rooted into statistical physics are those adopted in~\cite{bacilieri2023reconstructing}, where trade volumes between firms are inferred from the knowledge of the network topology, and in~\cite{ialongo2022reconstructing}, where the authors propose the so-called \emph{Stripe-Corrected Gravity Model} (SCGM), i.e. a method to reconstruct both the topology and the link weights from the knowledge of how much companies buy and sell within each industrial sector - in words, the sector-specific in-strength and out-strength of each firm.

The SCGM belongs to the family of Exponential Random Graphs (ERGs)~\cite{cimini2019statistical}. Algorithms of this kind are induced by the constrained maximisation of Shannon entropy~\cite{jaynes1957information,squartini2011analytical}: as a consequence, they are employed to generate configurations that preserve (on average) the value of a number of node-specific quantities, otherwise being maximally random. A very accurate reconstruction can be achieved upon constraining both degrees and strengths~\cite{mastrandrea2014enhanced,gabrielli2019}; in case the degrees are unknown, a \emph{fitness ansatz}~\cite{caldarelli2002scale} can be used to infer them from other, correlated properties~\cite{garlaschelli2004fitness,cimini2015systemic,cimini2015estimating}.

As models like the ones described here have been already successfully employed to reconstruct economic and financial networks~\cite{squartini2018reconstruction}, here we focus on the four variants named \emph{Stripe-Corrected Gravity Model} (SCGM)~\cite{ialongo2022reconstructing}, \emph{Input-Output Gravity Model} (IOGM), \emph{Density-Corrected Gravity Model} (DCGM)~\cite{cimini2015systemic} and \emph{Stripe-Corrected MaxEnt Model} (SCMM) and use them to reconstruct the Ecuadorian production network - specifically, the systemicness of each firm. As our results show, reconstruction models preserving the sector-specific in-strength and out-strength of each firm (i.e. the SCGM and, only approximately, the IOGM) are capable of generating reliable, synthetic production networks, characterised by a firm-specific `risk profile' (evaluated by computing the ESRI values~\cite{diem2022quantifying}) closely matching the empirical one.

\section*{Reconstruction of production networks}

\subsection*{Data description}

We construct the Ecuadorian production network from VAT data collected by the Internal Revenues Service (IRS) of Ecuador~\cite{astudillo2021towards}, covering the year 2008. Data are cleaned (see Supplementary Materials S1) to represent a weighted, directed network with $2.189.066$ transactions, i.e. links, and $60.488$ firms, i.e. nodes, classified into $387$ sectors at the ISIC 4 digits level. Such a network is, then, filtered by keeping only the links whose weight exceeds a threshold of $22.300\:\$$: upon doing so, the number of connections reduces to $130.044$, established among $29.089$ firms, in turn classified into $371$ sectors (see Methods). Since the only information about the production of a firm is represented by the corresponding industrial code, we assume the output of each company to belong to the same sector - in other words, to be a `single sector' one.

\subsection*{The Stripe-Corrected Gravity Model (SCGM)}

As any firm needs a specific set of inputs to output its products, the information about which products are traded between firms is an important one to consider. Since, however, this information is not available to us, here we assume that each firm sells products within a single sector, i.e. the one corresponding to its ISIC code. The SCGM~\cite{ialongo2022reconstructing} incorporates this information by assigning to the link $i\rightarrow j$ the weight

\begin{equation}
w_{i\to j}^\text{SCGM}=\frac{s_i^\text{out}s_{g_i\to j}}{W_{g_i}^\text{out}\:p_{i\to j}^\text{SCGM}},\:\:\text{with probability\:\:}p_{i\to j}^\text{SCGM}=\frac{z_{g_i}s_i^\text{out}s_{g_i\to j}}{1+z_{g_i}s_i^\text{out}s_{g_i\to j}};
\label{eq:SCGM}
\end{equation}
naturally, $w_{i\to j}^\text{SCGM}=0$ with probability $1-p_{i\to j}^\text{SCGM}$. Notice that $s_i^\text{out}$ is the out-strength of firm $i$, related to its ISIC sector $g_i$; $s_{g_i\to j}=\sum_{k\in g_i}w_{k\to j}$ is the in-strength by sector of firm $j$, i.e. the contribution to the in-strength of firm $j$ coming from the sector to which firm $i$ belongs, i.e. $g_{i}$ - in words, how much $j$ buys from the suppliers belonging to sector $g_i$ (a quantity that has been called \emph{stripe} in~\cite{ialongo2022reconstructing}, whence the name of the method); $W_{g_i}^\text{out}=\sum_{k\in g_i}s_k^\text {out}=\sum_j s_{g_i\to j}$ is the outgoing flux of sector $g_i$. The parameter $z_{g_i}$ is determined by imposing

\begin{equation}
\langle L_{g_i}\rangle=\sum_{i\in g}\sum_{j(\neq i)}p_{i\to j}^\text{SCGM}=L^*_{g_i},
\end{equation}
i.e. that the expected number of links within sector $g_i$ matches the empirical one. The formulation of the SCGM ensures that it preserves (on average) the out-strength and the in-strength by sector of each firm, i.e. $\langle s_i^\text{out}\rangle=s_{i}^\text{out}$ and $\langle s_{g_i\to j}\rangle=s_{g_i\to j}$, thus reproducing the productive structure of each firm (see Methods). The SCGM can be easily generalised to accommodate firms having `multiple sectors' outputs (see Methods and ~\cite{ialongo2022reconstructing} for further details).

\subsection*{The Input-Output Gravity Model (IOGM)}

The SCGM can be reformulated by employing input/output flows between sectors. In this case, the in-strength by sector of firm $j$ is replaced by the expression

\begin{equation}
s^{\text{IOGM}}_{g_i\to j}=s^{\text{in}}_j\frac{s_{g_i\to g_j}}{W^{\text{in}}_{g_j}}=s^{\text{in}}_j\frac{s_{g_i\to g_j}}{\sum_{g_i} s_{g_i\to g_j}}
\label{eq:IO_stripe}
\end{equation}
where $s_{g_i\to g_j}=\sum_{k\in g_i}\sum_{l\in g_j}w_{k\to l}$ is the total flux from sector $g_i$ to sector $g_j$ - in words, how much $g_j$ requires from $g_i$ - and $W_{g_j}^\text{in}=\sum_{g_i}s_{g_i\to g_j}=\sum_{g_i}\sum_{k\in g_i}\sum_{l\in g_j}w_{k\to l}$ is the incoming flux of sector $g_j$ - in words, how much $g_j$ requires from (the totality of) the other sectors. The expression above splits the in-strength of node $j$ into sector-specific contributions, the $i$-th one being precisely the fraction of input provided by sector $g_i$. The IOGM has the same functional form of the SCGM, i.e.

\begin{equation}
w_{i\to j}^\text{IOGM}=\frac{s_i^\text{out}s^{\text{IOGM}}_{g_i\to j}}{W_{g_i}^\text{out}\:p_{ij}^\text{IOGM}},\:\:\text{with probability\:\:}p_{i\to j}^\text{IOGM}=\frac{z_{g_i}s_i^\text{out}s^{\text{IOGM}}_{g_i\to j}}{1 + z_{g_i}s_i^\text{out}s^{\text{IOGM}}_{g_i\to j}};
\label{eq:IOscgm}
\end{equation}
naturally, $w_{i\to j}^\text{IOGM}=0$ with probability $1-p_{i\to j}^\text{IOGM}$. The parameter $z_{g_i}$ is determined by imposing

\begin{equation}
\langle L_{g_i}\rangle=\sum_{i\in g}\sum_{j(\neq i)}p_{i\to j}^\text{IOGM}=L^*_{g_i},
\end{equation}
i.e. that the expected number of links within sector $g_i$ matches the empirical one. The model preserves (on average) the out-strength and the in-strength of each firm and the total flux from sector $g_i$ to sector $g_j$, i.e. $\langle s_i^\text{out}\rangle=s_i^\text{out}$, $\langle s_i^\text{in}\rangle=s_i^\text{in}$ and $\langle s_{g_i\to g_{j}}\rangle=s_{g_{i}\to g_{j}}$, but approximates the in-strength by sector of each firm, reading $s_{g_{i}\to j}$, with $s_{g_{i}\to j}^\text{IOGM}$ (see Methods for further details).

\subsection*{The Density-Corrected Gravity Model (DCGM)}

The DCGM~\cite{cimini2015systemic} derives from the Directed Binary Configuration Model (DBCM), in turn induced by the in- and out-degree sequences. As the latter ones are often unavailable, the DCGM relies on a fitness ansatz that employs the strengths as parameters to set the values of the degrees. Such a position returns quite accurate results for a variety of economic and financial networks~\cite{cimini2015systemic}, including production networks~\cite{ialongo2022reconstructing}. The DCGM is defined by imposing

\begin{equation}
w_{i\to j}^\text{DCGM}=\frac{s_i^\text{out}s_j^\text{in}}{W\:p_{i\to j}^\text{DCGM}},\:\:\text{with probability\:\:}p_{i\to j}^\text{DCGM}=\frac{zs_i^\text{out}s_j^\text{in}}{1+zs_i^\text{out}s_j^\text{in}};
\end{equation}
naturally, $w_{i\to j}^\text{DCGM}=0$ with probability $1-p_{i\to j}^\text{DCGM}$. Notice that the DCGM does not employ sector-specific quantities, as $s_i^\text{out}=\sum_{j(\neq i)}w_{i\to j}$ is the out-strength of node $i$, $s_j^\text{in}=\sum_{i(\neq j)}w_{i\to j}$ is the in-strength of node $j$ and $W=\sum_{i}s_{i}^\text{out}=\sum_{i}s_{i}^\text{in}$ is the total weight of the network. The parameter $z$ is determined by imposing

\begin{equation}
\langle L \rangle=\sum_i\sum_{j(\neq i)}p_{i\to j}^\text{DCGM}=L^*.
\end{equation}
i.e. that the expected number of links of the network matches the empirical one. The model preserves (on average) the out-strength and the in-strength of each firm, i.e. $\langle s_i^\text{out}\rangle^\text{DCGM}=s_i^\text{out}$ and $\langle s_i^\text{in}\rangle^\text{DCGM}=s_i^\text{in}$ (see~\cite{cimini2015systemic} for further details).

\subsection*{The Stripe-Corrected MaxEnt Model (SCMM)}

As a last benchmark, let us introduce the SCMM, i.e. a topology-free model prescribing to pose

\begin{equation}
w_{i\to j}^\text{SCMM}=\frac{s_i^\text{out}s_{g_i\to j}}{W_{g_i}^\text{out}};
\label{eq:weight_sME}
\end{equation}
the SCMM represents the deterministic version of the SCGM, as it generates a single, reconstructed network rather than an entire ensemble of reconstructed configurations. The model preserves the same quantities of the SCGM but induces a topology which is unrealistically dense, since $w_{i\to j}^\text{SCMM}$ equals zero only if firm $i$ has no output or firm $j$ does not receive any input from sector $g_i$.

\section*{Results}

\begin{figure*}[t!]
\centering\begin{subfigure}{0.02\textwidth}
    \textbf{a)}
\end{subfigure}
\begin{subfigure}[t]{0.30\textwidth}
\includegraphics[width=\textwidth,valign=t]{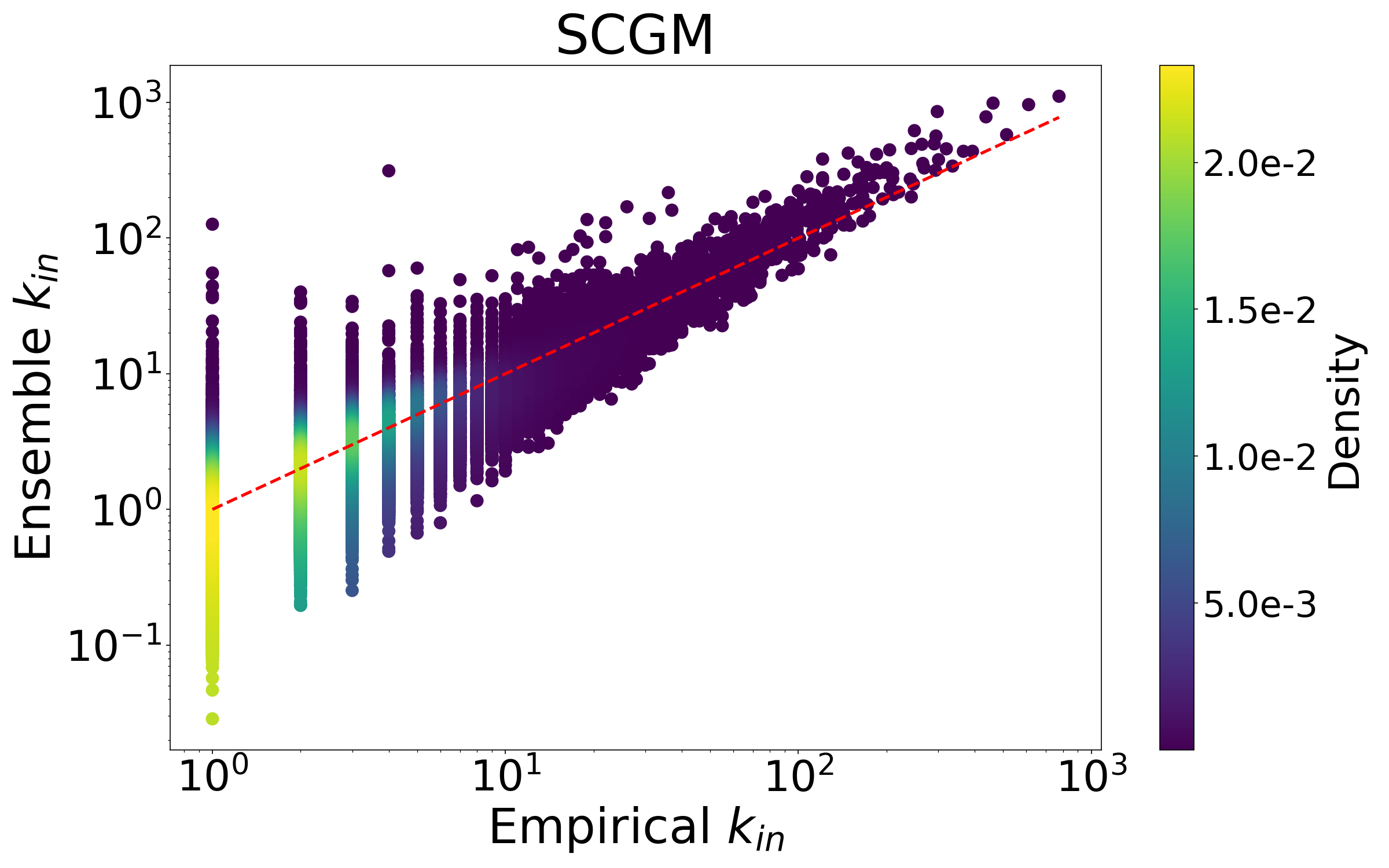}
\end{subfigure}
\begin{subfigure}{0.02\textwidth}
    \textbf{b)}
\end{subfigure}
\begin{subfigure}[t]{0.30\textwidth}
\includegraphics[width=\textwidth,valign=t]{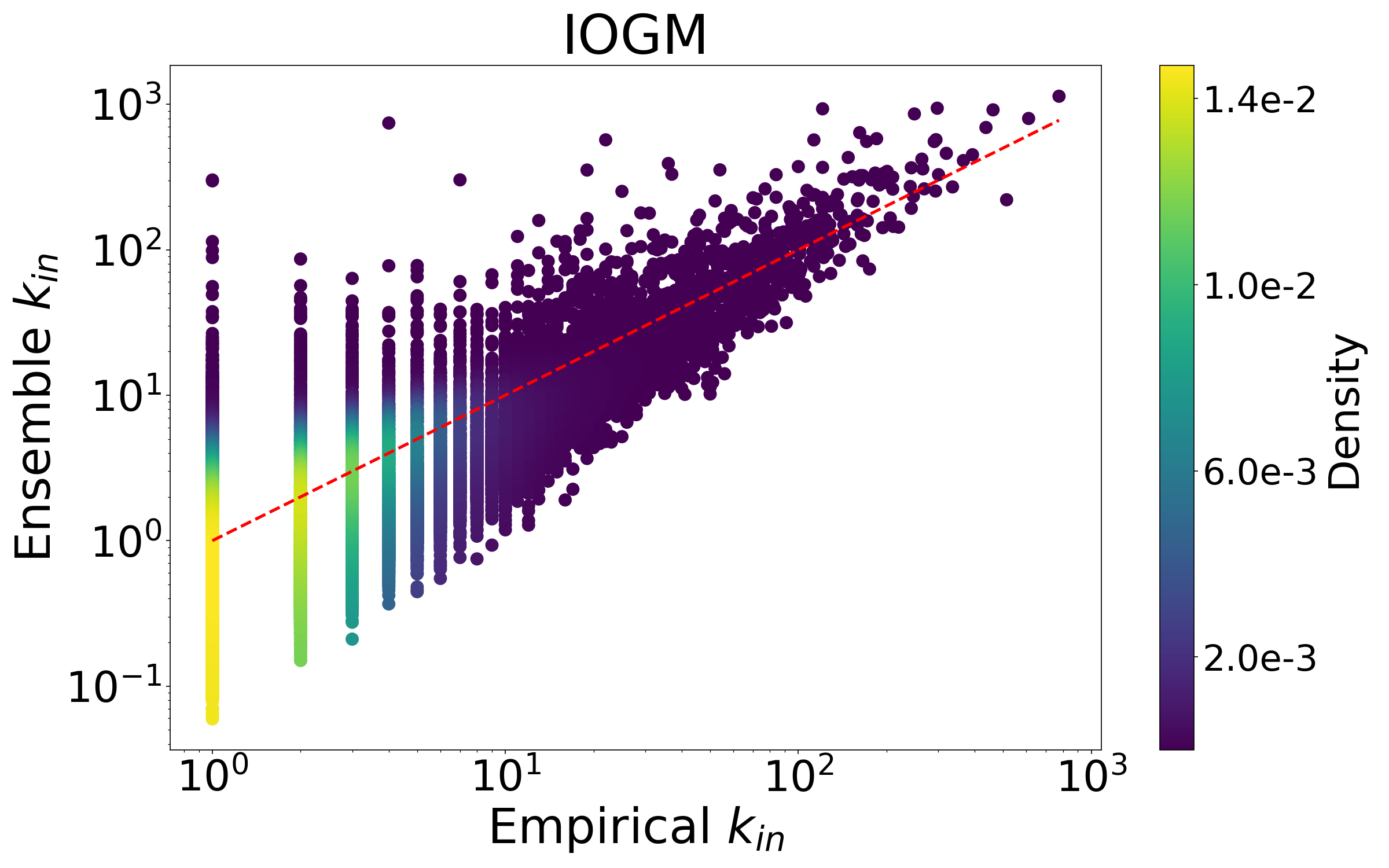}
\end{subfigure}
\begin{subfigure}{0.02\textwidth}
    \textbf{c)}
\end{subfigure}
\begin{subfigure}[t]{0.30\textwidth}
\includegraphics[width=\textwidth,valign=t]{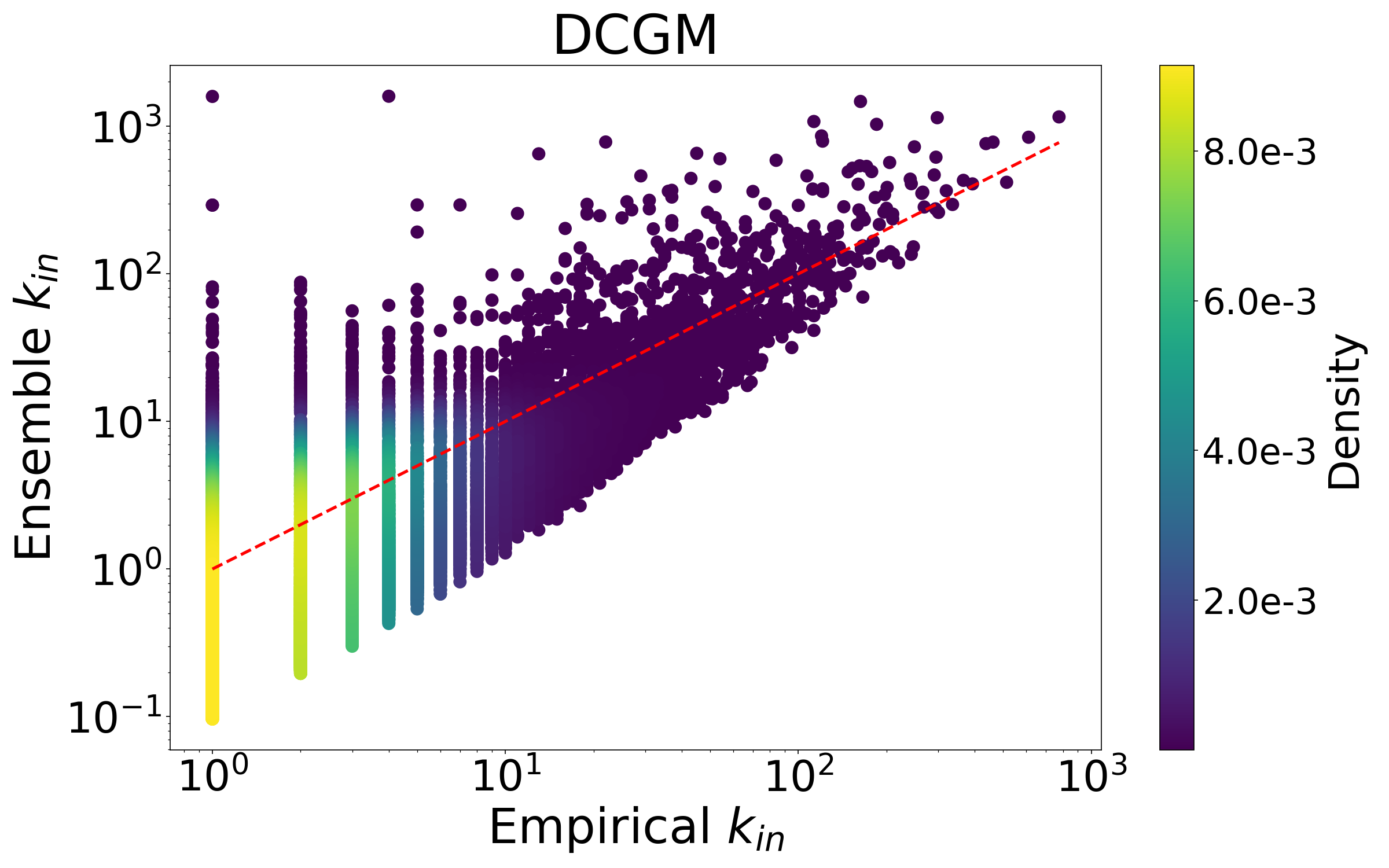}
\end{subfigure}
\begin{subfigure}{0.02\textwidth}
    \textbf{d)}
\end{subfigure}
\begin{subfigure}[t]{0.30\textwidth}
\includegraphics[width=\textwidth,valign=t]{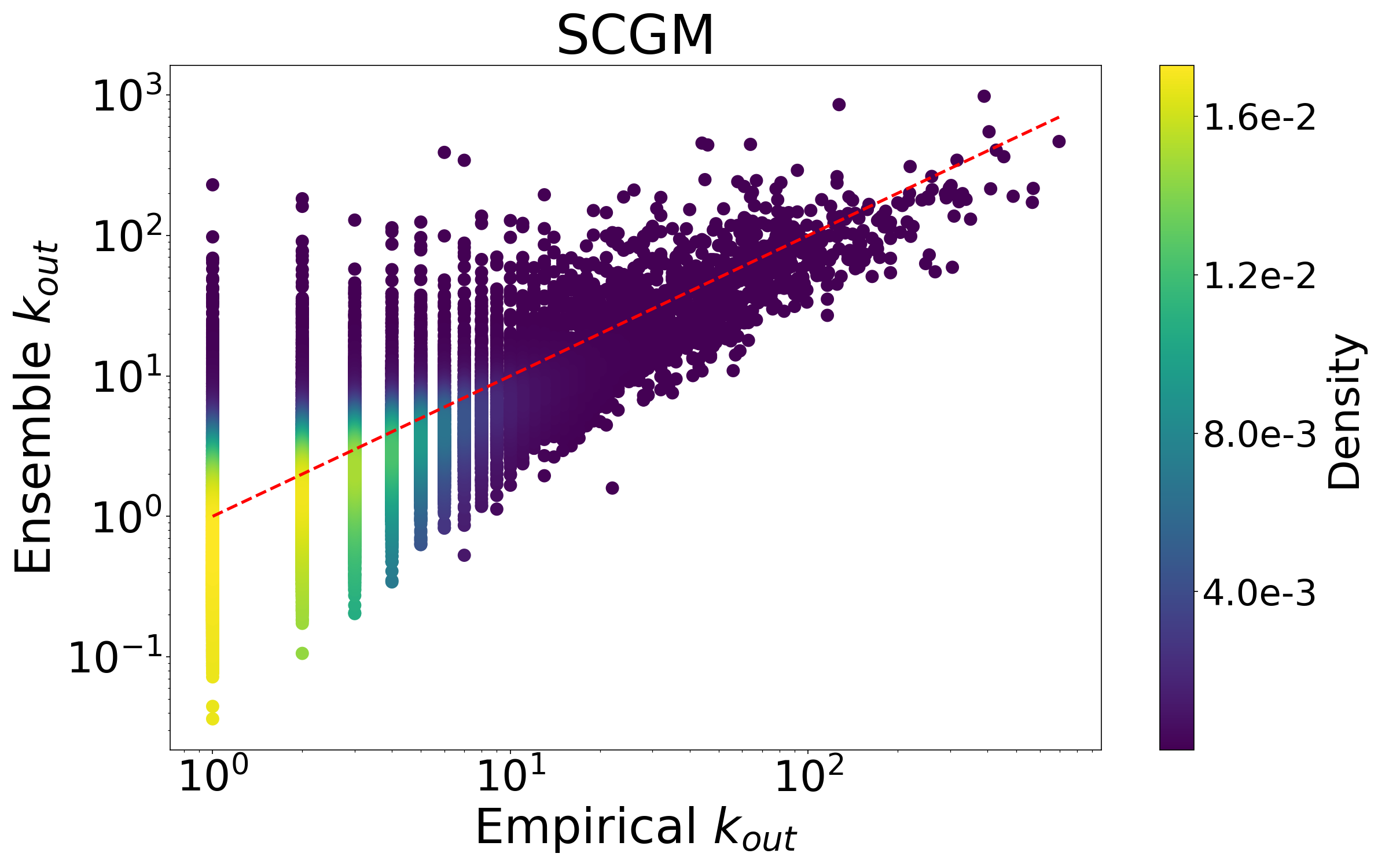}
\end{subfigure}
\begin{subfigure}{0.02\textwidth}
    \textbf{e)}
\end{subfigure}
\begin{subfigure}[t]{0.30\textwidth}
\includegraphics[width=\textwidth,valign=t]{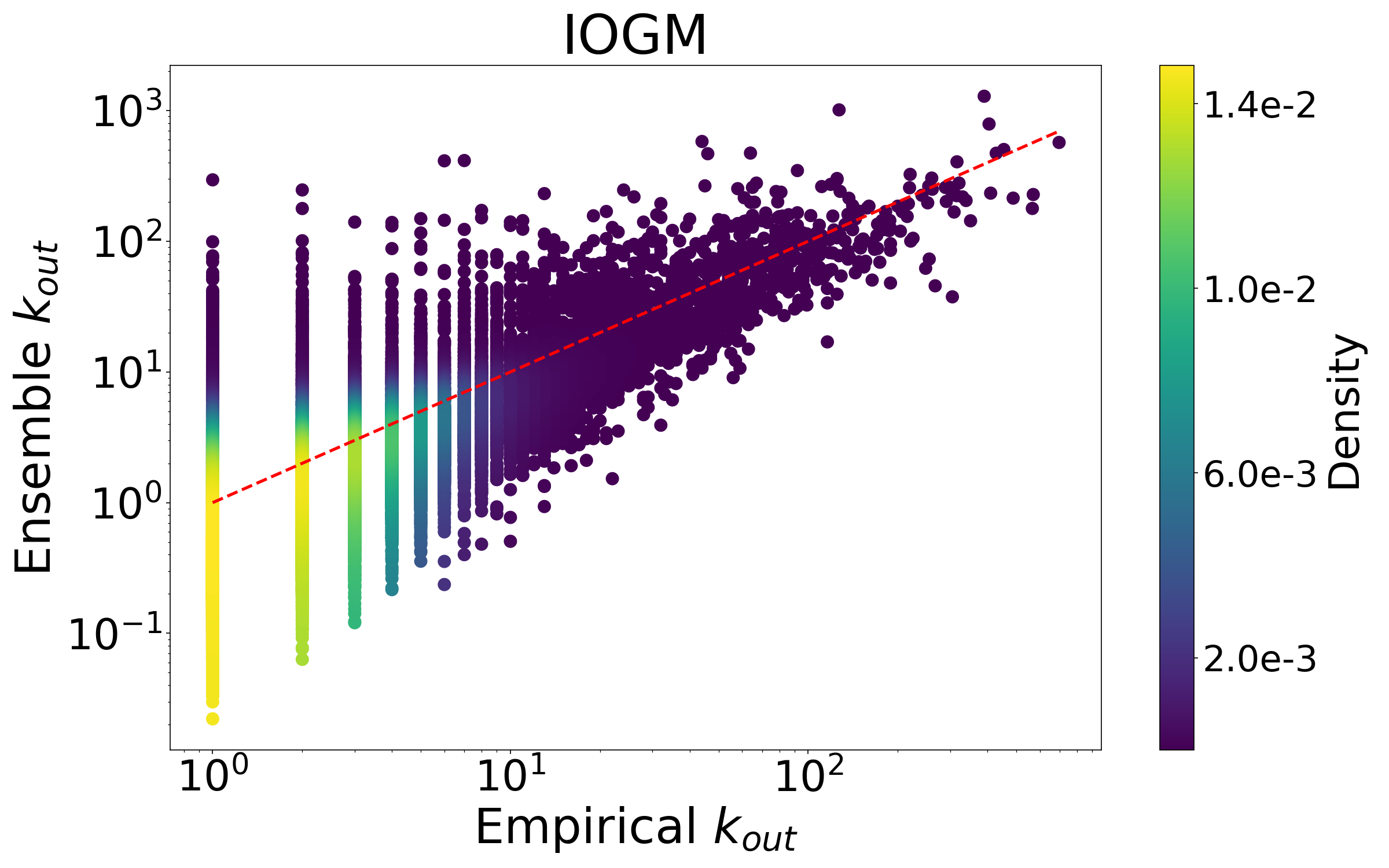}
\end{subfigure}
\begin{subfigure}{0.02\textwidth}
    \textbf{f)}
\end{subfigure}
\begin{subfigure}[t]{0.30\textwidth}
\includegraphics[width=\textwidth,valign=t]{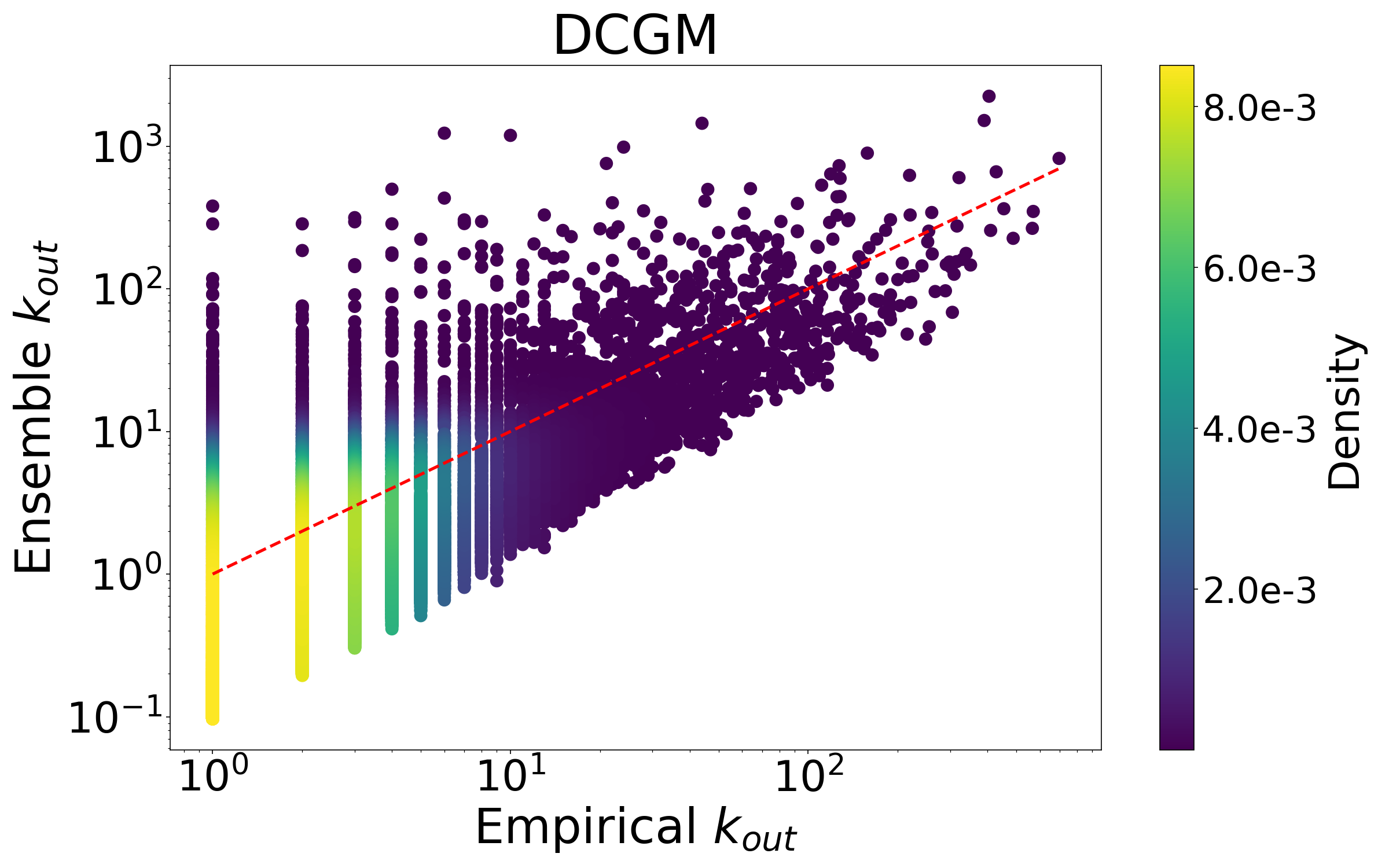}
\end{subfigure}
\caption{Scattering the empirical in-degrees (top) and out-degrees (bottom) versus the reconstructed ones reveals that the SCGM (a,d) yields the best agreement, closely followed by the IOGM (b,e). The DCGM (c,f) is the one scoring worst, being characterised by the largest dispersion around the identity line. The SCMM has not been considered as it does not preserve any topological information about the empirical network.}
\label{fig1}
\end{figure*}

\subsection*{Reconstruction of structural features}

Let us start by assessing to what extent the purely structural properties of the Ecuadorian production network are reproduced by the four reconstruction models considered in the present paper (see Supplementary Materials S2 for the definition of the considered quantities).

First, let us focus on the in- and out-degrees. Although they are not preserved by any model, scattering the empirical degrees versus the reconstructed ones, i.e. $k^\text{in}_i$ versus $\langle k^\text{in}_i\rangle$ and $k^\text{out}_i$ versus $\langle k^\text{out}_i\rangle$, reveals that the various models perform quite satisfactorily in reproducing them. More specifically, the SCGM yields the best agreement, as its cloud of points is the one characterised by the smallest dispersion around the identity line. While the IOGM closely follows the SCGM (see Figure \ref{fig1}), the SCMM has not been considered as it does not preserve any topological information about the empirical network.

\begin{figure*}[t!]
\centering
\begin{subfigure}{0.02\textwidth}
    \textbf{a)}
\end{subfigure}
\begin{subfigure}[t]{0.46\textwidth}
\includegraphics[width=\textwidth, valign=t]{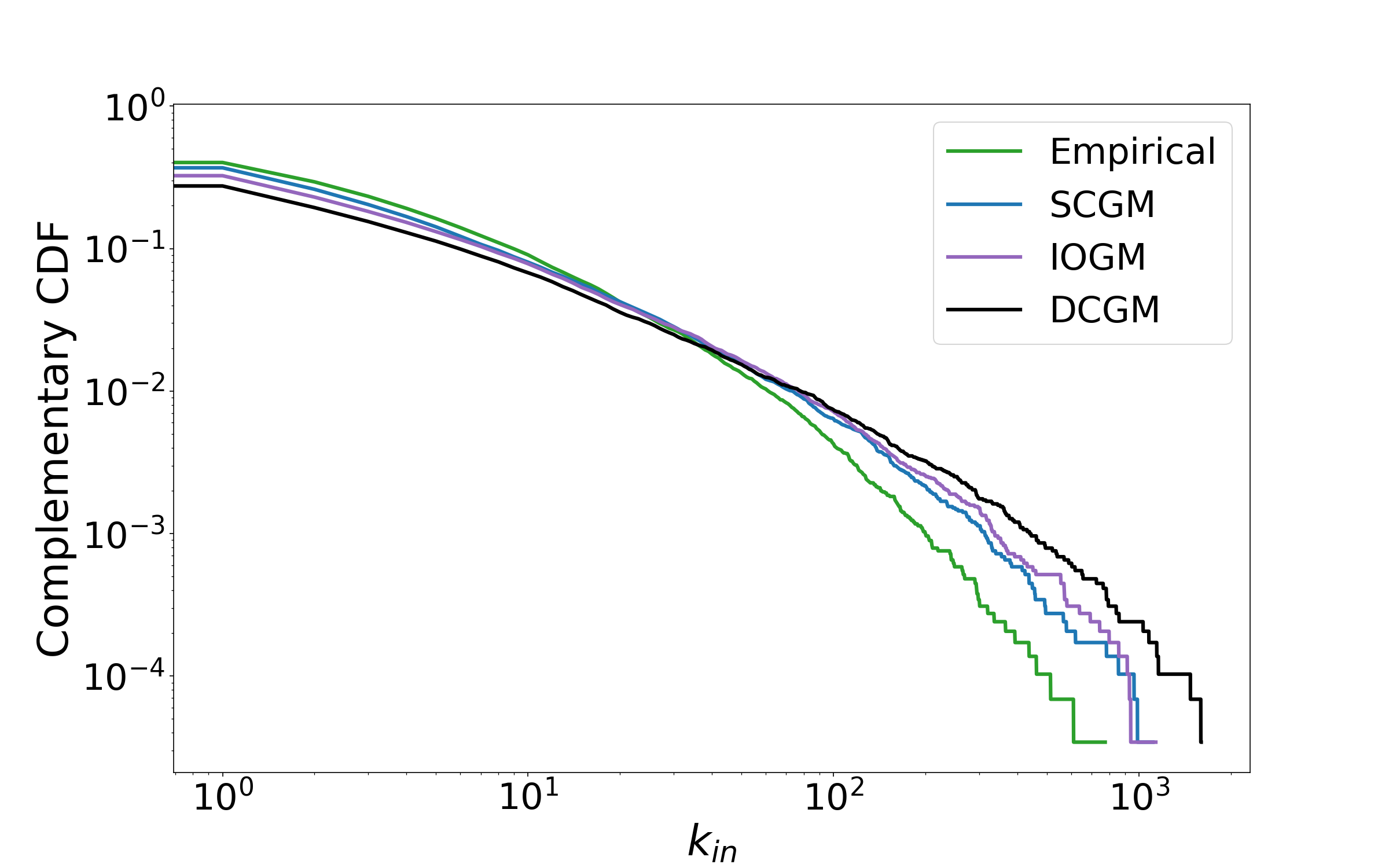}
\end{subfigure}
\begin{subfigure}{0.02\textwidth}
    \textbf{b)}
\end{subfigure}
\begin{subfigure}[t]{0.46\textwidth}
\includegraphics[width=\textwidth, valign=t]{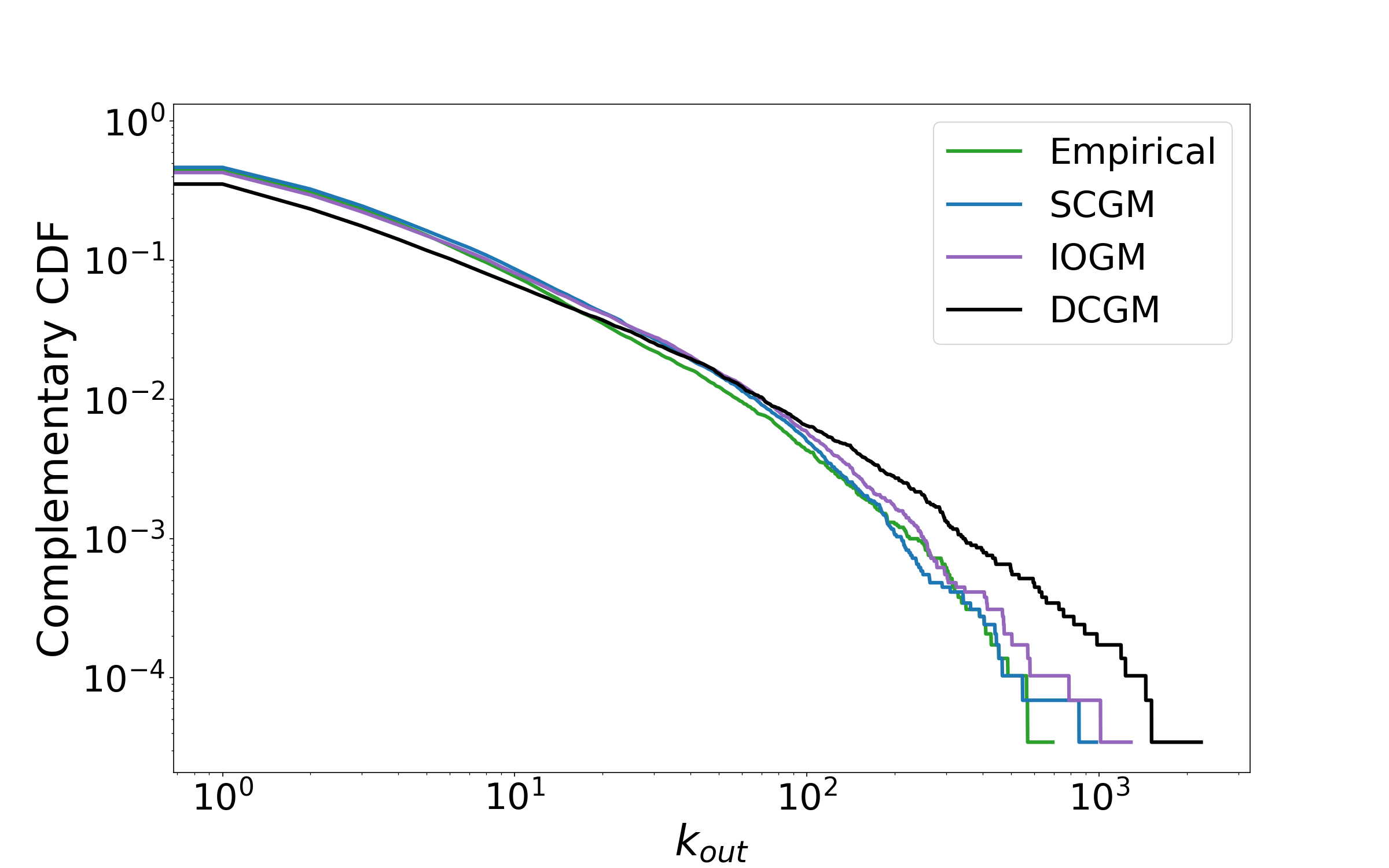}
\end{subfigure}
\caption{Complementary Cumulative Distributions of in-degrees (a) and out-degrees (b) for the Ecuadorian production network and the three reconstruction models not inducing a trivial topology.}
\label{fig2}
\end{figure*}

These results are further confirmed by comparing the distributions of the expected in- and out-degrees with those of the empirical in- and out-degrees (see Figure~\ref{fig2}). Still, all models return degree distributions with a fatter tail, a result indicating that large (small) degrees are overestimated (underestimated) by each of our reconstruction models. Moreover, the skewness of the distribution increases for models with less strict constraints: since the SCGM admits the presence of the link $i\to j$ only if $s_{g_i\to j}\neq 0$, the IOGM admits the presence of the link $i\to j$ only if $s_{g_i\to g_j}\neq 0$ and the DCGM admits the presence of the link $i\to j$ only if $s_i^\text{out}\neq 0$ and $s_j^\text{in}\neq 0$, any firm is potentially `more connected' under the DCGM than under the SCGM.

Moving to considering higher-order topological properties (see Supplementary Materials S3), the analysis of the average nearest neighbours degree (ANND) and of the average nearest neighbours strength (ANNS) reveals that the Ecuadorian production network is disassortative - i.e. firms having more customers (suppliers) tend to be connected with firms having fewer suppliers (customers) and viceversa. The SCGM is able to capture these trends, including the fluctuations of highly central firms, closely followed by the IOGM. Besides, the analysis of the binary, triangular and square clustering coefficients reveals a hierarchical organisation of the Ecuadorian production network, a feature that both the SCGM and the IOGM capture as well.

\subsection*{Reconstruction of the Economic Systemic Risk Index}

ESRI is a firm-level systemic risk indicator that has been recently introduced~\cite{diem2022quantifying}. Briefly speaking, the systemicness of firm $i$ is evaluated by \emph{a)} removing $i$ from the network; \emph{b)} running two, iterative processes accounting for both the upstream - i.e. concerning the demand - and downstream - i.e. concerning the supply - components of the propagating shock; \emph{c)} computing the output reduction experienced by the whole production network as the sum of the output reductions experienced by the individual firms, after the convergence of the two shockwaves (see Methods).

\begin{figure*}[t!]
\centering
\begin{subfigure}{0.02\textwidth}
    \textbf{a)}
\end{subfigure}
\begin{subfigure}[t]{0.47\textwidth}
\includegraphics[width=\textwidth,valign=t]{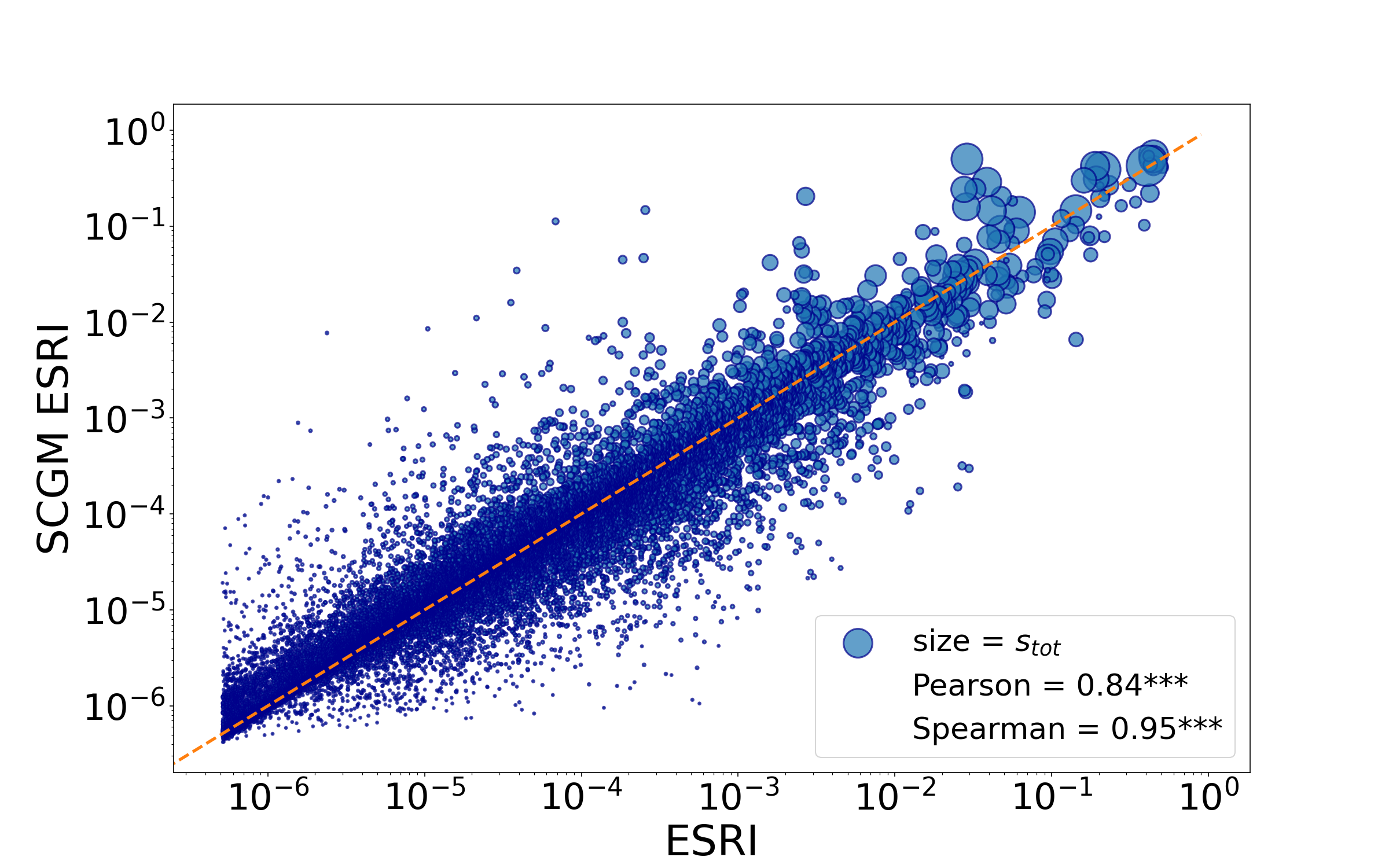}
\end{subfigure}
\begin{subfigure}{0.02\textwidth}
    \textbf{b)}
\end{subfigure}
\begin{subfigure}[t]{0.47\textwidth}
\includegraphics[width=\textwidth,valign=t]{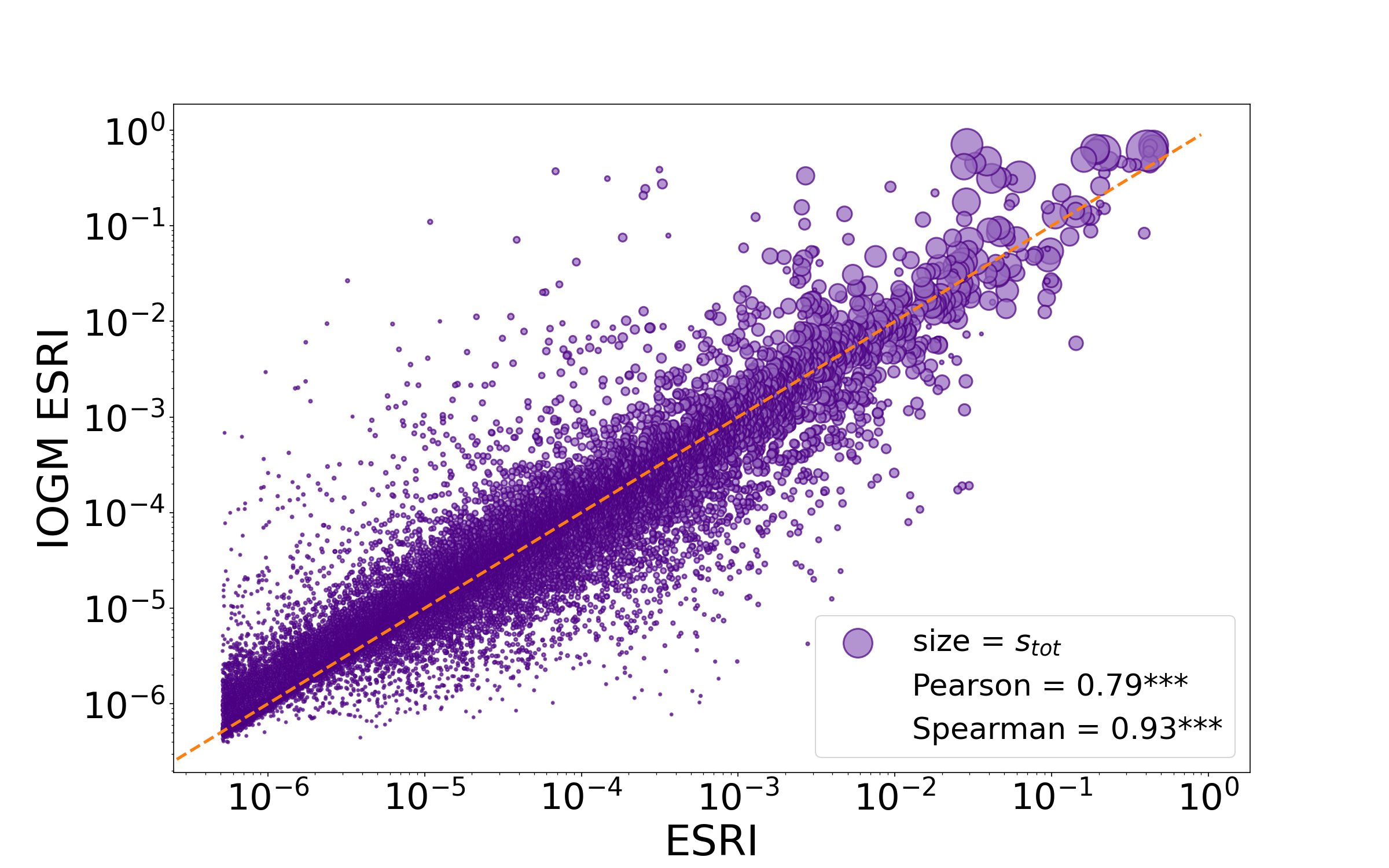}
\end{subfigure}
\begin{subfigure}{0.02\textwidth}
    \textbf{c)}
\end{subfigure}
\begin{subfigure}[t]{0.47\textwidth}
\includegraphics[width=\textwidth,valign=t]{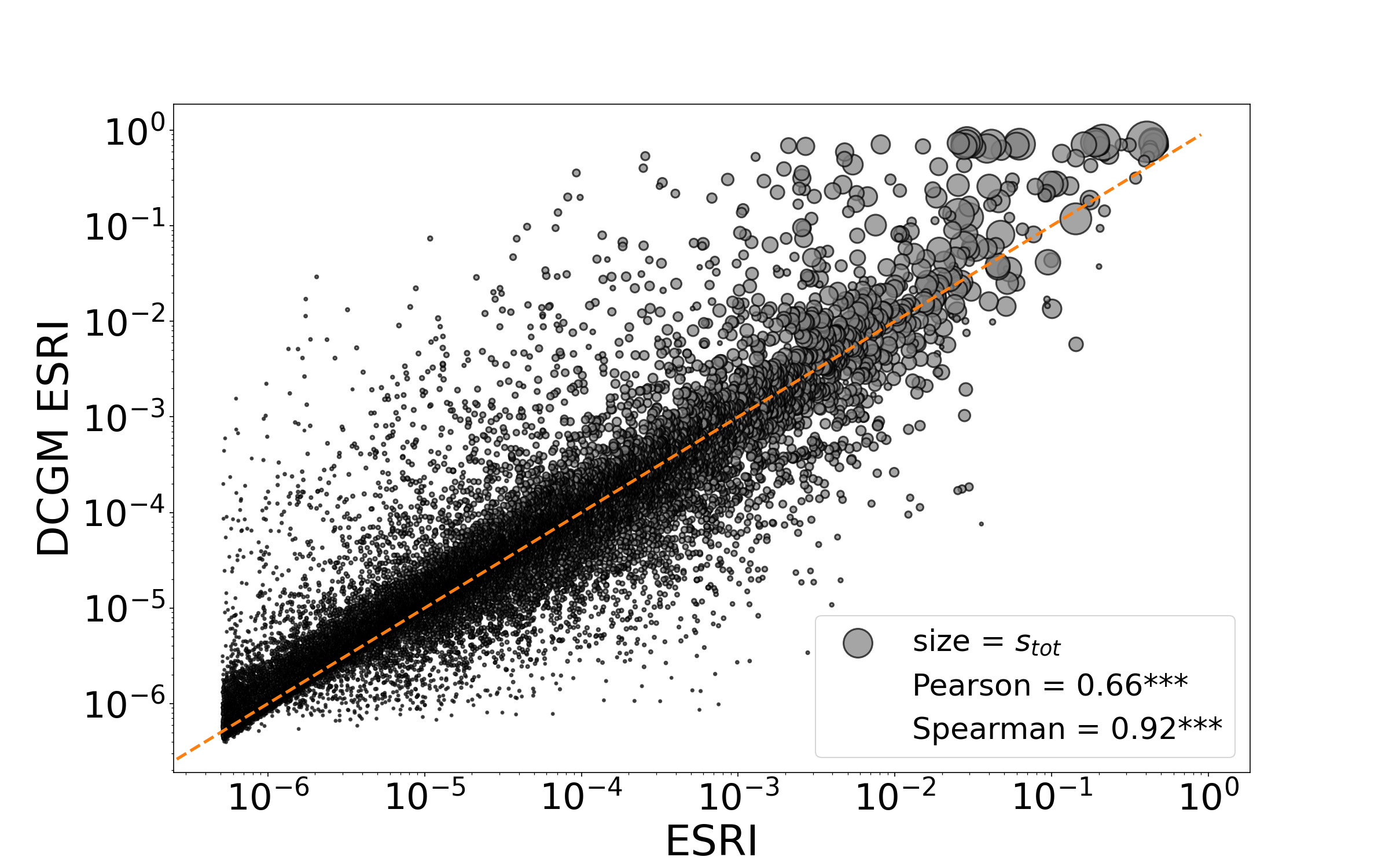}
\end{subfigure}
\begin{subfigure}{0.02\textwidth}
    \textbf{d)}
\end{subfigure}
\begin{subfigure}[t]{0.47\textwidth}
\includegraphics[width=\textwidth,valign=t]{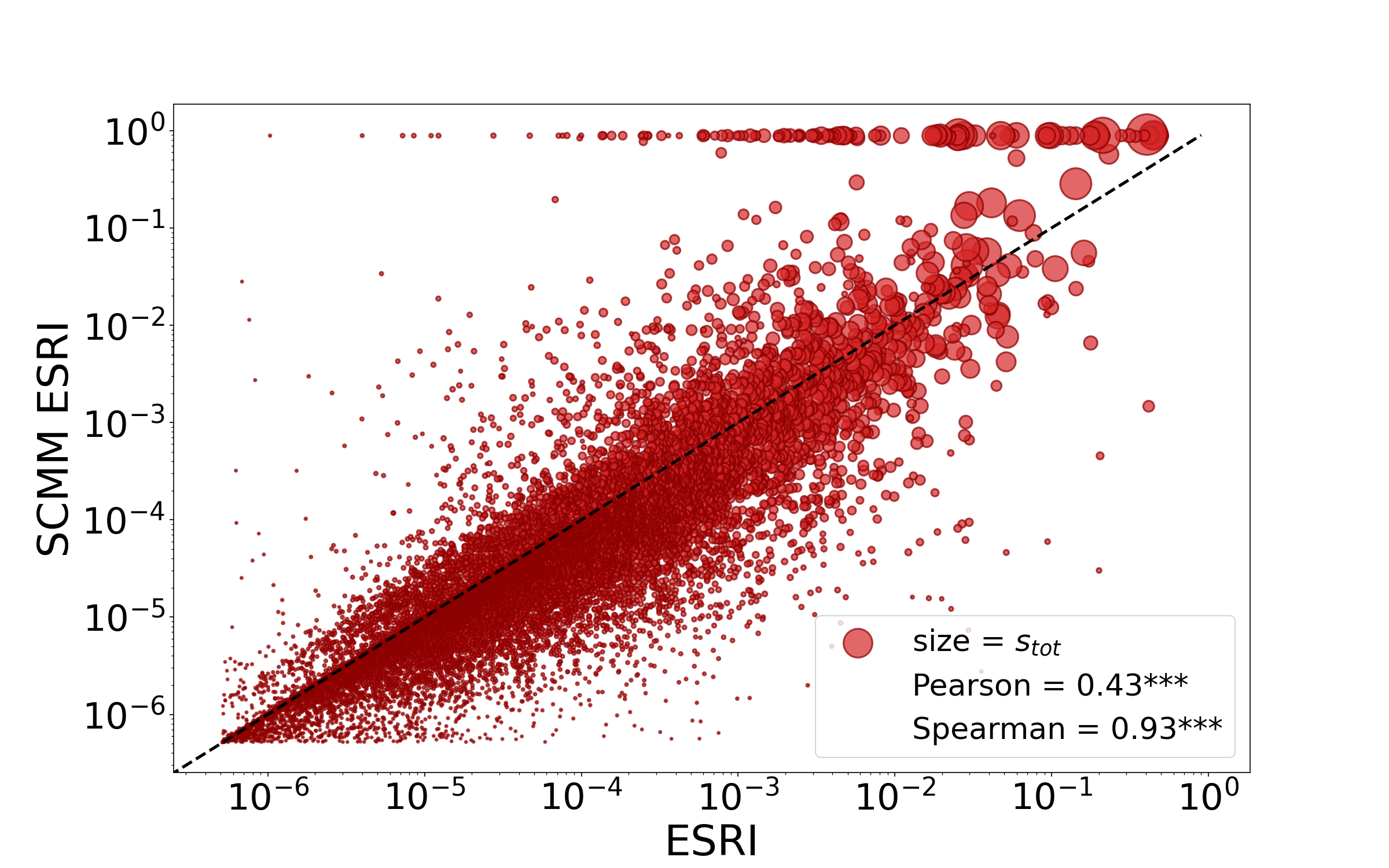}
\end{subfigure}
\caption{Scattering the empirical ESRI values versus the expected ESRI values reveals that the SCGM (a) and the IOGM (b) perform satisfactorily in recovering the former ones: in fact, the Pearson correlation coefficient amounts at $\simeq0.8$ for both models. The DCGM (c) and the SCMM (d), instead, perform worse, being characterised by a larger dispersion of values: in fact, the Pearson correlation coefficient amounts at $\simeq0.7$ for the DCGM and at $\simeq0.4$ for the SCMM.}
\label{fig3}
\end{figure*}

As Figure \ref{fig3} shows, both the ESRI values expected under the SCGM and the ESRI values expected under the IOGM (both evaluated numerically, by averaging over $10^3$ sampled configurations) display a very good agreement with the empirical ESRI values (the Pearson correlation coefficient amounts at $\simeq0.8$ for both models). The DCGM and the SCMM perform worse, being characterised by a larger dispersion of values (the Pearson correlation coefficient amounts at $\simeq0.7$ for the DCGM and $\simeq0.4$ for the SCMM); this is particularly evident in the case of the SCMM, since it `assigns' the maximum ESRI value to a large set of firms whose empirical ESRI value is, instead, quite small.

\begin{figure*}[t!]
\centering
\begin{subfigure}{0.02\textwidth}
    \textbf{a)}
\end{subfigure}
\begin{subfigure}[t]{0.47\textwidth}
\includegraphics[width=\textwidth,valign=t]{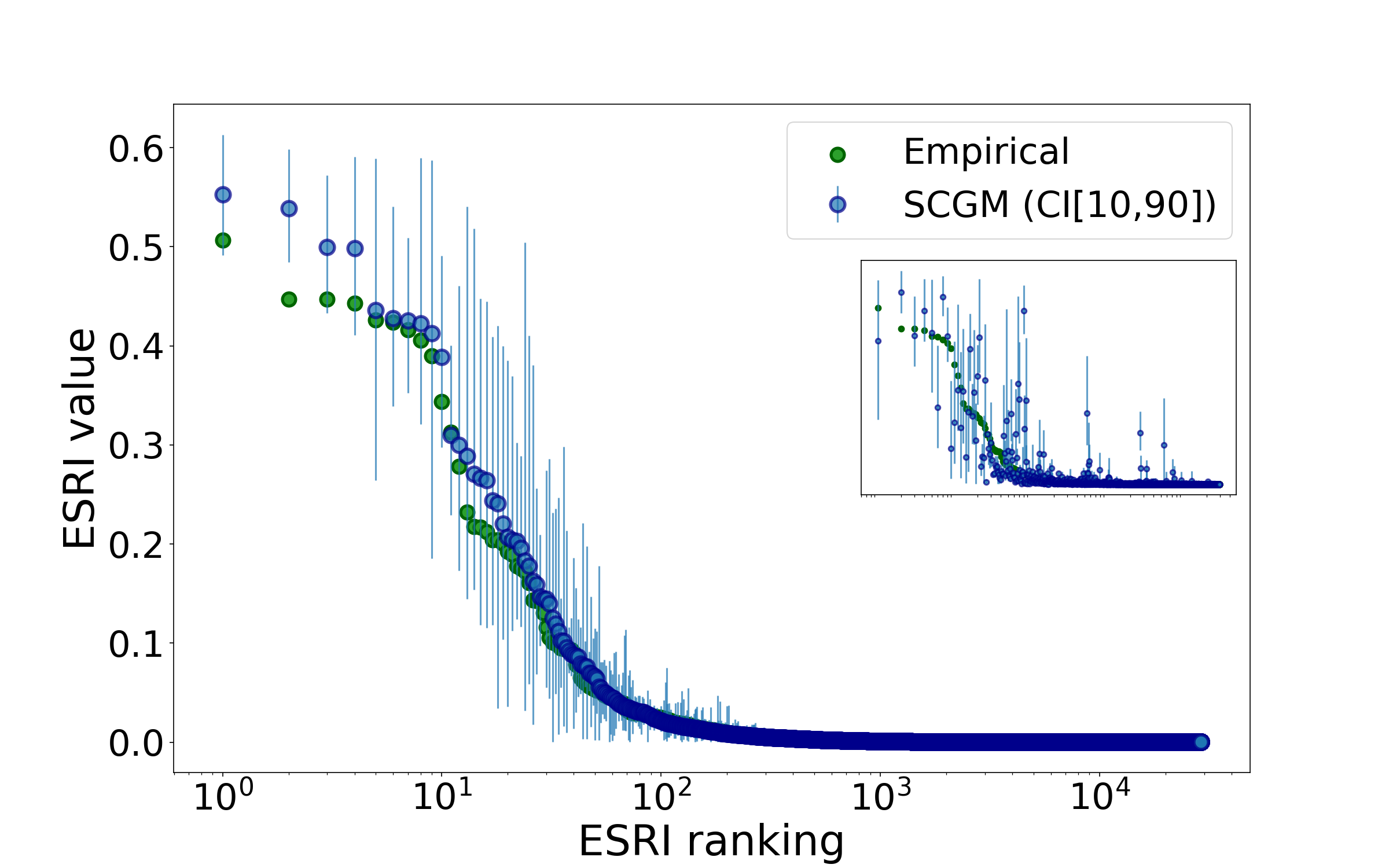}
\end{subfigure}
\begin{subfigure}{0.02\textwidth}
    \textbf{b)}
\end{subfigure}
\begin{subfigure}[t]{0.47\textwidth}
\includegraphics[width=\textwidth,valign=t]{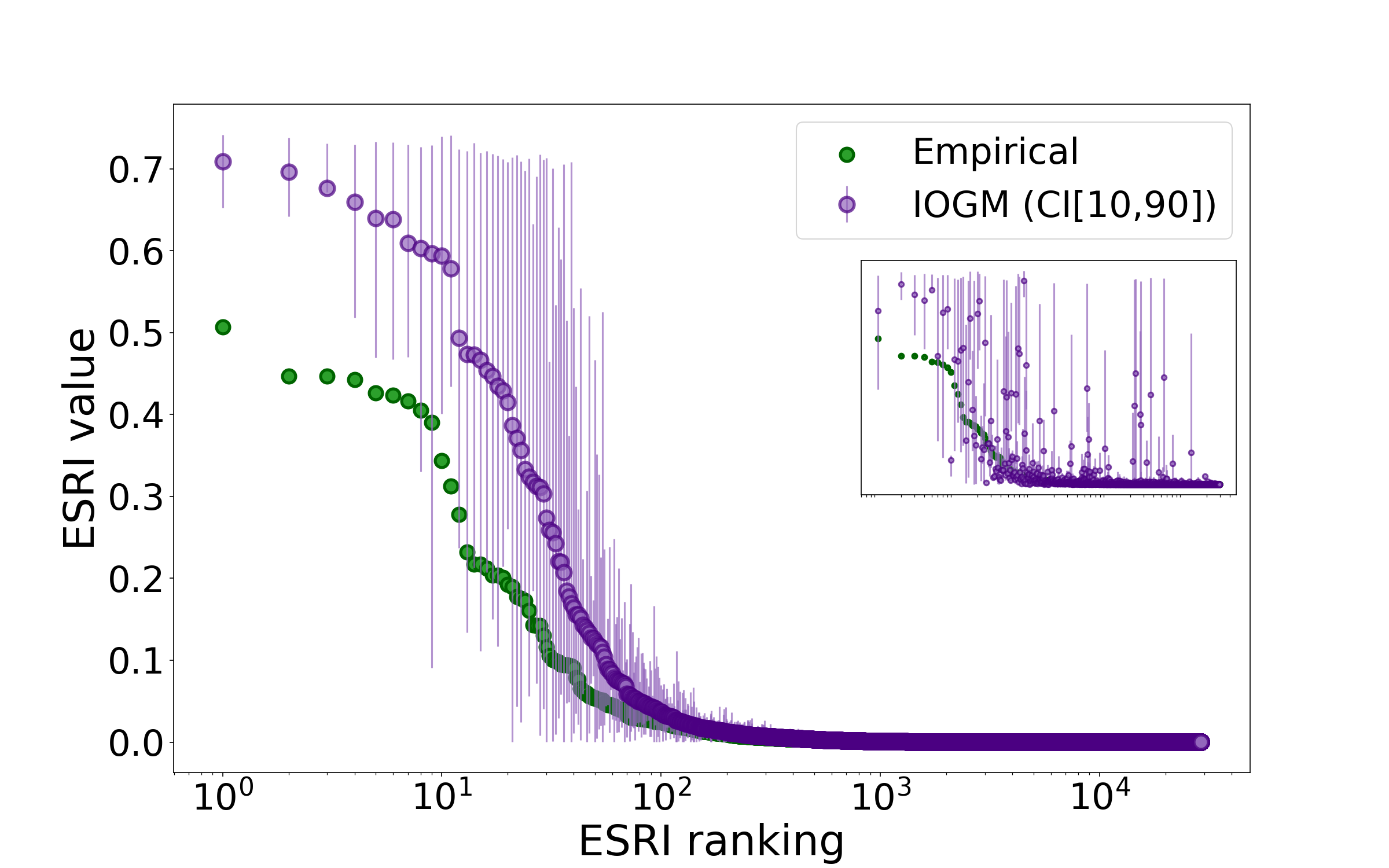}
\end{subfigure}
\begin{subfigure}{0.02\textwidth}
    \textbf{c)}
\end{subfigure}
\begin{subfigure}[t]{0.47\textwidth}
\includegraphics[width=\textwidth,valign=t]{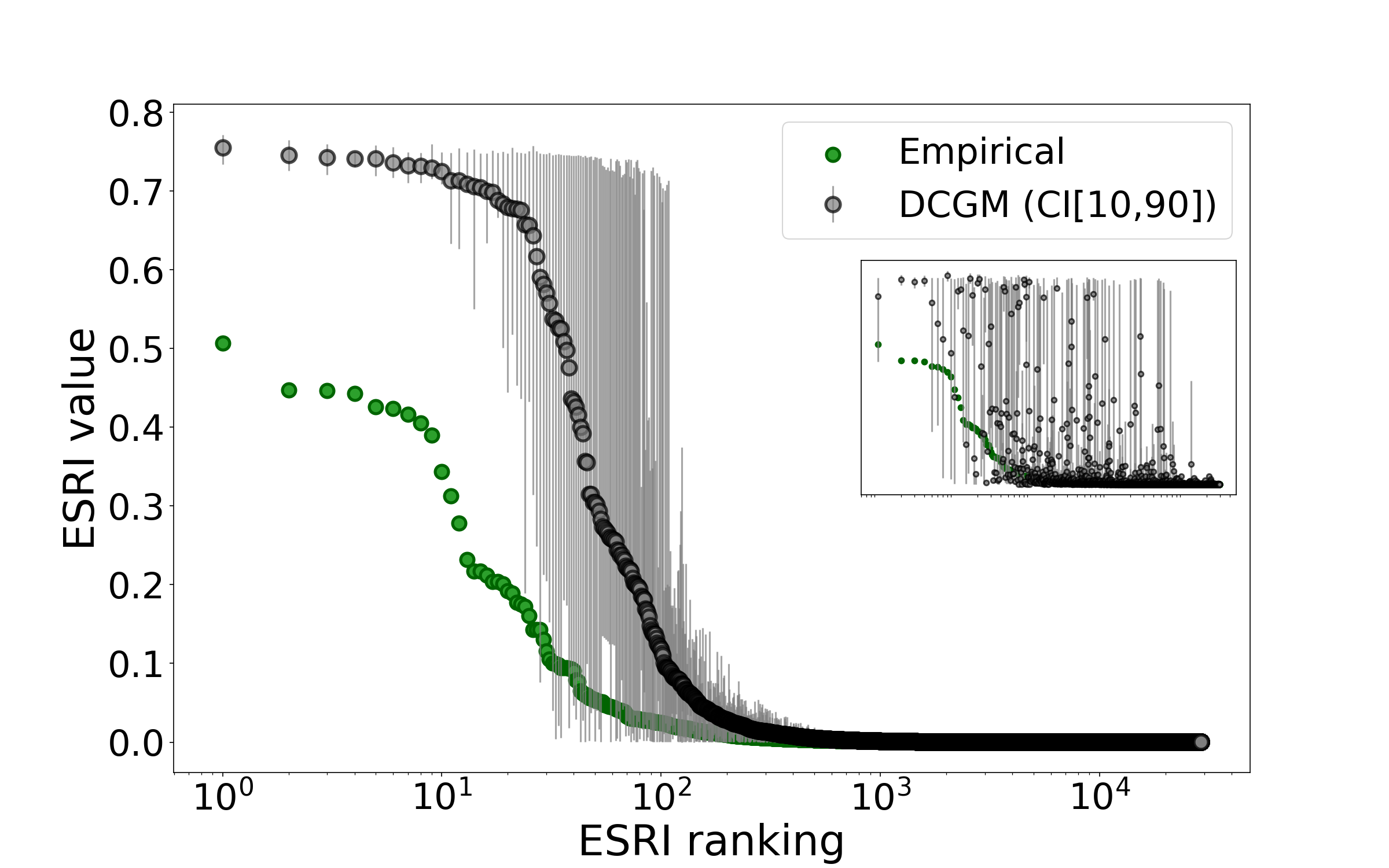}
\end{subfigure}
\begin{subfigure}{0.02\textwidth}
    \textbf{d)}
\end{subfigure}
\begin{subfigure}[t]{0.47\textwidth}
\includegraphics[width=\textwidth,valign=t]{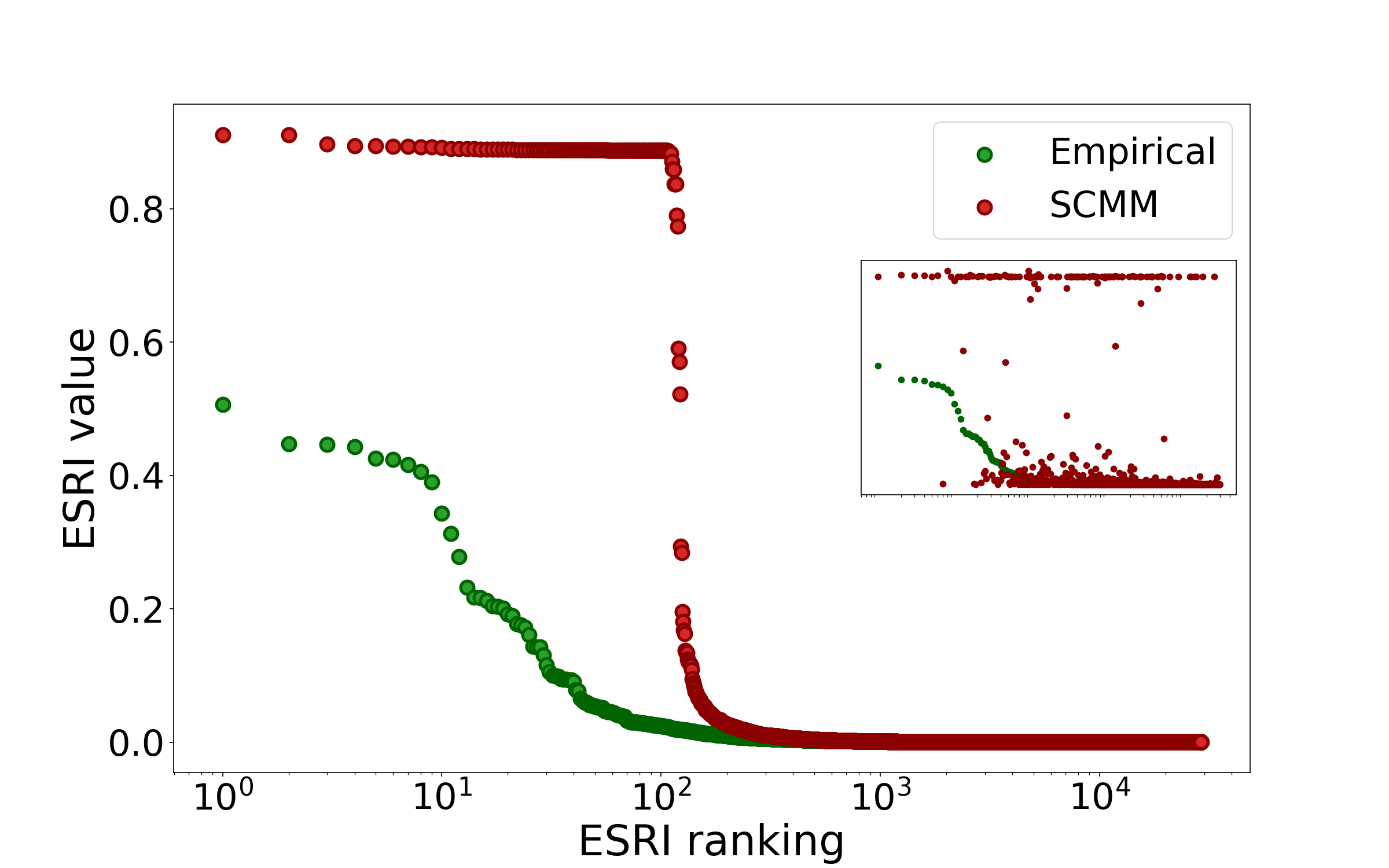}
\end{subfigure}
\caption{Scattering the empirical and the reconstructed ESRI values versus the corresponding rankings reveals that all models reproduce the plateau induced by the riskiest firms and the subsequent, steep decrease of the trend. In fact, the Spearman correlation coefficient is larger than $\simeq0.9$ for both the SCGM (a) and the IOGM (b) and $\simeq0.9$ for both the DCGM (c) and the SCMM (d). The insets show the reconstructed ESRI values, reordered according to the empirical ranking (each abscissa correspond to a firm): the expected ESRI values under the SCGM are scattered around the empirical ones, while the other models overestimate (sometimes even substantially) the ESRI values of all firms.}
\label{fig4}
\end{figure*}

Scattering the empirical and the reconstructed ESRI values versus the corresponding rankings (see Figure \ref{fig4}), reveals that all models correctly recover the plateau induced by the riskiest firms and the subsequent, steep decrease of the trend - even though points with the same abscissa do not necessarily represent the same firm\footnote{The empirical ESRI ranking of a firm can be attributed, by our reconstruction models, to a different firm.}, rankings are highly correlated, as confirmed by the Spearman correlation coefficient between the empirical and the expected ESRI values, larger than $0.9$ for all the models. Still, the height and size of the plateau increase for models with less strict constraints, i.e. less informed models overestimate both the ESRI value of highly risky firms and the number of such firms in the network. Such a problem can be (at least, partially) imputed to the density of the sampled configurations on which ESRI is computed: while the SCGM and the IOGM are designed to preserve the empirical link density \emph{on average}, specific samples can contain disconnected nodes, hence causing the density of the connected component to be larger than the empirical one~\cite{gabrielli2024critical}; this, in turn, leads to the aforementioned overestimation (see Supplementary Materials S5). Consistently, the density of samples reaches its maximum with the SCMM, which admits the presence of all, compatible links with the constraints defining the SCGM.

The superior performance of the SCGM is even more evident when reordering the ESRI values of the models according to the empirical ranking (see the insets of Figure \ref{fig4}): while the expected ESRI values under the SCGM are scattered around the empirical ones, the other models substantially  overestimate the ESRI value of all firms.

\subsubsection*{Analysis of the plateau}

\begin{figure*}[t!]
\centering
\begin{subfigure}{0.02\textwidth}
    \textbf{a)}
\end{subfigure}
\begin{subfigure}[t]{0.47\textwidth}
\includegraphics[width=\textwidth,valign=t]{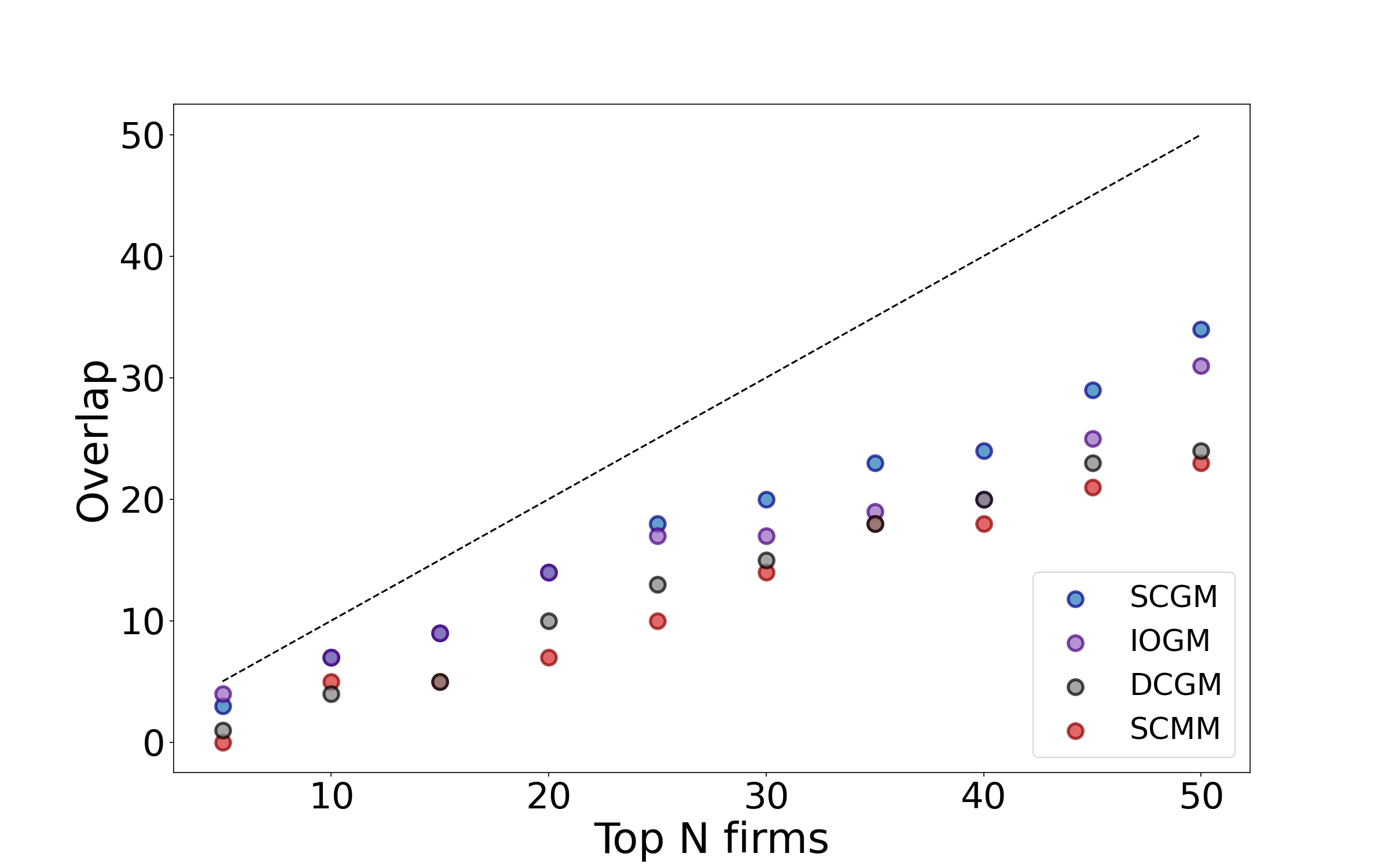}
\end{subfigure}
\begin{subfigure}{0.02\textwidth}
    \textbf{b)}
\end{subfigure}
\begin{subfigure}[t]{0.47\textwidth}
\includegraphics[width=\textwidth,valign=t]{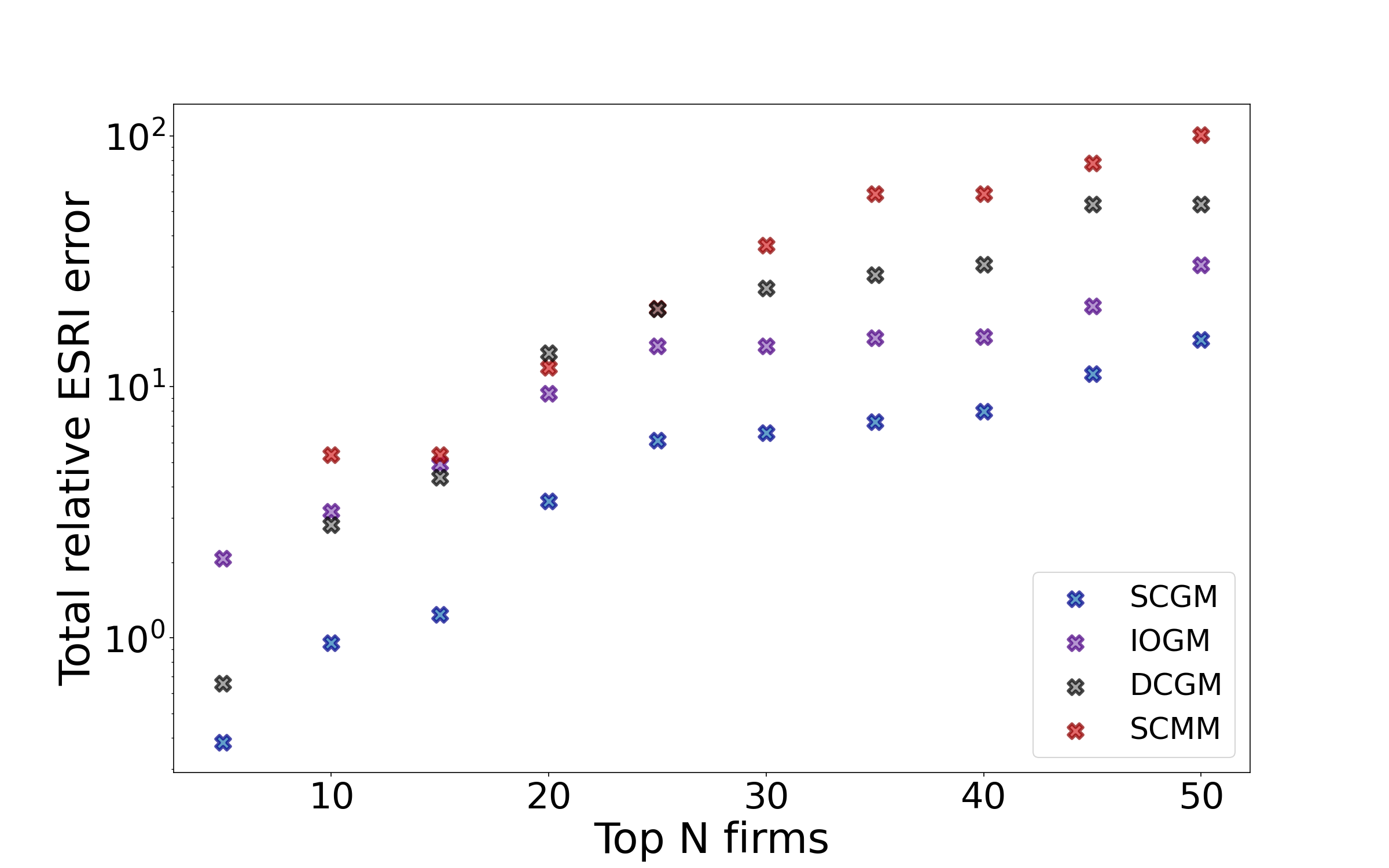}
\end{subfigure}
\caption{Both the SCGM and the IOGM correctly identify practically all the riskiest firms, according to the ranking provided by the empirical ESRI values, up to $N=10$. Although the number of such firms decreases as the ranking is browsed, the SCGM provides the largest overlap for all values of $N$ (a). The total relative error about the ESRI values confirms such a picture, indicating the SCGM as the model scoring best in recovering both the ranking of firms and their ESRI values. The IOGM, instead, performs better in recovering the ranking of the firms than their ESRI values.}
\label{fig5}
\end{figure*}

As the failure of the firms belonging to the plateau (i.e. the 10 firms with highest ESRI values) would induce an output reduction amounting at the $40\%-50\%$ of the total value of the Ecuadorian economy, their identification represents an important test for the performance of our reconstruction models. Overall, the number of highly risky firms (i.e. the ones in the top positions of the ESRI ranking) that are correctly spotted by the models decreases as we browse the ranking; however, both the SCGM and the IOGM correctly identify 7 out of the top 10 (9 out of the top 15) firms, with the SCGM providing the largest overlap for bigger values of $N$ (see Figure \ref{fig5}a). More quantitatively, we can introduce the total relative error about the ESRI values (TRE), reading

\begin{equation}
\text{TRE}_m^N=\sum_{r=1}^N\frac{|\text{ESRI}_r-\langle\text{ESRI}_r\rangle_m|}{\text{ESRI}_r},
\end{equation}
with $r$ indicating the $r$-th ranked firm according to the empirical ESRI values, $N$ indicating the number of firms that is being considered and $m$ indicating the model that is being tested. The superior performance of the SCGM becomes even more evident when scattering $\text{TRE}_m^N$ versus $N$, for each model in our basket (see Figure \ref{fig5}b); in particular, the plot confirms that the IOGM performs better in recovering the ranking of the firms than in reproducing their ESRI values.

\subsubsection*{Analysis of the sector-level ESRI values}

\begin{figure*}[t!]
\centering
\begin{subfigure}{0.02\textwidth}
    \textbf{a)}
\end{subfigure}
\begin{subfigure}[t]{0.47\textwidth}
\includegraphics[width=\textwidth,valign=t]{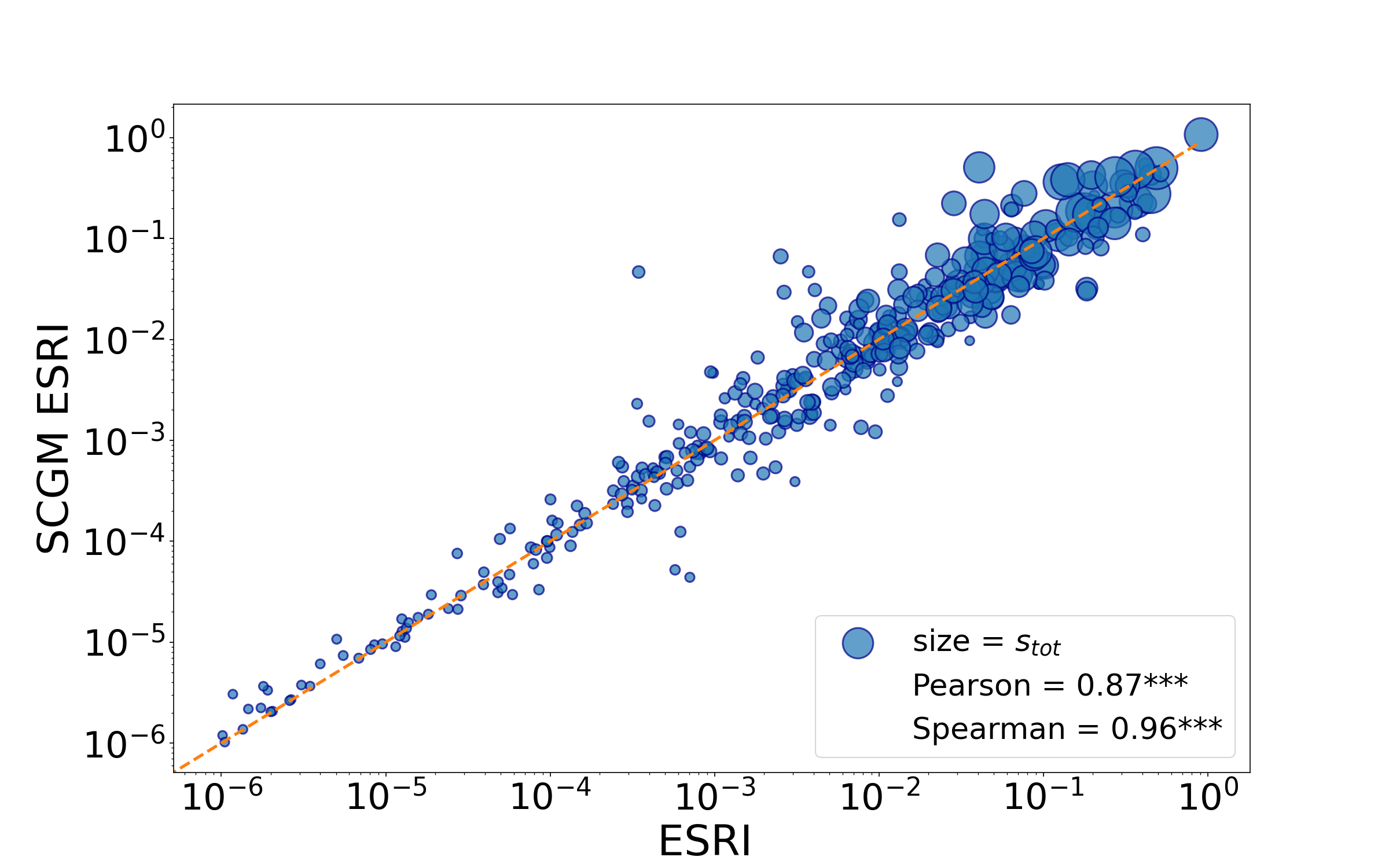}
\end{subfigure}
\begin{subfigure}{0.02\textwidth}
    \textbf{b)}
\end{subfigure}
\begin{subfigure}[t]{0.47\textwidth}
\includegraphics[width=\textwidth,valign=t]{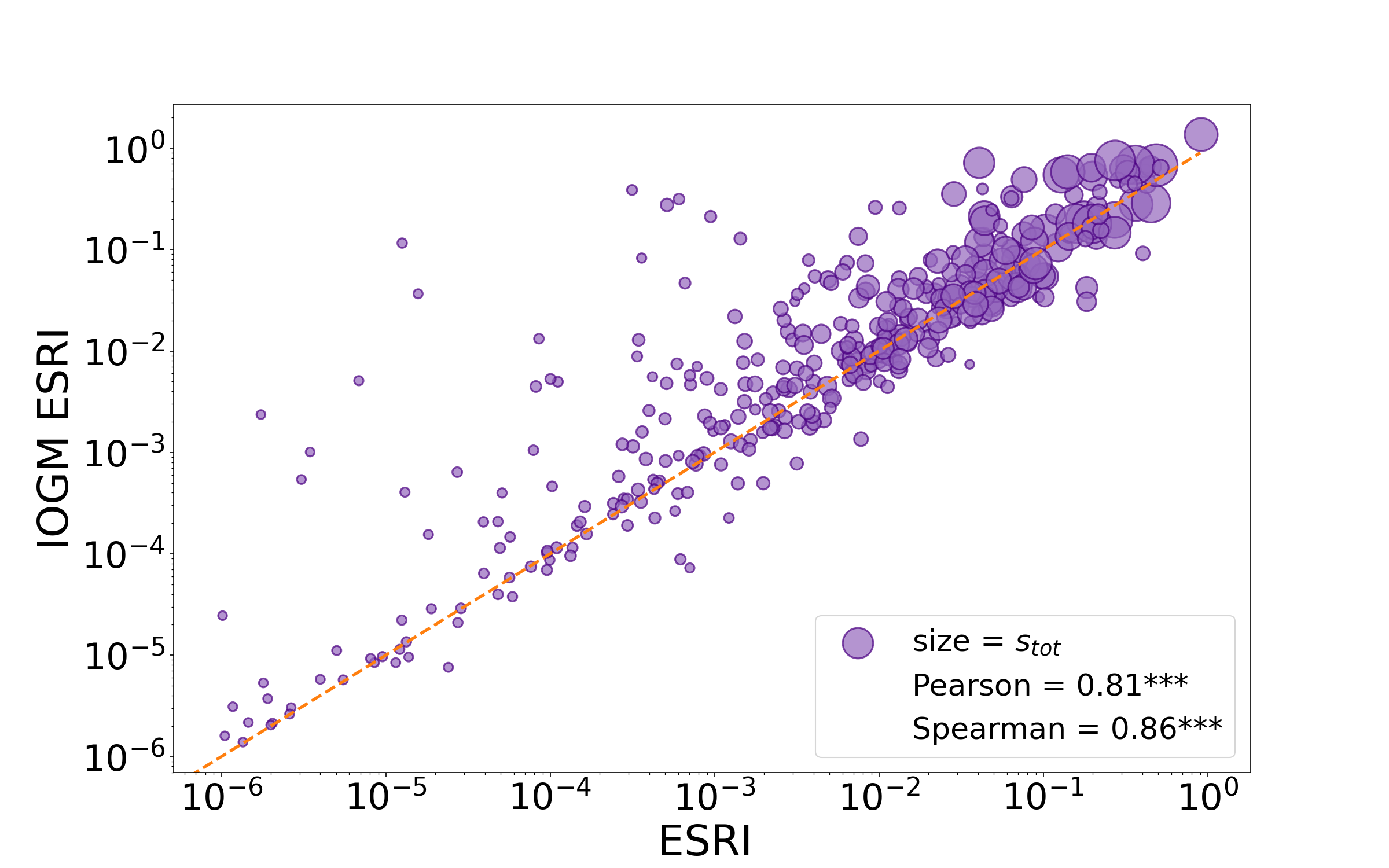}
\end{subfigure}
\begin{subfigure}{0.02\textwidth}
    \textbf{c)}
\end{subfigure}
\begin{subfigure}[t]{0.47\textwidth}
\includegraphics[width=\textwidth,valign=t]{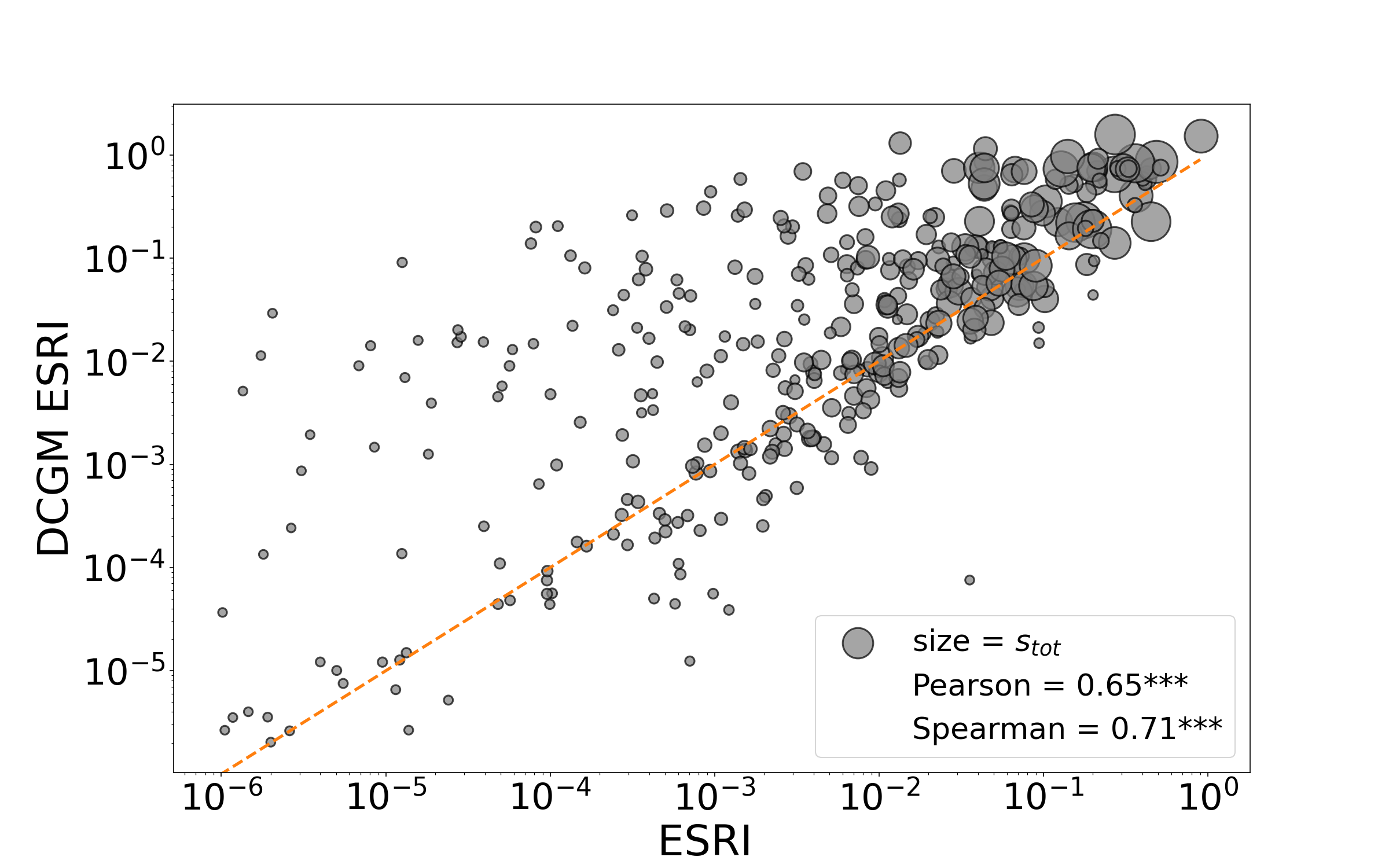}
\end{subfigure}
\begin{subfigure}{0.02\textwidth}
    \textbf{d)}
\end{subfigure}
\begin{subfigure}[t]{0.47\textwidth}
\includegraphics[width=\textwidth,valign=t]{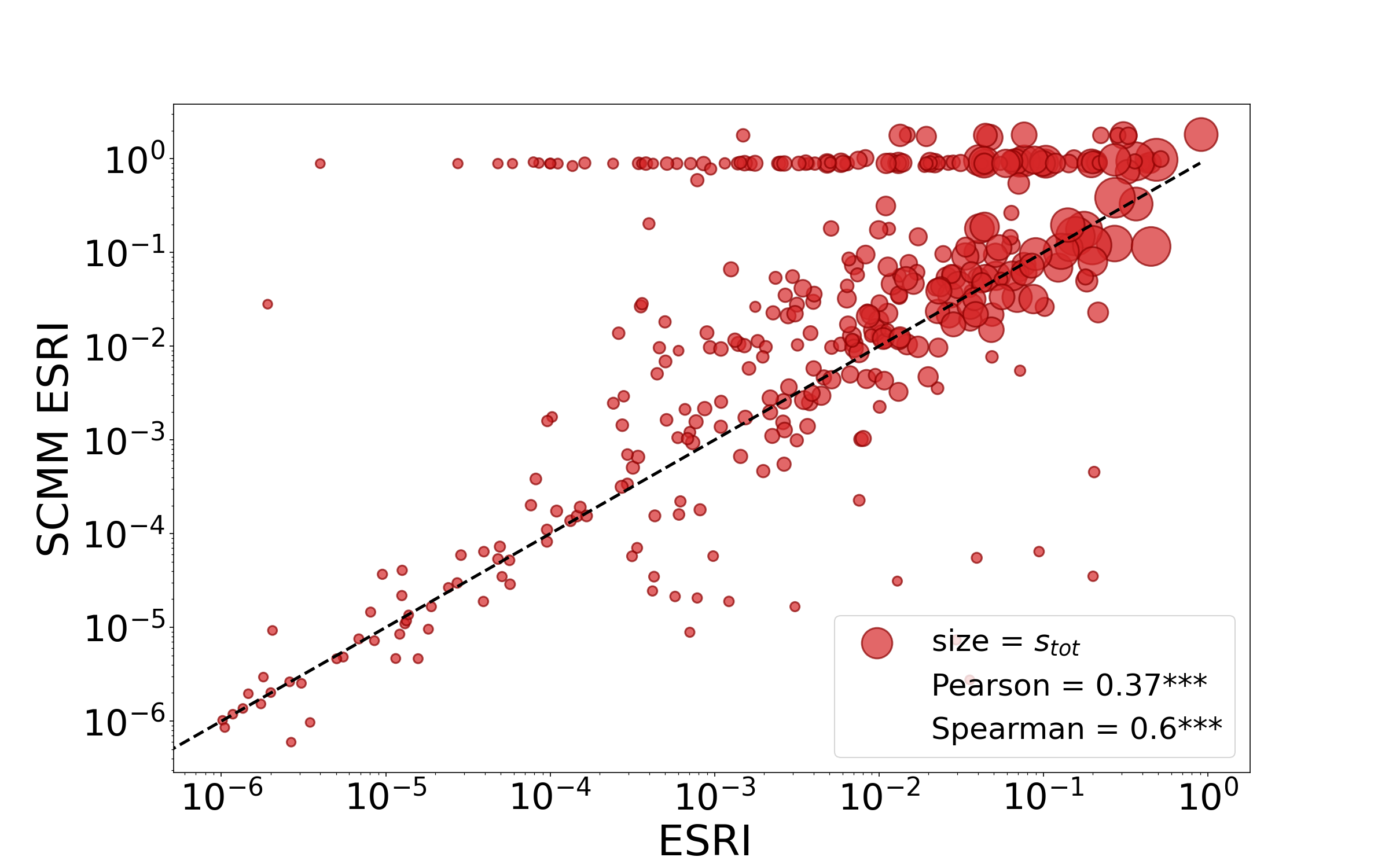}
\end{subfigure}
\caption{Scattering the empirical sector-level ESRI values versus the expected sector-level ESRI values allows us to understand how our reconstruction models perform at the aggregate level constituted by the $371$ industrial sectors returned by the ISIC 4 digits classification. Overall, the sector-level performance of the SCGM (a) is even better than the firm-level performance of the SCGM while the predictive power of the IOGM (b), the DCGM (c) and the SCMM (d) experiences a significant drop.}
\label{fig6}
\end{figure*}

In order to understand how our reconstruction models perform at the sector level, we have also computed the ESRI values of the $371$ industrial sectors, defined as the sum of the ESRI values of the firms they encompass (see Supplementary Materials S1). Since the ESRI of a sector should be computed upon shocking all the firms belonging to that sector simultaneously, our measure can be understood as representing a proxy of it, useful to quantify the performance of our models when aggregating individual firms. As Figure \ref{fig6} shows, scattering the empirical sector-level ESRI values versus the expected sector-level ESRI values returns a quite different picture from the one provided by Figure \ref{fig3}: while the SCGM displays an even better performance, the predictive power of all the other models experiences a significant drop - beside confirming the superior performance of the SCGM, such a result highlights the importance of achieving an accurate reconstruction at the micro-level, in order to be able to achieve an accurate reconstruction at the aggregate level as well.

\section*{Discussion}

Due to the lack of large-scale, publicly available data hindering the possibility of carrying out proper, empirical analyses of firm-level data, the role played by techniques to reconstruct economic networks from partial information is rapidly gaining importance. Although several methods have been proposed so far, their performance has been tested on the ability of correctly predicting the existence of supply links~\cite{brintrup2018predicting,mungo2023reconstructing,mungo2024reconstructing,kosasih2021machine,wichmann2020extracting}, reconstructing purely structural properties~\cite{ialongo2022reconstructing,reisch2022monitoring} and macroscopic, economic quantities in a static framework~\cite{bacilieri2023reconstructing}: however, an explicit assessment about their capability in reproducing the firm-level systemic risk is still lacking~\cite{acemoglu2012net,diem2024estimating,inoue2019firm}. Our contributions aims at filling this gap, by testing the four, different models named \emph{Stripe-Corrected Gravity Model}~\cite{ialongo2022reconstructing}, \emph{Input-Output Gravity Model}, \emph{Density-Corrected Gravity Model}~\cite{cimini2015systemic} and \emph{Stripe-Corrected MaxEnt Model} on the reproduction of the ESRI values of the firms constituting the Ecuadorian production network. What we find is that the models reproducing the empirical in-strength by sector of firms (i.e. the SCGM and, to a lesser extent, the IOGM) lead to an overall accurate estimate of the ESRI values; remarkably, both models are able to correctly identify most of the riskiest companies in the network, proving to be reliable tools to analyze the systemicness of a national production network in absence of topological information. When aggregating ESRI values at the sector level, instead, the SCGM outperforms any other model, signaling the need for a very accurate reconstruction at the micro level in order to retain meaningful aggregate scenarios: this result sheds light on the minimal amount of empirical information that should be disclosed in order to obtain reliable reconstructed configurations.

\section*{Methods}

\subsection*{Data reporting threshold}

VAT-derived data concerning production networks often come with a reporting threshold, $t$, indicating that only transactions exceeding $t$ appear in the data. Such a procedure can be useful to sparsify the network, hence making it computationally easier to handle~\cite{bacilieri2022firm}. In order to evaluate the effects induced by a certain reporting threshold on the network structure, we consider the four, different values $t_1=0\:\$$, $t_2=10.000\:\$$, $t_3=22.300\:\$$ and $t_4=50.000\:\$$. Table \ref{tab:2} reports the main network statistics for each value of $t$. Basically, the largest differences are observed when passing from $t_1$ to $t_2$; similar conclusions can be drawn for what concerns the performance of our reconstruction models (see Supplementary Materials S4): in a nutshell, the performance of the SCGM greatly improves when passing from $t_1$ to $t_2$ but improves only marginally when adopting larger thresholds. This result and our concerns about computational feasibility drove us towards the choice of filtering the data with $t_3$.

\begin{table}[b!]
\centering
\begin{tabular}{lccccc}
\hline
 & Link density & $\langle k\rangle$ & $\langle s\rangle$ & Percentage of total weight & Number of sectors \\
\hline
Full network & $6,0\cdot 10^{-4}$  & $\simeq 36$ & 772.964 & 100\% & 387 \\
$t_2=10.000\:\$$ & $2,0\cdot 10^{-4}$ & $\simeq 6$ & 1.246.587 & 96\% & 375 \\
$\mathbf{t_3=22.300\:\$}$ & $\mathbf{1,5\cdot 10^{-4}}$ & $\mathbf{\simeq 4}$ & \textbf{1.492.342} & \textbf{93\%} & \textbf{371} \\
$t_4=50.000\:\$$ & $1,5\cdot 10^{-4}$ & $\simeq 3$ & 1.934.368 & 89\% & 361 \\
\hline
\end{tabular}
\caption{Basic network statistics for different values of the data reporting threshold.}
\label{tab:2}
\end{table}

\subsection*{Quantities preserved by our reconstruction models}

Let us, now, inspect in more detail the properties of the SCGM and the IOGM. The proofs below are based on the \emph{i)} firm-level relationships $\langle s^\text{out}_i\rangle=\sum_{j(\neq i)}\langle w_{i\to j}\rangle$ and $\langle s^\text{in}_j\rangle=\sum_{i(\neq j)}\langle w_{i\to j}\rangle$; \emph{ii)} the sector-level relationships $\langle s_{g_i\to j}\rangle=\sum_{k\in g_i}\langle w_{k\to j}\rangle$, $\langle s_{i\to g_j}\rangle=\sum_{l\in g_j}\langle w_{i\to l}\rangle$ and $\langle s_{g_i\to g_j}\rangle=\sum_{k\in g_i}\sum_{l\in g_j}\langle w_{k\to l}\rangle$; \emph{iii)} the global relationships $\langle W_{g_i}^\text{out}\rangle=\sum_{g_i}\sum_{k\in g_i}\sum_{l\in g_j}\langle w_{k\to l}\rangle$ and $\langle W_{g_j}^\text{in}\rangle=\sum_{g_j}\sum_{k\in g_i}\sum_{l\in g_j}\langle w_{k\to l}\rangle$.\\

According to Equation \eqref{eq:SCGM}, we have 

\begin{equation}
\langle w_{i\to j}^\text{SCGM}\rangle=\frac{s_i^\text{out}s_{g_i\to j}}{W_{g_i}^\text{out}};
\end{equation}
hence, the following quantities are preserved (on average):

\begin{align}
\langle s_i^\text{out}\rangle&=\sum_{l(\neq i)}\langle w_{il}^\text{SCGM}\rangle=s_i^\text{out}\sum_{l(\neq i)}\frac{s_{g_i\to l}}{W_{g_i}^\text{out}}=s_i^\text{out};\\
\langle s_j^\text{in}\rangle&=\sum_{k(\neq j)}\langle w_{kj}^\text{SCGM}\rangle=\sum_{g_i}s_{g_i\to j}\sum_{\substack{k\in g_i\\k(\neq j)}}\frac{s_k^\text{out}}{W_{g_i}^\text{out}}=s_j^\text{in};\\
\langle s_{g_i\to j}\rangle&=\sum_{\substack{k\in g_i\\k(\neq j)}}\langle w_{kj}^\text{SCGM}\rangle=s_{g_i\to j}\sum_{\substack{k\in g_i\\k(\neq j)}}\frac{s_k^\text{out}}{W_{g_i}^\text{out}}=s_{g_{i}\to j};\\
\langle s_{g_i\to g_j}\rangle&=\sum_{k\in g_i}\sum_{\substack{l\in g_j\\l(\neq k)}}\langle w_{kl}^\text{SCGM}\rangle=
\sum_{k\in g_i}\frac{s_k^\text{out}}{W_{g_i}^\text{out}}\sum_{\substack{l\in g_j\\l(\neq k)}}s_{g_i\to l}=s_{g_i\to g_j}.
\end{align}

According to Equation \eqref{eq:IOscgm}, we have 

\begin{equation}
\langle w_{i\to j}^\text{IOGM}\rangle=\frac{s_i^\text{out}s^{\text{IOGM}}_{g_i\to j}}{W_{g_i}^\text{out}}=\frac{s_i^\text{out}s^{\text{in}}_{j}}{W_{g_i}^\text{out}W_{g_j}^\text{in}}s_{g_i\to g_j};
\end{equation}
hence, the following quantities are preserved (on average):

\begin{align}
\langle s_i^\text{out}\rangle&=\sum_{l(\neq i)}\langle w_{il}^\text{IOGM}\rangle=s_i^\text{out}{}\sum_{g_j}\frac{s_{g_i\to g_j}}{W_{g_i}^\text{out}}\sum_{\substack{l\in g_j\\l(\neq i)}}\frac{s_l^\text{in}}{W_{g_j}^\text{in}}=s_i^\text{out};\\
\langle s_j^\text{in}\rangle&=\sum_{k(\neq j)}\langle w_{kj}^\text{IOGM}\rangle=s_j^\text{in}\sum_{g_i}\frac{s_{g_i\to g_j}}{W_{g_j}^\text{in}}\sum_{\substack{k\in g_i\\k(\neq j)}}\frac{s_k^\text{out}}{W_{g_i}^\text{out}}=s_j^\text{in};\\
\langle s_{g_i\to j}\rangle&=\sum_{\substack{k\in g_i\\k(\neq j)}}\langle w_{kj}^\text{IOGM}\rangle=s_{g_i\to j}^\text{IOGM}\sum_{\substack{k\in g_i\\k(\neq j)}}\frac{s_k^\text{out}}{W_{g_i}^\text{out}}=s_{g_{i}\to j}^\text{IOGM};\\
\langle s_{g_i\to g_j}\rangle&=\sum_{k\in g_i}\sum_{\substack{l\in g_j\\l(\neq k)}}\langle w_{kl}^\text{IOGM}\rangle=s_{g_i\to g_j}
\sum_{k\in g_i}\frac{s_k^\text{out}}{W_{g_i}^\text{out}}\sum_{\substack{l\in g_j\\l(\neq k)}}\frac{s_l^\text{in}}{W_{g_j}^\text{in}}=s_{g_i\to g_j}.
\end{align}

\subsection*{The Economic Systemic Risk Index}

The \emph{Economic Systemic Risk Index} (ESRI), introduced in~\cite{diem2022quantifying}, is intended to quantify the systemicness of a firm by evaluating the output reduction experienced by the whole production network - i.e. the sum of the out-strengths of all nodes - in case of its failure. More in detail, after removing firm $i$ from the network, an upstream and a downstream shocks are propagated to any other firm $j(\neq i)$ through the two, iterative equations

\begin{equation}
x_{i}^{d}(t+1)=\min\left\{\min_{k\in\mathcal{I}_{i}^\text{Ess}}\left\{\frac{1}{\alpha_{ik}}\sum_{j=1}^{n}W_{ji}h_{j}^{d}(t)\delta_{p_{j},k}\right\},\beta_{i}+\frac{1}{\alpha_{i}}\sum_{k\in\mathcal{I}_{i}^\text{Not Ess}}\sum_{j=1}^{n}W_{ji}h_{j}^{d}(t)\delta_{p_{j},k}\right\}
\end{equation}
\label{eq:esri_downstream}
\begin{equation}
x_{i}^{u}(t+1)=\sum_{j=1}^{n}W_{ij}h_{j}^{u}(t)
\end{equation}
\label{eq:esri_upstream}
where $t$ indicates the time step of the propagation, $x_{i}^{d}(t)$ is the out-strength of firm $i$ at time $t$, following the propagation of the downstream shock and $x_{i}^{u}(t)$ is the out-strength of firm $i$ at time $t$, following the propagation of the upstream shock, with $x_i^d(0)=x_i^u(0)=s_i^\text{out}$, $h_i^{d}(t)=x_i^d(t)/s_i^\text{out}$, $h_i^u(t)=x_i^u(t)/s_i^\text{out}$. The shock propagation is framed within the picture provided by a generalised Leontief production function: inputs from \emph{essential} sectors (i.e. the sectors $k\in\mathcal{I}_{i}^\text{Ess}$) set a hard constraint on the output of firm $i$ (enforced through the presence of the minimum) while the inputs from \emph{non-essential} sectors (i.e. the sectors $k\in\mathcal{I}_{i}^\text{Not Ess}$) are treated in a linear way. The technical coefficients of the production function are calibrated on the empirical network: specifically, $\alpha_{ik}=\frac{\sum_{j=1}^nW_{ji}\delta_{p_j,k}}{\sum_{l=1}^nW_{il}}$ and $\alpha_i=\frac{\sum_{j=1}^nW_{ji}}{\sum_{l=1}^nW_{il}}$ while $\beta_{i}=\frac{[\sum_{k\in\mathcal{I}_{i}^\text{Ess}}\sum_{j=1}^nW_{ji}\delta_{p_j,k}}{\sum_{j=1}^nW_{ji}}\sum_{l=1}^nW_{il}$ represents the fraction of output of firm $i$ that can be produced with essential inputs only. The distinction between \emph{essential} and \emph{non-essential} inputs is derived from~\cite{pichler2021and}. After the two, independent shocks have converged at $t=t^*$, the residual fraction of output of firm $i$ is computed as $h_i(t^*)=\min\{h_i^{d}(t^*),h_i^{u}(t^*)\}$ and the ESRI value of firm $i$ is given by

\begin{equation}
\text{ESRI}_i=\sum_{j(\neq)i}\frac{s_j^\text{out}}{\sum_ks_k^\text{out}}(1 - h_j(t^*)).
\end{equation}

\bibliography{references}

\section*{Acknowledgements}
This work is supported by: the European Union - NextGenerationEU - National Recovery and Resilience Plan (Piano Nazionale di Ripresa e Resilienza, PNRR), project `SoBigData.it - Strengthening the Italian RI for Social Mining and Big Data Analytics' - Grant IR0000013 (n. 3264, 28/12/2021); `NetRes - Network analysis of economic and financial resilience', Italian DM n. 289, 25-03-2021 (PRO3 Scuole), CUP D67G22000130001 (\url{https://netres.imtlucca.it}); `RENet - Reconstructing economic networks: from physics to machine learning and back', MUR PRIN 2022MTBB22 funded by European Union – Next Generation EU; `C2T - From Crises to Theory: towards a science of resilience and recovery for economic and financial systems', MUR PRIN PNRR P2022E93B8 funded by European Union – Next Generation EU. DG acknowledges support from the Dutch Econophysics Foundation (Stichting Econophysics, Leiden, the Netherlands) and the Netherlands Organization for Scientific Research (NWO/OCW).
The authors thank the Centro de Estudios Fiscales of Ecuador’s Servicio de Rentas Internas (SRI), which provided the data for research purposes.
The authors thank the Centro de Estudios Fiscales of Ecuador’s Servicio de Rentas Internas (SRI), which provided the data for research purposes.

\section*{Author contributions statement}

Study conception and design: MF, GC, TS, ST, DG. Data collection: MF, PE, ST. Data analysis: MF. Discussion and interpretation of results: MF, GC, TS, PE, ST, DG. Draft manuscript preparation: MF, GC, TS, PE, ST, DG.

\section*{Competing interests}

The authors declare no competing interests.

\clearpage

\section*{\huge Supplementary Materials}

\renewcommand{\thesection}{S\arabic{section}}
\renewcommand{\thetable}{S\arabic{table}}
\renewcommand{\thefigure}{S\arabic{figure}}

\bigskip

\bigskip

\section{Data cleaning}

Raw data concerning the Ecuadorian production network in 2008 contains $8.467.187$ transactions among $4.270.735$ firms, classified into $425$ sectors at the ISIC 4 digits level. For what concerns the nodes: $3.942.095$ firms are labelled as \emph{personas naturales}, i.e. `private people'; $85$ firms ($59$ of which fall into the previous category) have an ISIC sector reported as \emph{NOTI?}, i.e. `not identified'; $72$ firms ($38$ of which fall into the first category) have an ISIC sector reported as \emph{99999}, i.e. `not specified'. For what concerns the links: $668.621$ transactions involve a firm listed as \emph{ND}, i.e. `foreign company'; $20.128$ transactions are self-loops; $92.439$ transactions have a reported tax value equal to $0$. Upon erasing all these nodes and links, we are left with $2.189.081$ transactions between $60.515$ firms, corresponding to $387$ sectors at the ISIC 4 digits level. Finally, upon focusing on the largest, weakly connected component, we are left with $2.189.066$ transactions between $60.488$ firms. Table \ref{tab:1} reports the names of the $22$ sectors at the ISIC 1 digit level, together with a brief description and the corresponding percentage of firms.

\begin{table}[h!]
\centering
\begin{tabular}{lll}
\hline
ISIC 1 digit level &                                        Description & \% firms \\
\hline
A      &                  Agriculture, forestry and fishing & 4.53\% \\
B      &                               Mining and quarrying & 0.70\% \\
C      &                                      Manufacturing & 7.00\% \\
D      &  Electricity, gas, steam and air conditioning supply & 0.16\% \\
E      &  Water supply; sewerage, waste management and remediation activities & 0.23\% \\
F      &                                       Construction & 5.06\% \\
G      &  Wholesale and retail trade; repair of motor vehicles and motorcycles & 21.8\% \\
H      &                         Transportation and storage & 5.57\% \\
I      &          Accommodation and food service activities & 1.74\% \\
J      &                      Information and communication & 2.24\% \\
K      &                 Financial and insurance activities & 3.93\% \\
L      &                             Real estate activities & 8.29\% \\
M      &  Professional, scientific and technical activities & 9.17\% \\
N      &      Administrative and support service activities & 5.38\% \\
O      &  Public administration and defence; compulsory social security & 5.88\% \\
P      &                                          Education & 5.31\% \\
Q      &            Human health and social work activities & 4.69\% \\
R      &                 Arts, entertainment and recreation & 1.13\% \\
S      &                           Other service activities & 7.01\% \\
T      &  Activities of households as employers & 0.003\% \\
U      &  Activities of extraterritorial organizations and bodies & 0.09\% \\
V      &                               No economic activity & 0.04\% \\
\hline
\end{tabular}
\caption{Percentages of firms in the Ecuadorian production network belonging to the $22$ sectors at the ISIC 1 digit level. The wholesale trade is the most populated sector, accounting for $\simeq22\%$ of all companies.}
\label{tab:1}
\end{table}

\clearpage

\section{Network statistics}

\subsection*{Degrees and strengths}

Given a weighted, directed network described by the adjacency matrix $W_{ij}\equiv\{w_{i\to j}\}$, the in-degree and out-degree of node $i$ read

\begin{align}
k_i^\text{in} &=\sum_{j(\neq i)}a_{j\to i},\\
k_i^\text{out}&=\sum_{j(\neq i)}a_{i\to j}
\end{align}
where $a_{i\to j}=\Theta[w_{i\to j}]$. The corresponding expected values read

\begin{align}
\langle k_i^\text{in}\rangle    &=\sum_{j(\neq i)}p_{j\to i},\\
\langle k_{i}^\text{out}\rangle &=\sum_{j(\neq i)}p_{i\to j}.
\end{align}

Analogously, the in-strength and out-strength of node $i$ read

\begin{align}
s_i^\text{in}  &=\sum_{j(\neq i)}w_{j\to i},\\
s_i^\text{out} &=\sum_{j(\neq i)}w_{i\to j}
\end{align}
and the corresponding expected values read

\begin{align}
\langle s_i^\text{in}\rangle  &=\sum_{j(\neq i)}\langle w_{j\to i}\rangle,\\
\langle s_i^\text{out}\rangle &=\sum_{j(\neq i)}\langle w_{i\to j}\rangle.
\end{align}

\subsection*{Binary assortativity coefficients}

\begin{figure*}[t!]
\centering
\begin{subfigure}{0.02\textwidth}
    \textbf{a)}
\end{subfigure}
\begin{subfigure}[t]{0.47\textwidth}
\includegraphics[width=\textwidth,valign=t]{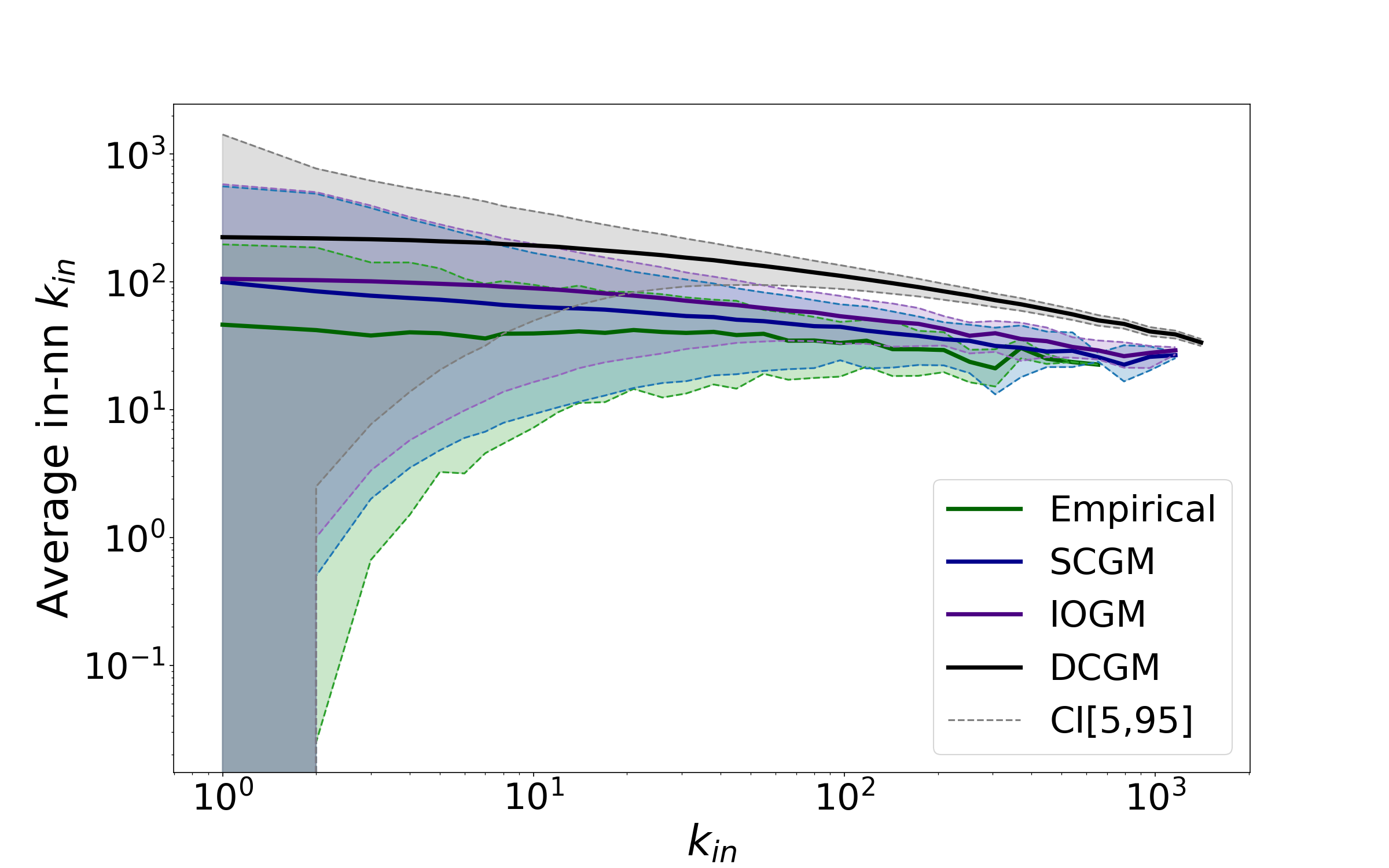}
\end{subfigure}
\begin{subfigure}{0.02\textwidth}
    \textbf{b)}
\end{subfigure}
\begin{subfigure}[t]{0.47\textwidth}
\includegraphics[width=\textwidth,valign=t]{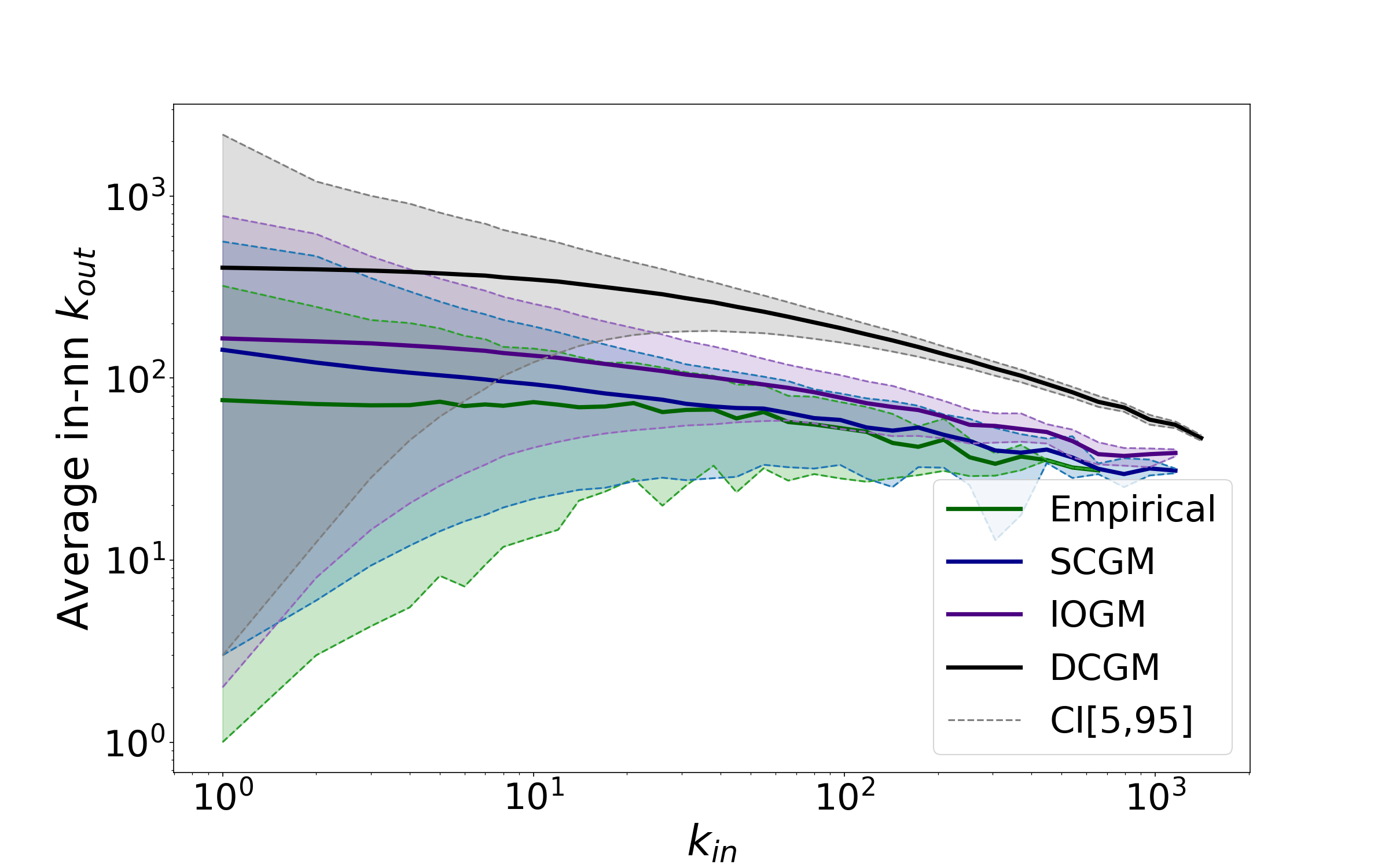}
\end{subfigure}
\begin{subfigure}{0.02\textwidth}
    \textbf{c)}
\end{subfigure}
\begin{subfigure}[t]{0.47\textwidth}
\includegraphics[width=\textwidth,valign=t]{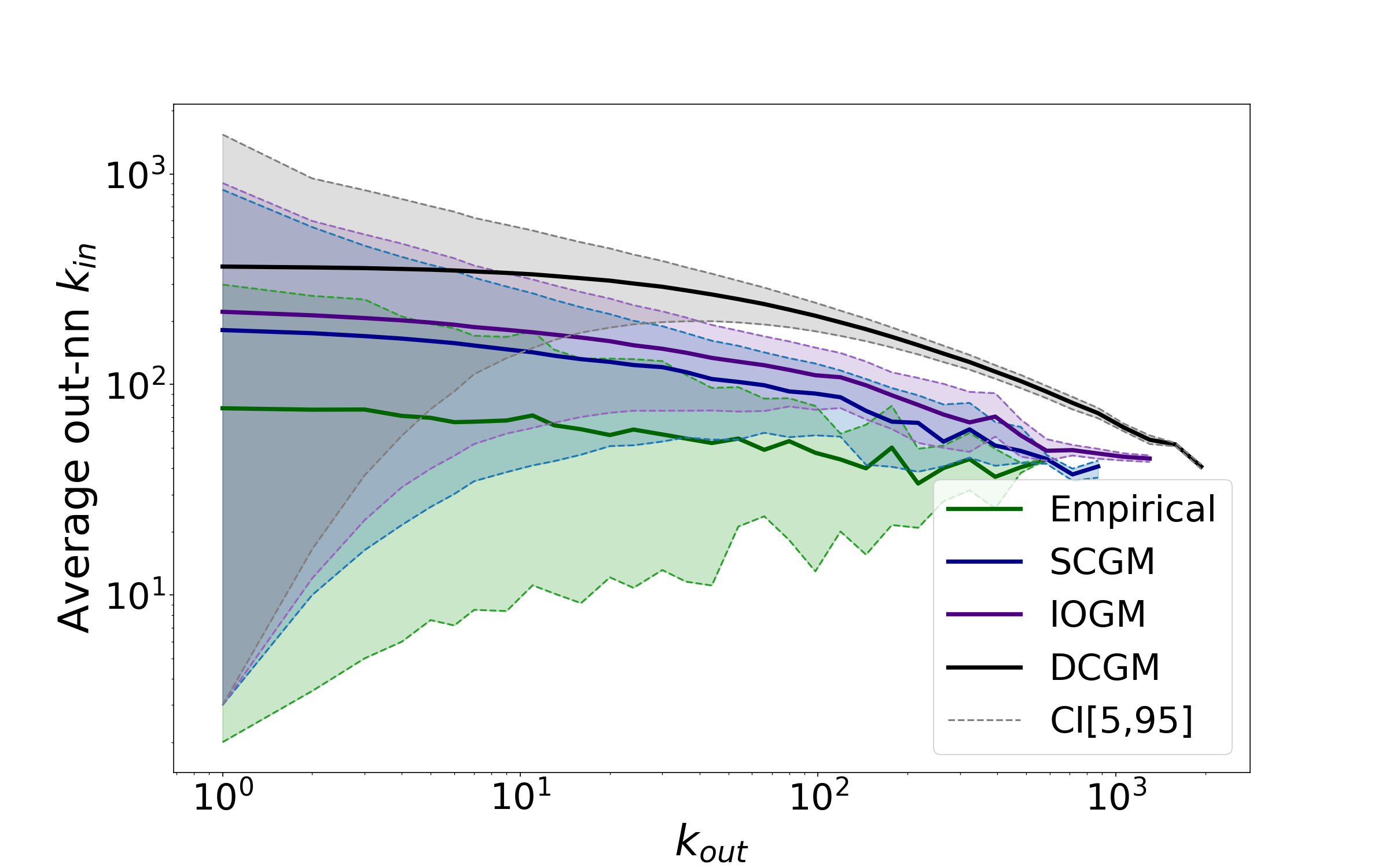}
\end{subfigure}
\begin{subfigure}{0.02\textwidth}
    \textbf{d)}
\end{subfigure}
\begin{subfigure}[t]{0.47\textwidth}
\includegraphics[width=\textwidth,valign=t]{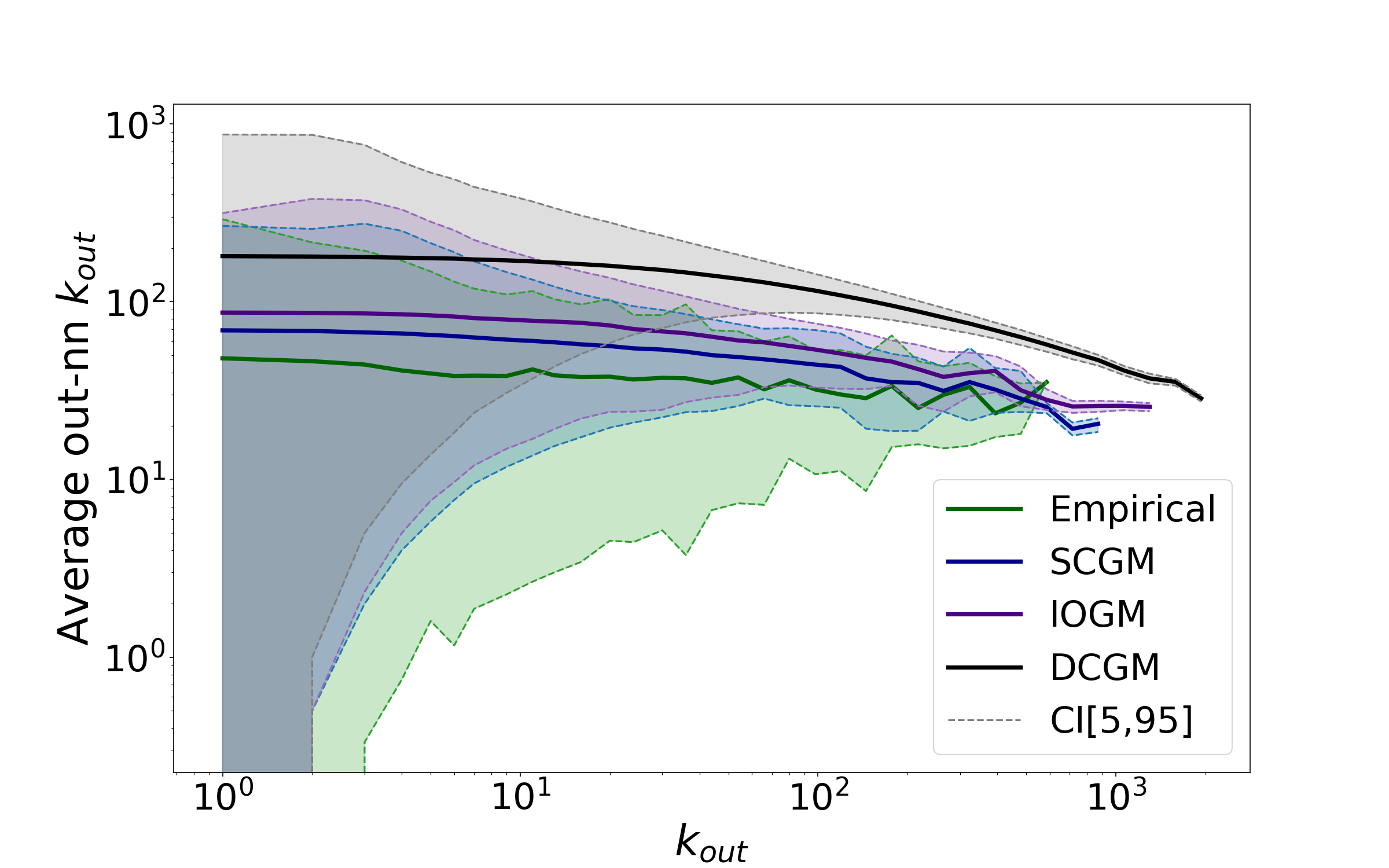}
\end{subfigure}
\caption{Empirical and expected $k_i^\text{in-in}$ (a) and $k_i^\text{in-out}$ (b) scattered versus $k_i^{in}$. Empirical and expected $k_i^\text{out-in}$ and (c) $k_i^\text{out-out}$ (d) scattered versus $k_i^{out}$.}
\label{fig7}
\end{figure*}

Let us, now, define the average nearest neighbours degree (ANND) of a node. In a directed network, we can define four, different measures. For instance, 

\begin{equation}
k_i^\text{out-out}=\frac{1}{k_i^\text{out}}\sum_{j(\neq i)}a_{i\to j}k_j^\text{out};
\end{equation}
its expected value reads

\begin{equation}
\langle k_i^\text{out-out}\rangle=\frac{1}{\langle k_i^\text{out}\rangle}\sum_{j(\neq i)}p_{i\to j}\langle k_j^\text{out}\rangle
\end{equation}
and similarly for the other quantities~\cite{squartini2011analytical}. The scatter plots of the aforementioned statistics are shown in Figure \ref{fig7}).

\clearpage

\subsection*{Weighted assortativity coefficients}

\begin{figure*}[t!]
\centering
\begin{subfigure}{0.02\textwidth}
    \textbf{a)}
\end{subfigure}
\begin{subfigure}[t]{0.47\textwidth}
\includegraphics[width=\textwidth,valign=t]{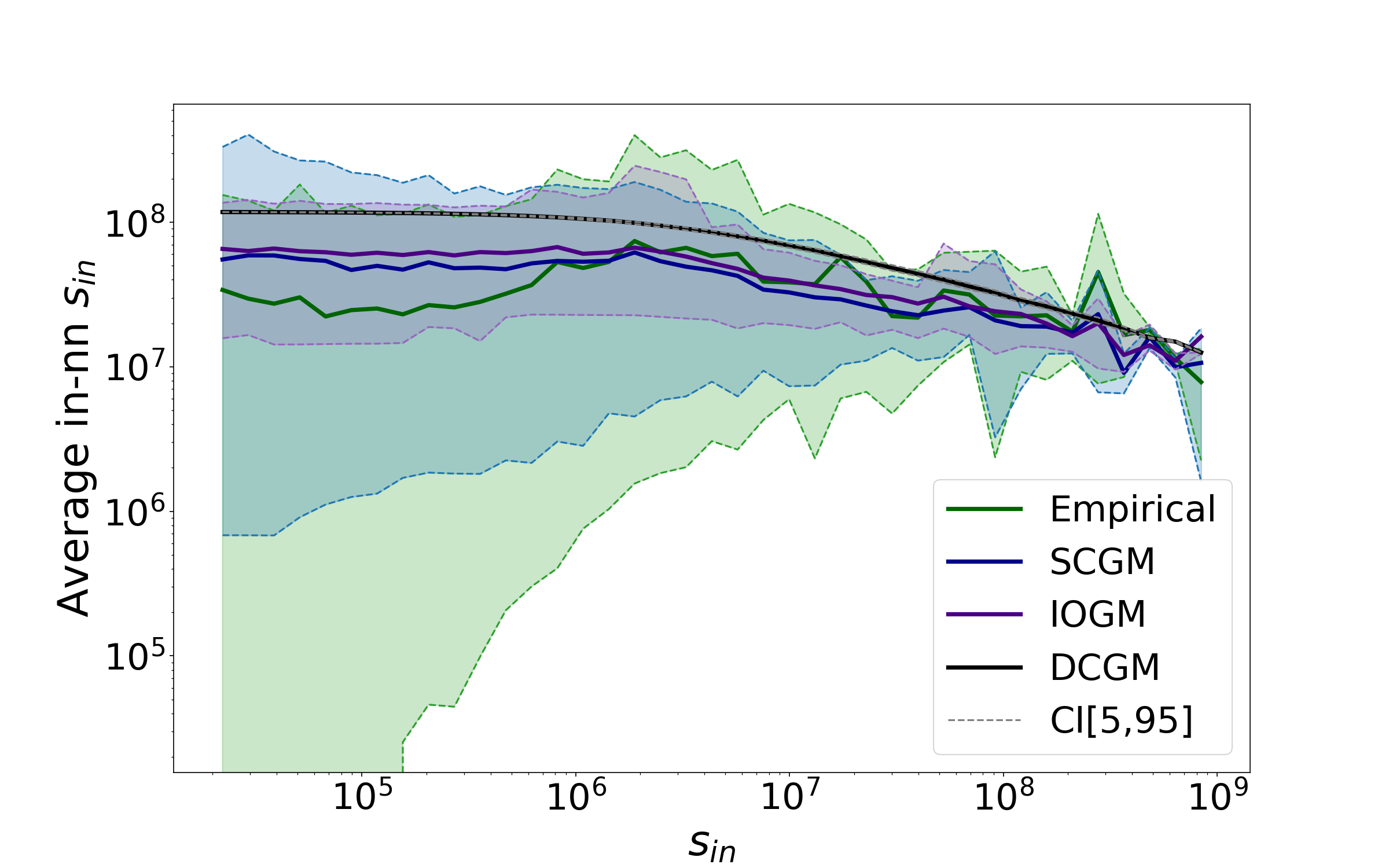}
\end{subfigure}
\begin{subfigure}{0.02\textwidth}
    \textbf{b)}
\end{subfigure}
\begin{subfigure}[t]{0.47\textwidth}
\includegraphics[width=\textwidth,valign=t]{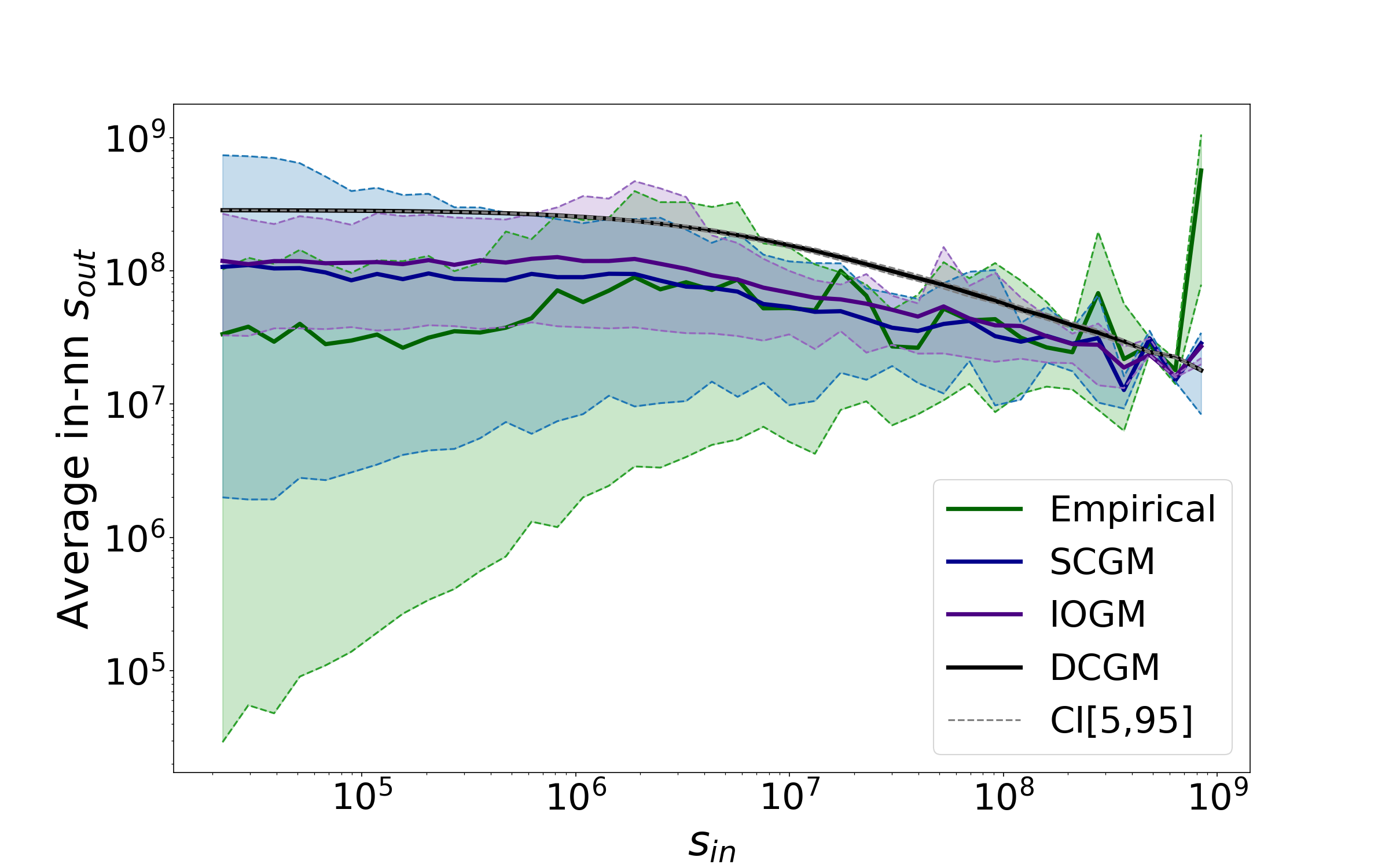}
\end{subfigure}
\begin{subfigure}{0.02\textwidth}
    \textbf{c)}
\end{subfigure}
\begin{subfigure}[t]{0.47\textwidth}
\includegraphics[width=\textwidth,valign=t]{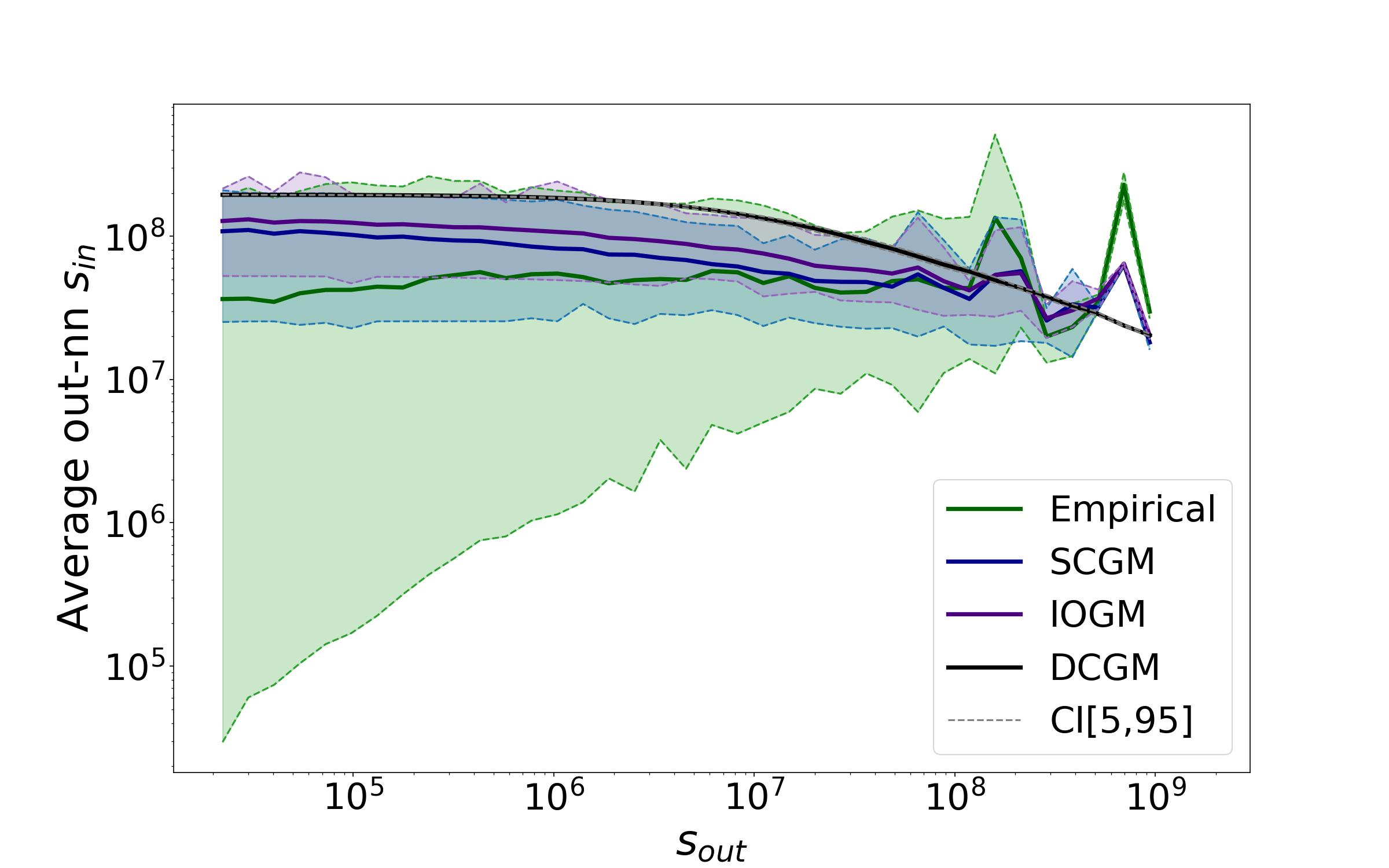}
\end{subfigure}
\begin{subfigure}{0.02\textwidth}
    \textbf{d)}
\end{subfigure}
\begin{subfigure}[t]{0.47\textwidth}
\includegraphics[width=\textwidth,valign=t]{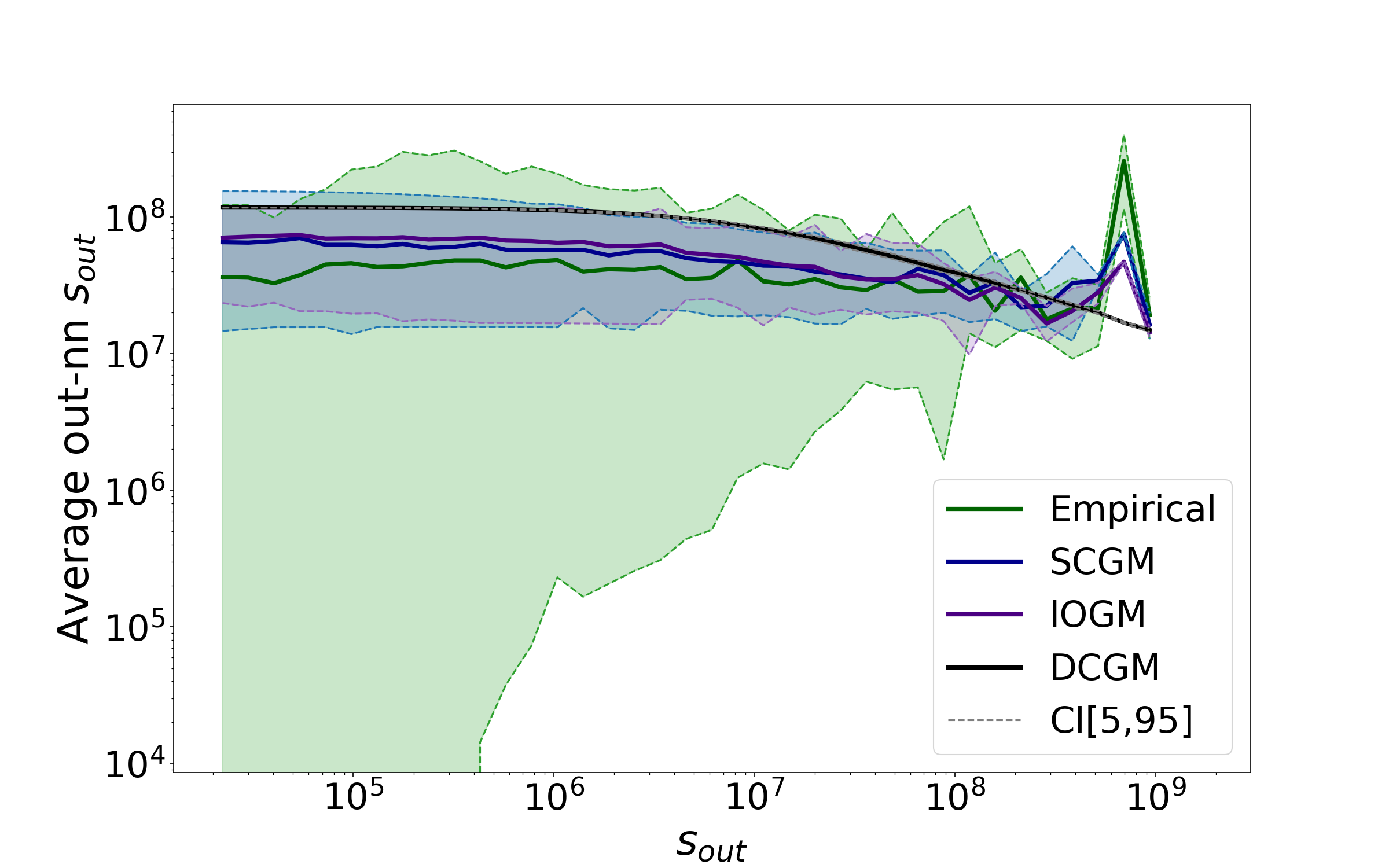}
\end{subfigure}
\caption{Empirical and expected $s_i^\text{in-in}$ (a) and $s_i^\text{in-out}$ (b) scattered versus $s_i^{in}$. Empirical and expected $s_i^\text{out-in}$ and (c) $s_i^\text{out-out}$ (d) scattered versus $s_i^{out}$.}
\label{fig8}
\end{figure*}

Similarly, we can define the average nearest neighbours strength (ANNS) of a node. In a directed network, we can define four, different measures. For instance,

\begin{equation}
s_i^\text{out-out}=\frac{1}{k_i^\text{out}}\sum_{j(\neq i)}a_{j\to i}s_j^\text{out};
\end{equation}
its expected value reads

\begin{equation}
\langle s_i^\text{out-out}\rangle=\frac{1}{\langle k_i^\text{out}\rangle}\sum_{j(\neq i)}p_{j\to i}\langle s_j^\text{out}\rangle
\end{equation}
and similarly for the other quantities~\cite{squartini2011analytical}. The scatter plots of the aforementioned statistics are shown in Figure \ref{fig8}).

\subsection*{Clustering coefficients}

\begin{figure*}[t!]
\centering
\begin{subfigure}{0.02\textwidth}
    \textbf{a)}
\end{subfigure}
\begin{subfigure}[t]{0.47\textwidth}
\includegraphics[width=\textwidth,valign=t]{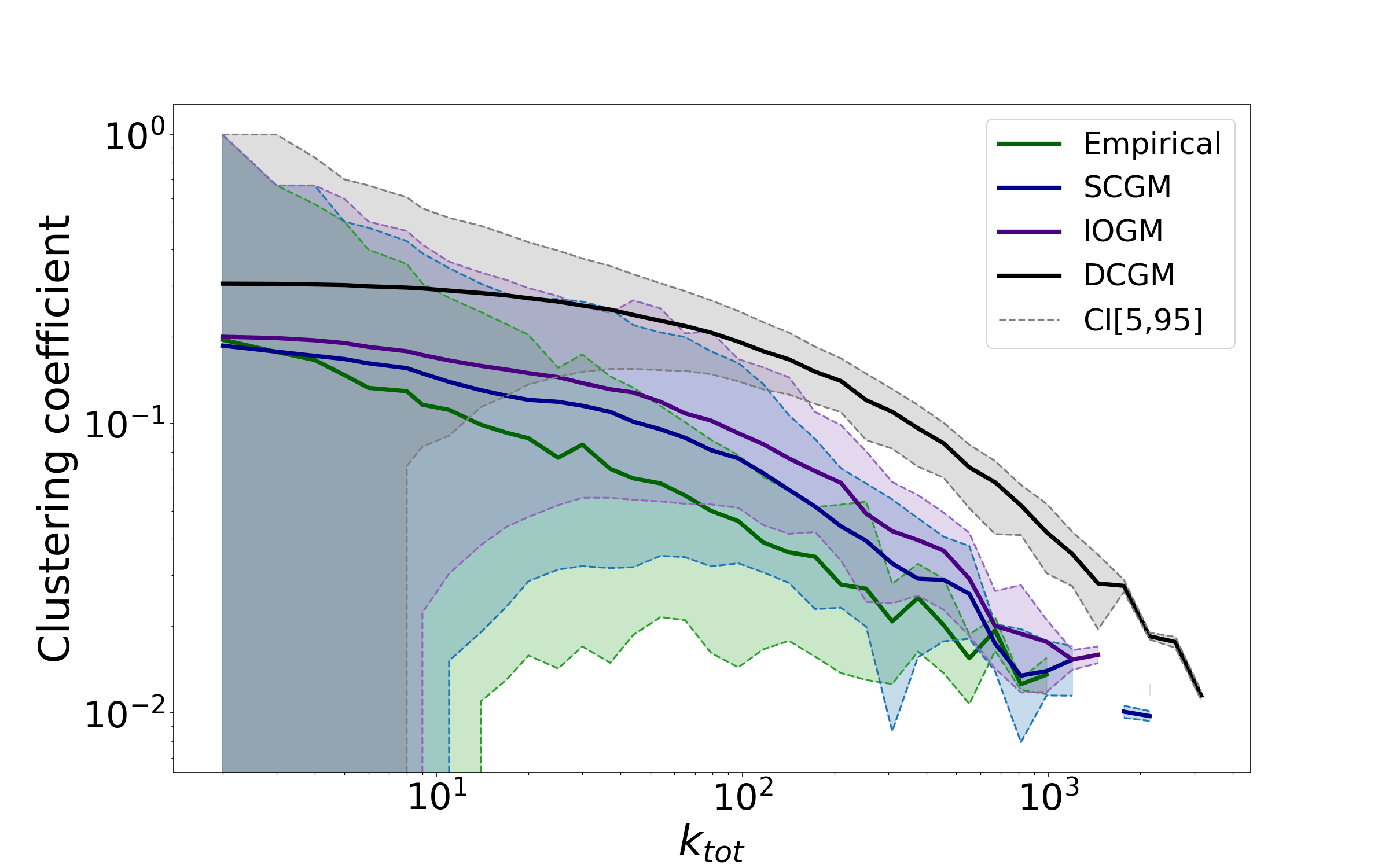}
\end{subfigure}
\begin{subfigure}{0.02\textwidth}
    \textbf{b)}
\end{subfigure}
\begin{subfigure}[t]{0.47\textwidth}
\includegraphics[width=\textwidth,valign=t]{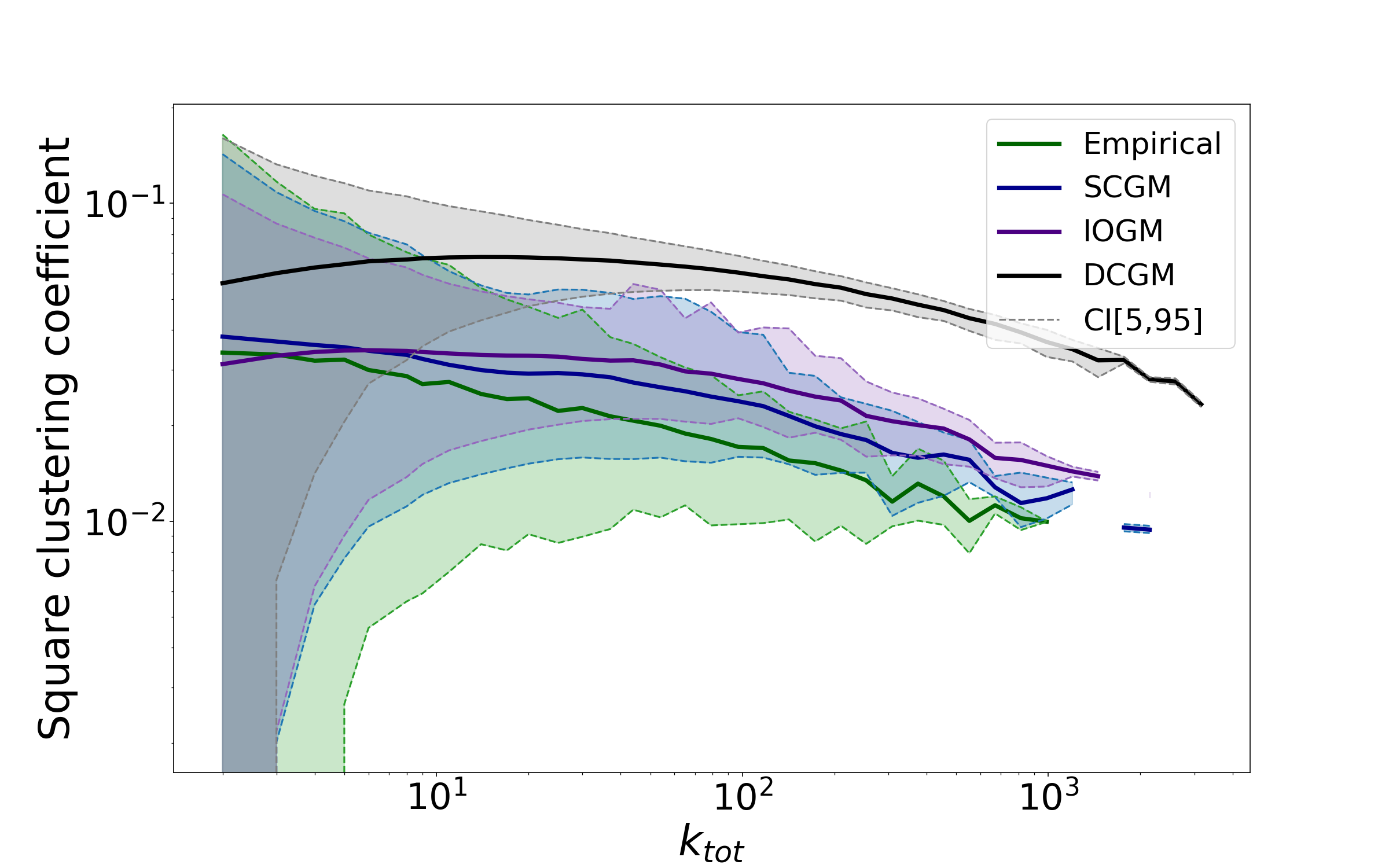}
\end{subfigure}
\caption{Empirical and expected triangular (a) and square (b) clustering coefficient scattered versus $k_i^\text{tot}$.}
\label{fig9}
\end{figure*}

Higher-order properties of the network can be inspected by considering the clustering coefficient(s). Here, we have considered the undirected version of the triangular clustering coefficient, reading

\begin{equation}
C^3_i=\frac{\sum_{j(\neq i)}\sum_{k(\neq i,j)}a_{ij}a_{jk}a_{ki}}{k_i(k_i-1)}
\end{equation}
and evaluating the fraction of closed triangles it participates in, and the undirected version of the square clustering coefficient, reading

\begin{equation}
C^4_i=\frac{\sum_m\sum_{l(>m)}q_i(m,l)}{\sum_m\sum_{l(>m)}[(k_m-1)+(k_l-1)-q_i(m,l)]}
\end{equation}
where $q_i(m,l)=\sum_{j(\neq i)}a_{ik}a_{il}a_{jm}a_{jl}$ and evaluating the fraction of cycles with four edges involving the common neighbours of $m$ and $l$ other than $i$~\cite{zhang2008clustering}. The scatter plots of the aforementioned statistics are shown in Figure \ref{fig9}).

\subsection*{Pooling}

Our reconstruction models give origin to sampled configurations characterised by several, disconnected nodes. In order to overcome such an inconvenient, we have computed the ensemble average of the network statistics of interest according to the procedure called \emph{pooling}, allowing us to obtain scatter plots where the role of independent variable is played by $k$. Given $10^3$ configurations sampled from model $m$ and a network statistics $f$, we have considered $f_i^s$ and $k_i^s$; then, for each value, $k$, of the degree, we have computed the average value $\langle f\rangle_m$ as

\begin{equation}
\langle f\rangle_{m}=\frac{1}{N_k}\sum_{s=1}^{10^3}\sum_{i:k_i^{s}=k}f_i^{s}
\end{equation}
where $N_k$ is the total number of nodes, across all samples, whose degree equals $k$.

\clearpage

\clearpage

\section{Comparing different data reporting thresholds}

As discussed in the main text, the choice of using a threshold to remove the weakest links was motivated by invoking both conceptual and computational reasons. In order to evaluate the effect of such a filtering procedure on the performance of the SCGM, we tested it for four, different values of the threshold, $t$:

\begin{itemize}
\item $t_1=0\:\$$, i.e. the full network;
\item $t_2=10.000\:\$$;
\item $t_3=22.300\:\$$, i.e. the selected threshold;
\item $t_4=50.000\$$.
\end{itemize}

\subsection*{Degree distributions}

The distributions of in-degrees and out-degrees (see Figure \ref{fig10} and Figure \ref{fig11}) clearly show that the performance of the SCGM improves as we increase the value of the threshold, pointing out that the weakest links are also the most difficult ones to reconstruct. While the reconstruction performance clearly improves when passing from $t_1$ to $t_2$, it seems to stabilise for bigger values of $t$.\\

Similar conclusions can be drawn upon scattering the empirical values of the in-degrees (see Figure \ref{fig12}) and of the out-degrees (see Figure \ref{fig13}) versus the reconstructed ones. The dispersion of the expected values of the in-degree (out-degree) for a given, empirical value of the in-degree (out-degree) shrinks from roughly six (eight) orders of magnitude on the full network to almost three (four) orders of magnitude for $t_2=10.000\:\$$ and remains stable for bigger values of $t$.

\subsection*{Binary and weighted assortativity}

\begin{figure*}[t!]
\centering
\begin{subfigure}{0.02\textwidth}
    \textbf{a)}
\end{subfigure}
\begin{subfigure}[t]{0.47\textwidth}
\includegraphics[width=\textwidth,valign=t]{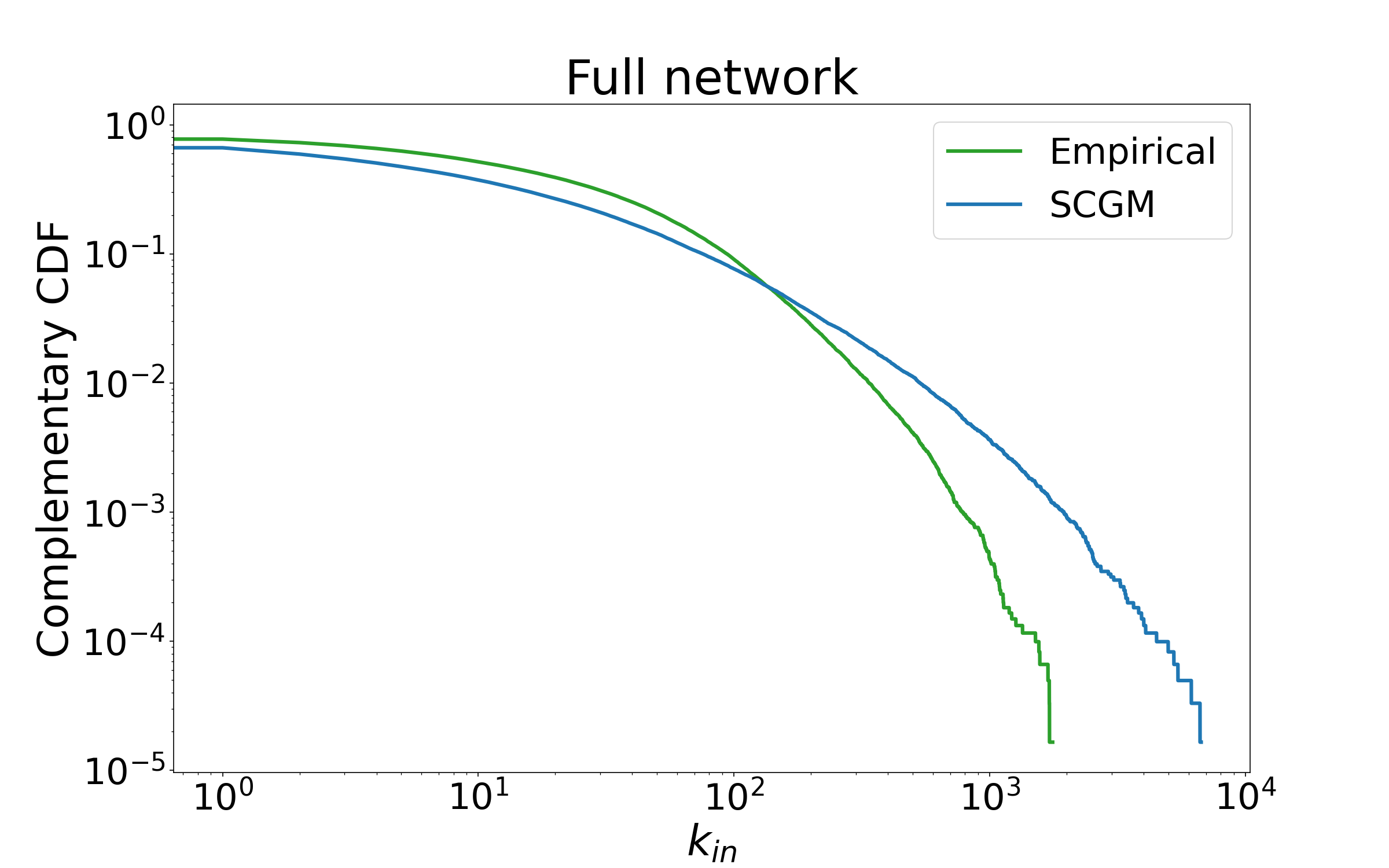}
\end{subfigure}
\begin{subfigure}{0.02\textwidth}
    \textbf{b)}
\end{subfigure}
\begin{subfigure}[t]{0.47\textwidth}
\includegraphics[width=\textwidth,valign=t]{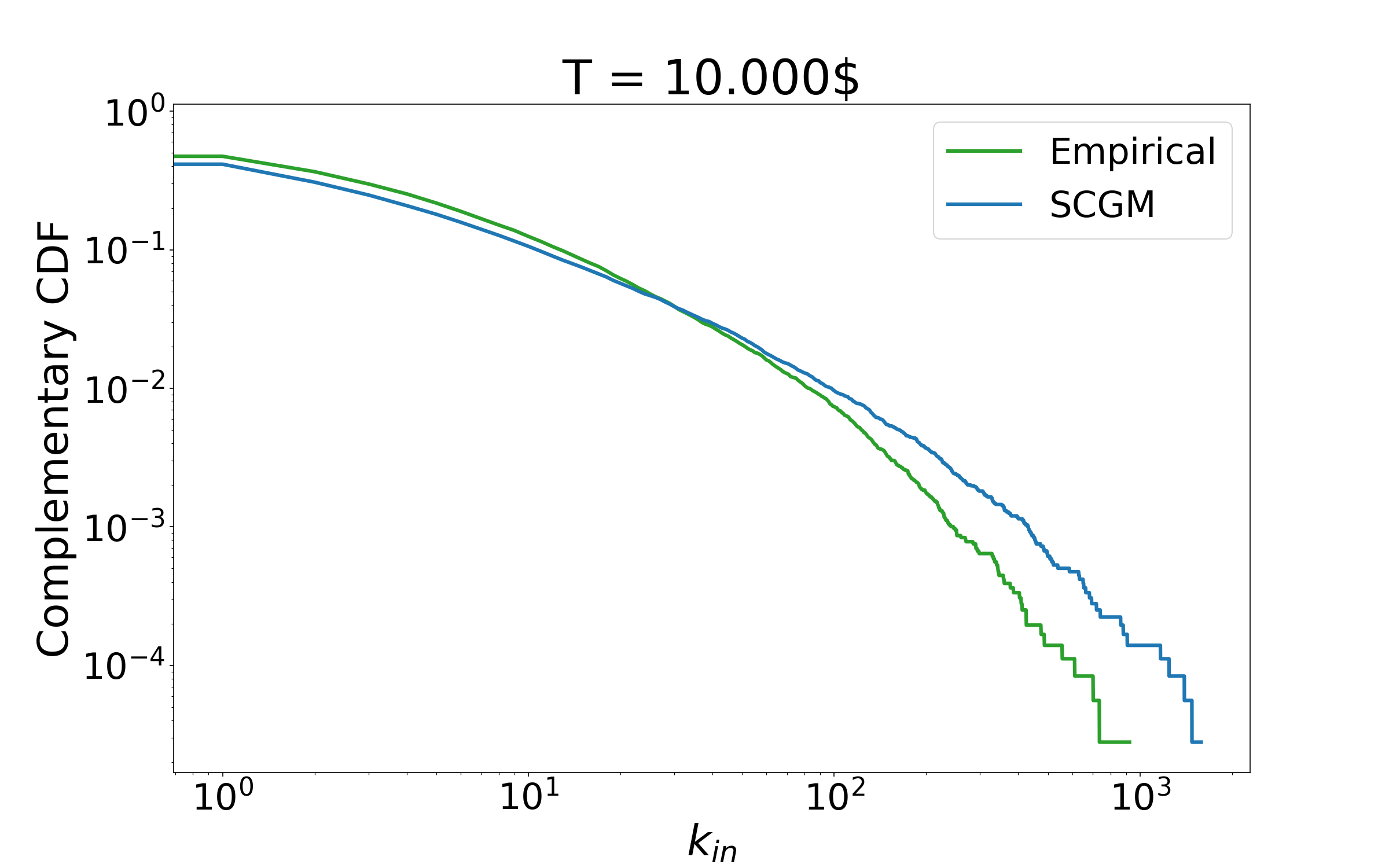}
\end{subfigure}
\begin{subfigure}{0.02\textwidth}
    \textbf{c)}
\end{subfigure}
\begin{subfigure}[t]{0.47\textwidth}
\includegraphics[width=\textwidth,valign=t]{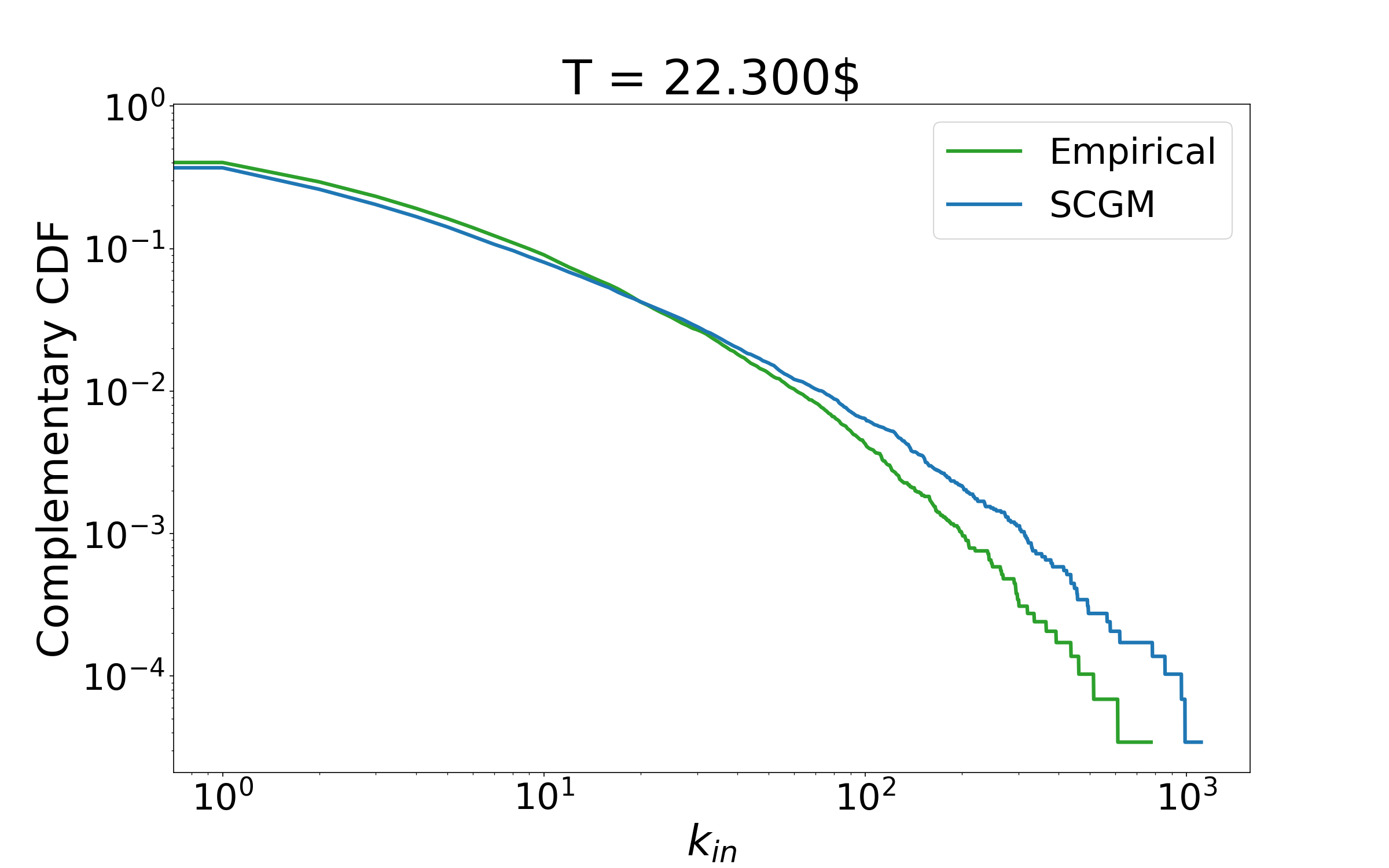}
\end{subfigure}
\begin{subfigure}{0.02\textwidth}
    \textbf{d)}
\end{subfigure}
\begin{subfigure}[t]{0.47\textwidth}
\includegraphics[width=\textwidth,valign=t]{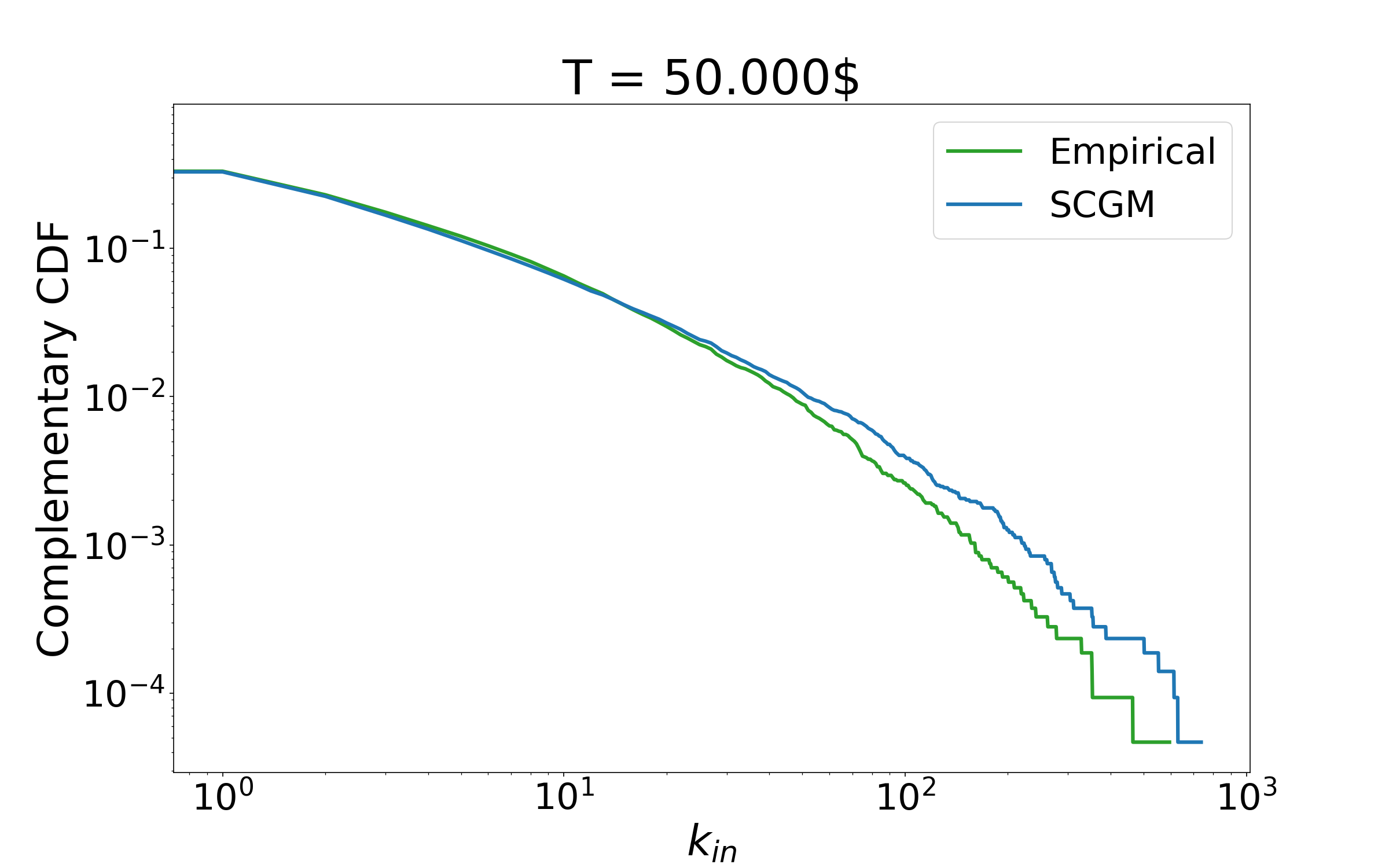}
\end{subfigure}
\caption{Comparison of the distributions of in-degrees for the four, different thresholds.}
\label{fig10}
\end{figure*}

\begin{figure*}[t!]
\centering
\begin{subfigure}{0.02\textwidth}
    \textbf{a)}
\end{subfigure}
\begin{subfigure}[t]{0.47\textwidth}
\includegraphics[width=\textwidth,valign=t]{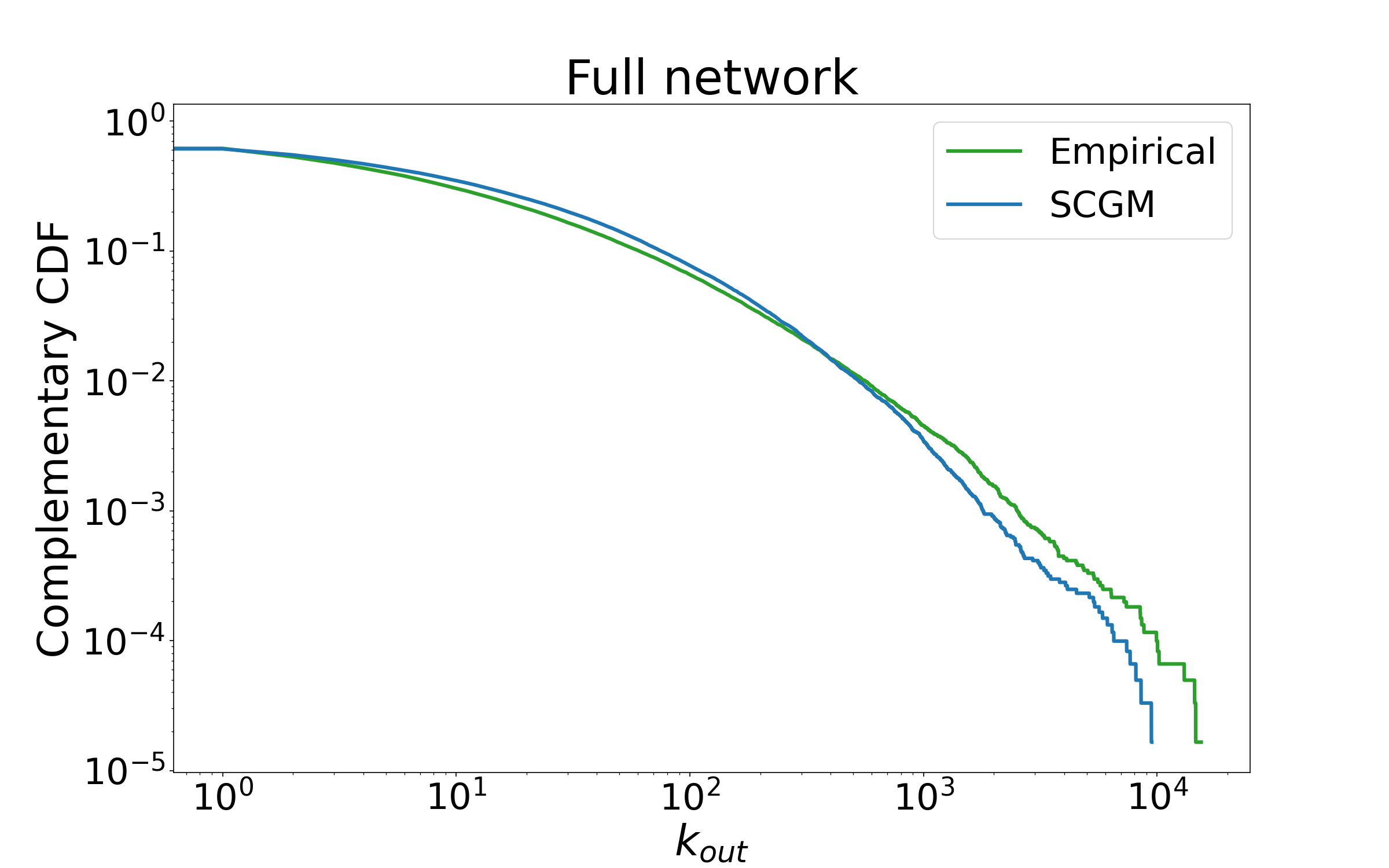}
\end{subfigure}
\begin{subfigure}{0.02\textwidth}
    \textbf{b)}
\end{subfigure}
\begin{subfigure}[t]{0.47\textwidth}
\includegraphics[width=\textwidth,valign=t]{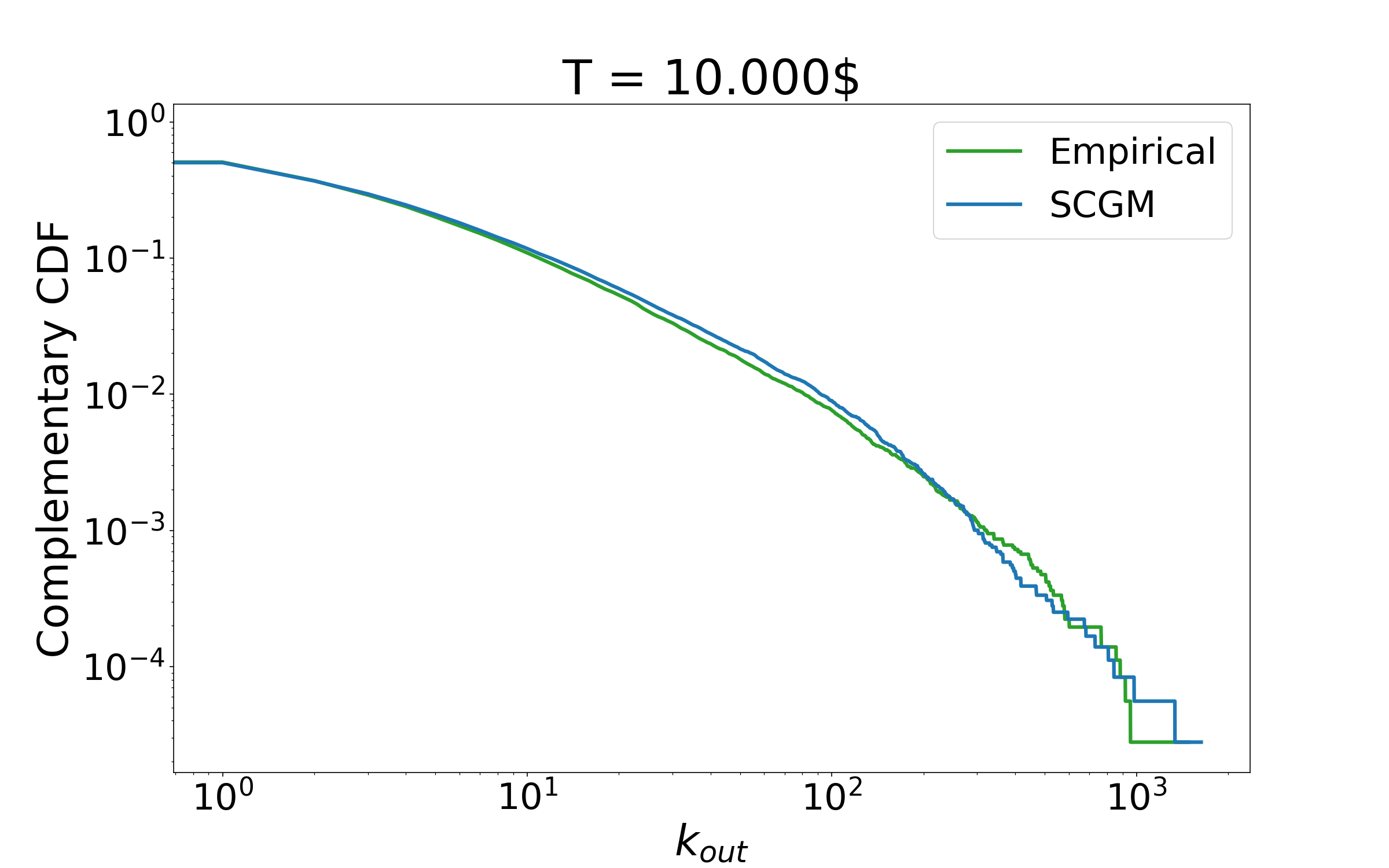}
\end{subfigure}
\begin{subfigure}{0.02\textwidth}
    \textbf{c)}
\end{subfigure}
\begin{subfigure}[t]{0.47\textwidth}
\includegraphics[width=\textwidth,valign=t]{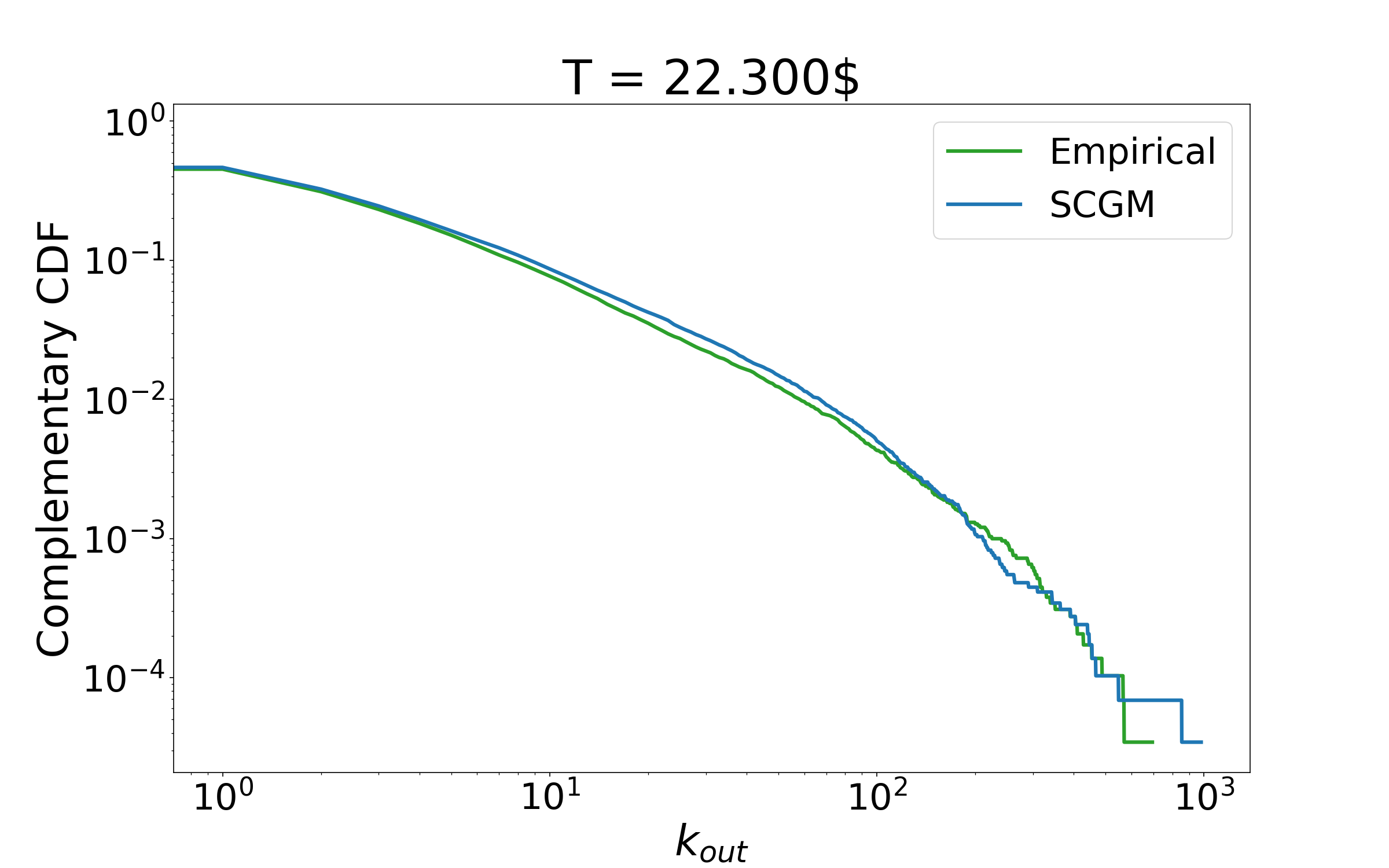}
\end{subfigure}
\begin{subfigure}{0.02\textwidth}
    \textbf{d)}
\end{subfigure}
\begin{subfigure}[t]{0.47\textwidth}
\includegraphics[width=\textwidth,valign=t]{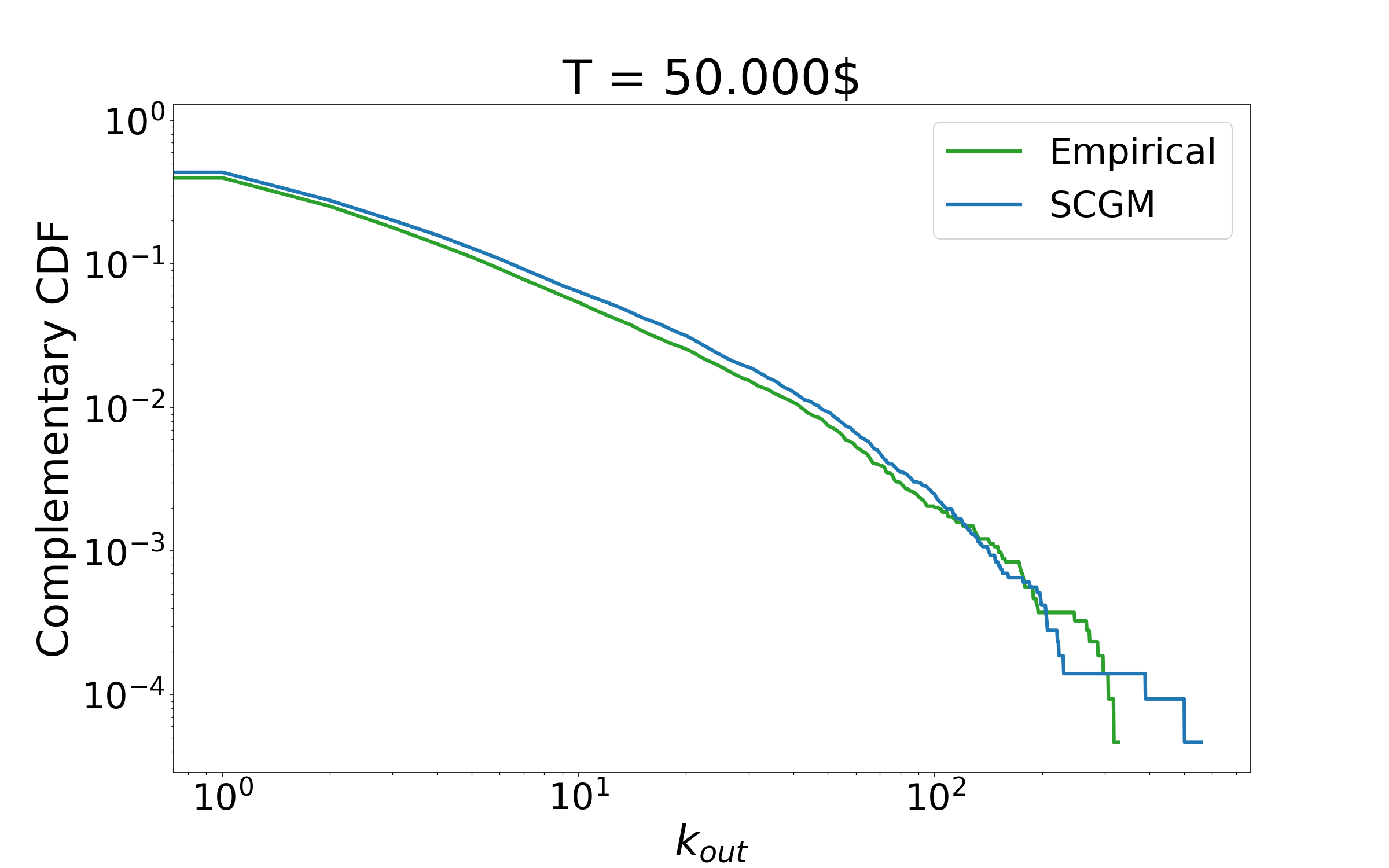}
\end{subfigure}
\caption{Comparison of the distributions of out-degrees for the four, different thresholds.}
\label{fig11}
\end{figure*}

\begin{figure*}[t!]
\centering
\begin{subfigure}{0.02\textwidth}
    \textbf{a)}
\end{subfigure}
\begin{subfigure}[t]{0.47\textwidth}
\includegraphics[width=\textwidth,valign=t]{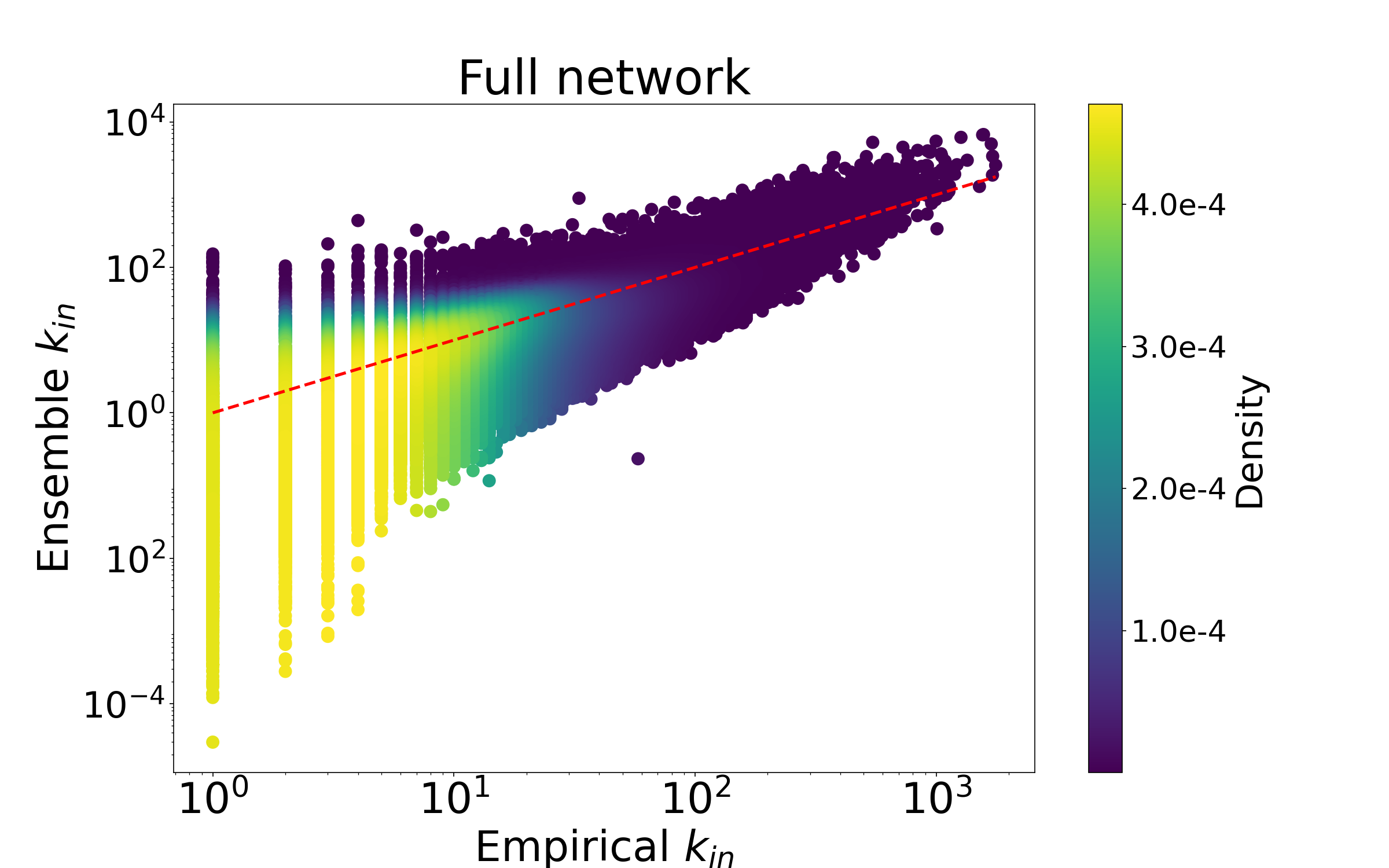}
\end{subfigure}
\begin{subfigure}{0.02\textwidth}
    \textbf{b)}
\end{subfigure}
\begin{subfigure}[t]{0.47\textwidth}
\includegraphics[width=\textwidth,valign=t]{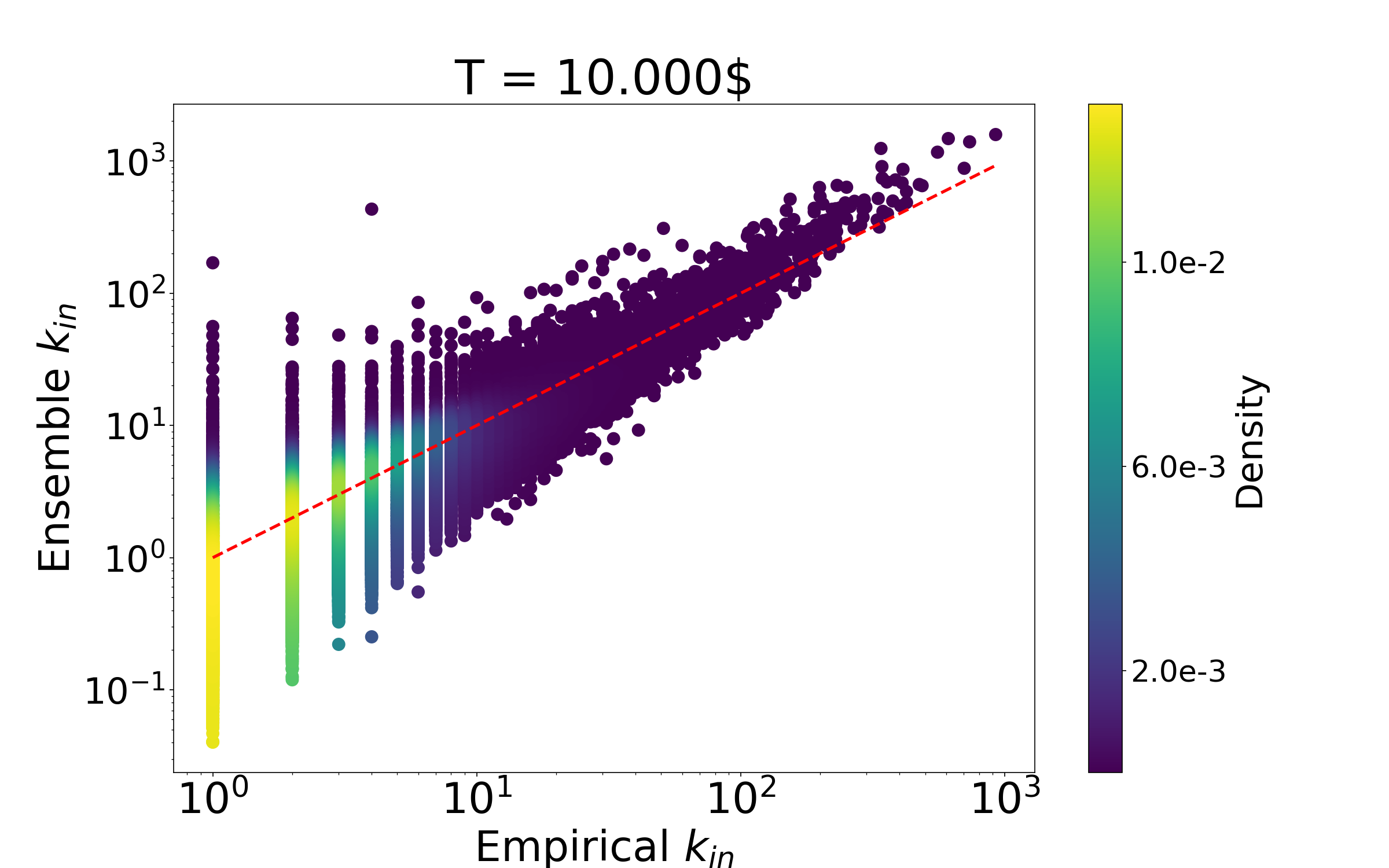}
\end{subfigure}
\begin{subfigure}{0.02\textwidth}
    \textbf{c)}
\end{subfigure}
\begin{subfigure}[t]{0.47\textwidth}
\includegraphics[width=\textwidth,valign=t]{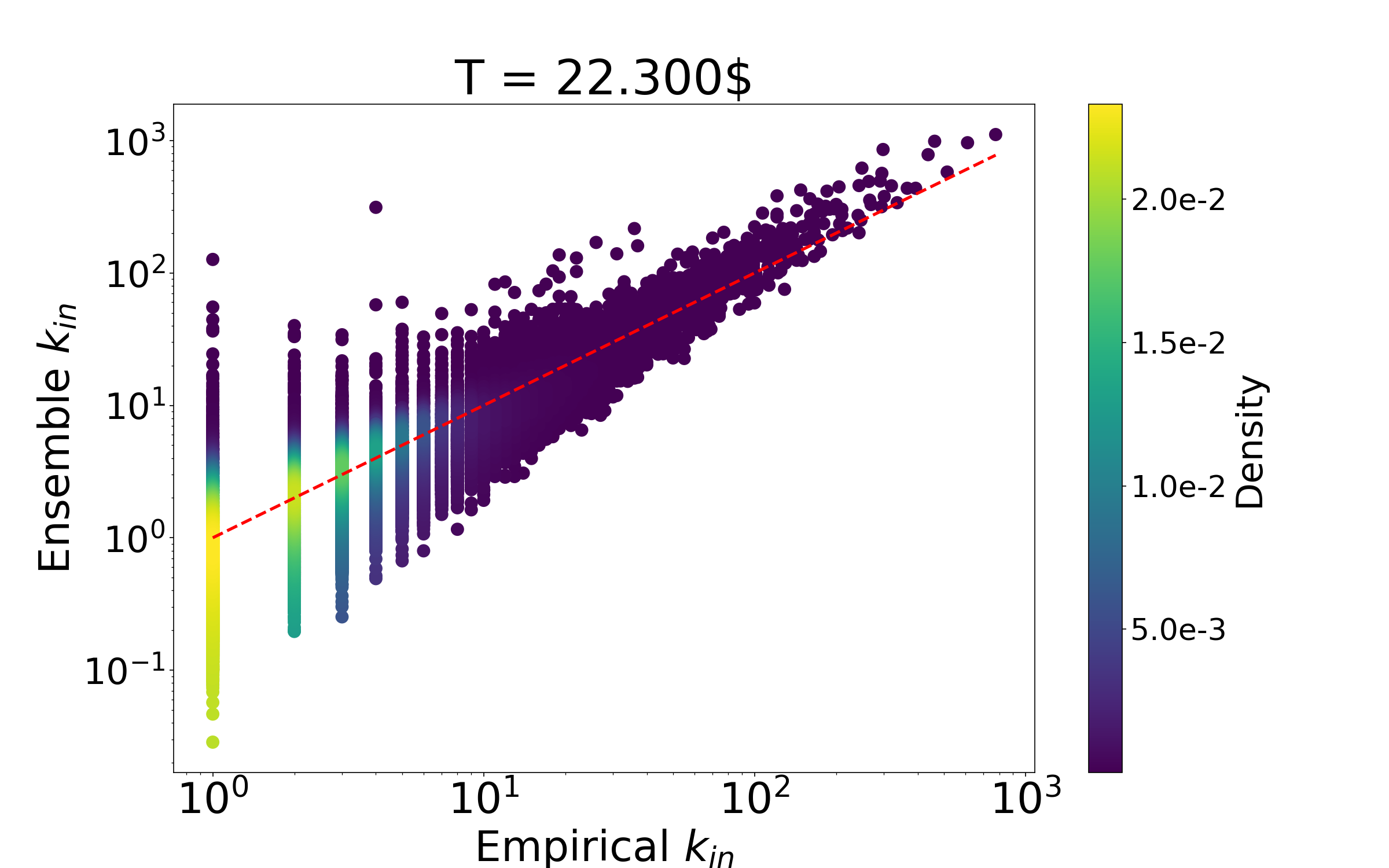}
\end{subfigure}
\begin{subfigure}{0.02\textwidth}
    \textbf{d)}
\end{subfigure}
\begin{subfigure}[t]{0.47\textwidth}
\includegraphics[width=\textwidth,valign=t]{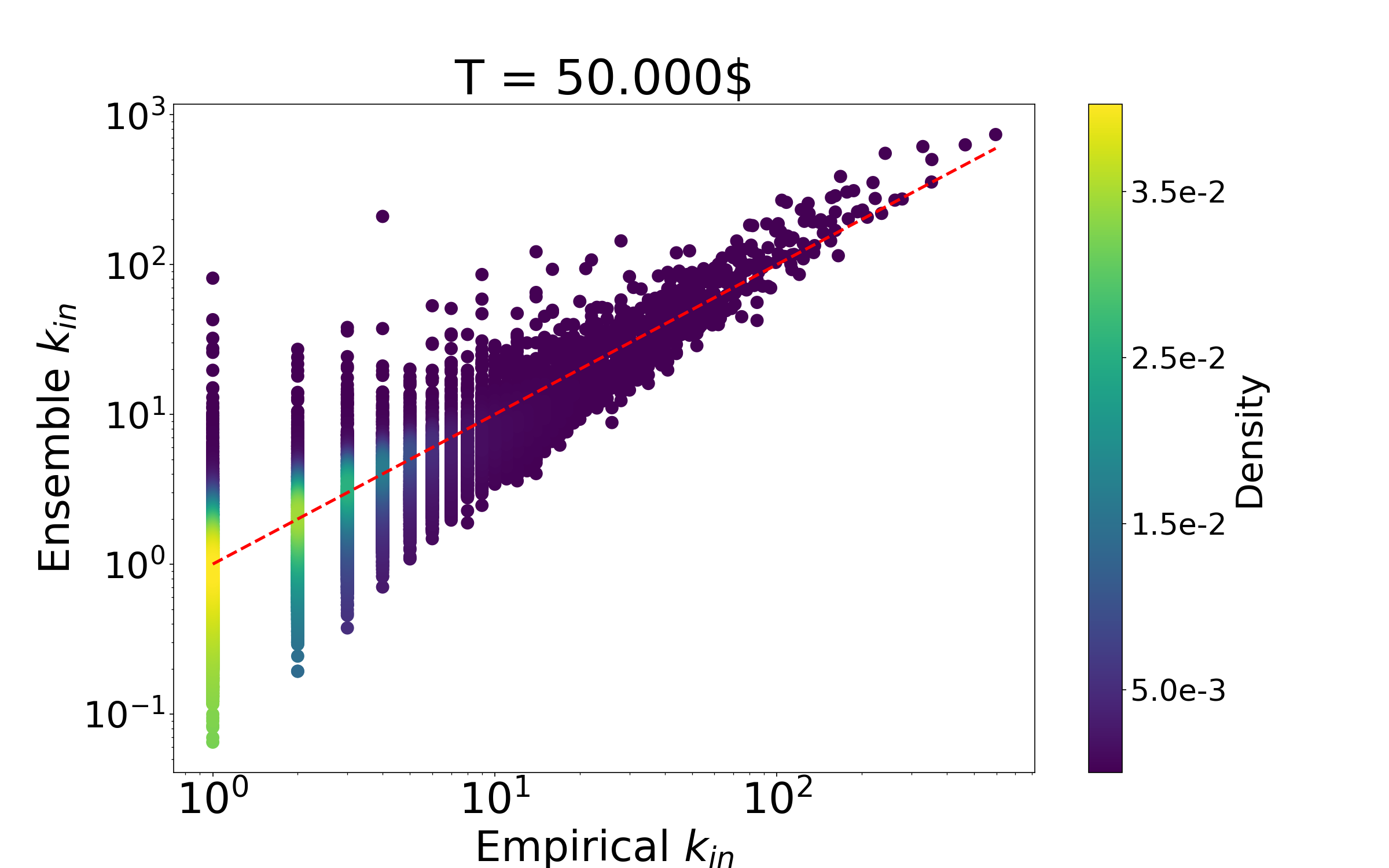}
\end{subfigure}
\caption{Empirical versus reconstructed in-degrees for the four, different thresholds.}
\label{fig12}
\end{figure*}

\begin{figure*}[t!]
\centering
\begin{subfigure}{0.02\textwidth}
    \textbf{a)}
\end{subfigure}
\begin{subfigure}[t]{0.47\textwidth}
\includegraphics[width=\textwidth,valign=t]{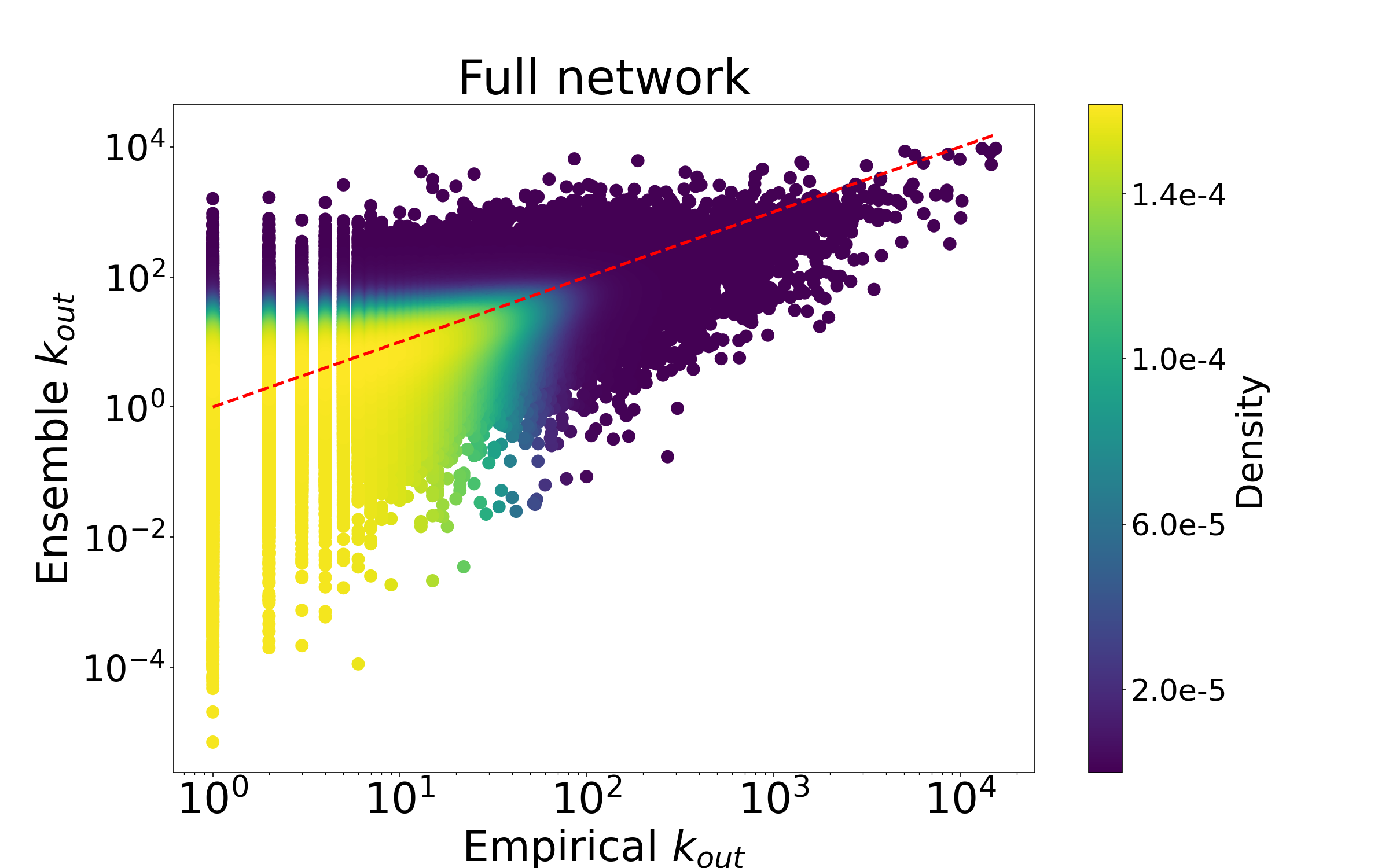}
\end{subfigure}
\begin{subfigure}{0.02\textwidth}
    \textbf{b)}
\end{subfigure}
\begin{subfigure}[t]{0.47\textwidth}
\includegraphics[width=\textwidth,valign=t]{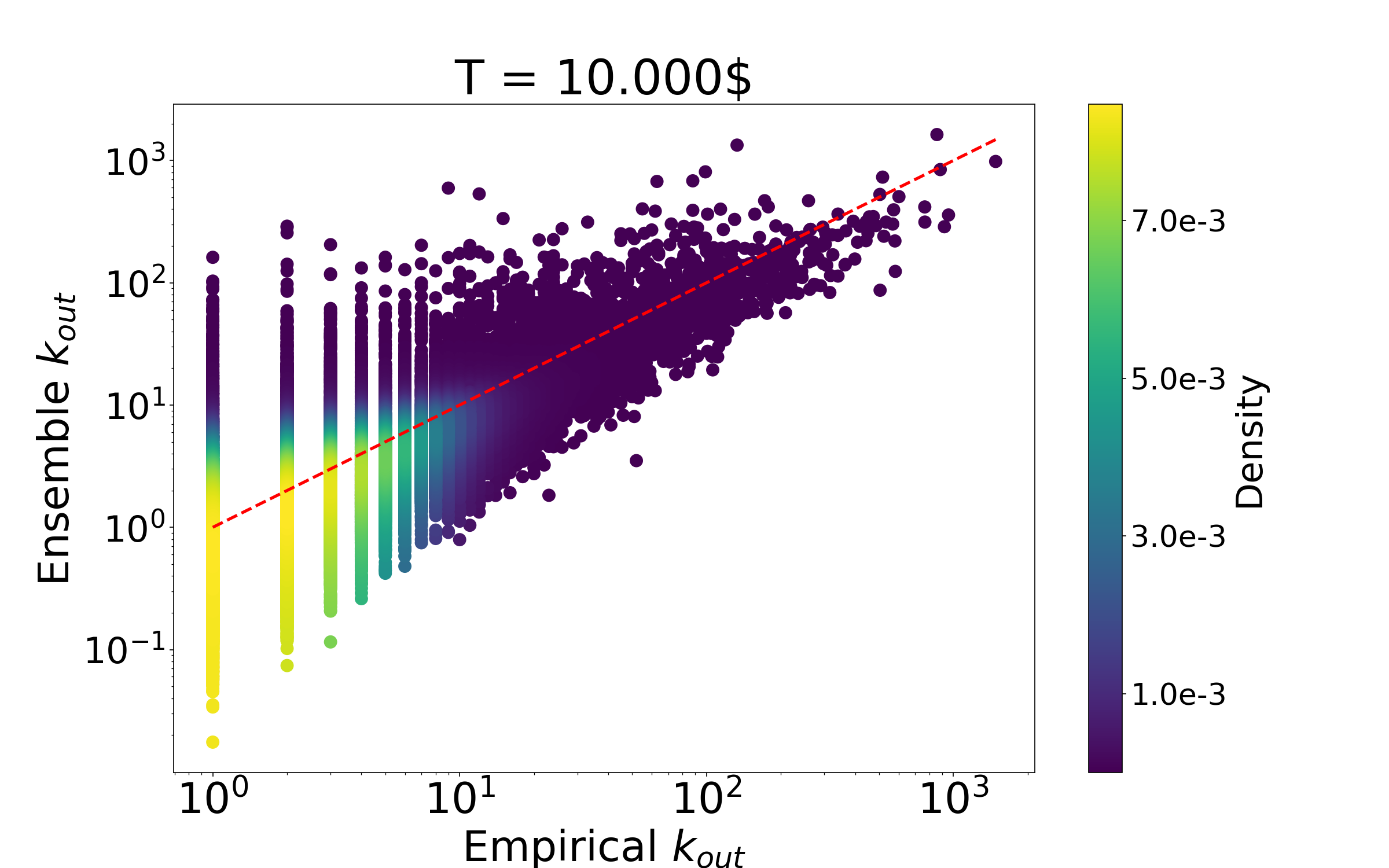}
\end{subfigure}
\begin{subfigure}{0.02\textwidth}
    \textbf{c)}
\end{subfigure}
\begin{subfigure}[t]{0.47\textwidth}
\includegraphics[width=\textwidth,valign=t]{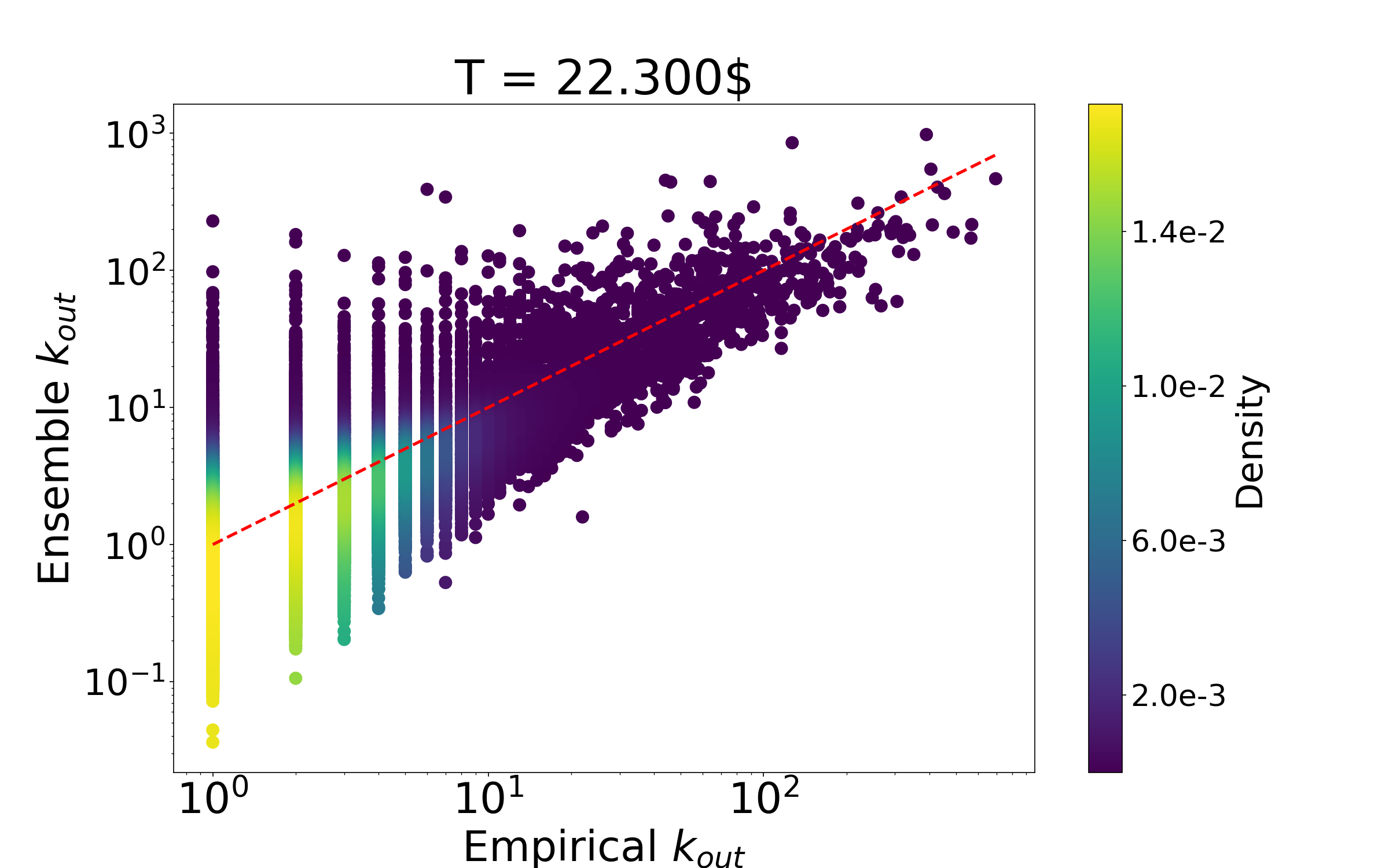}
\end{subfigure}
\begin{subfigure}{0.02\textwidth}
    \textbf{d)}
\end{subfigure}
\begin{subfigure}[t]{0.47\textwidth}
\includegraphics[width=\textwidth,valign=t]{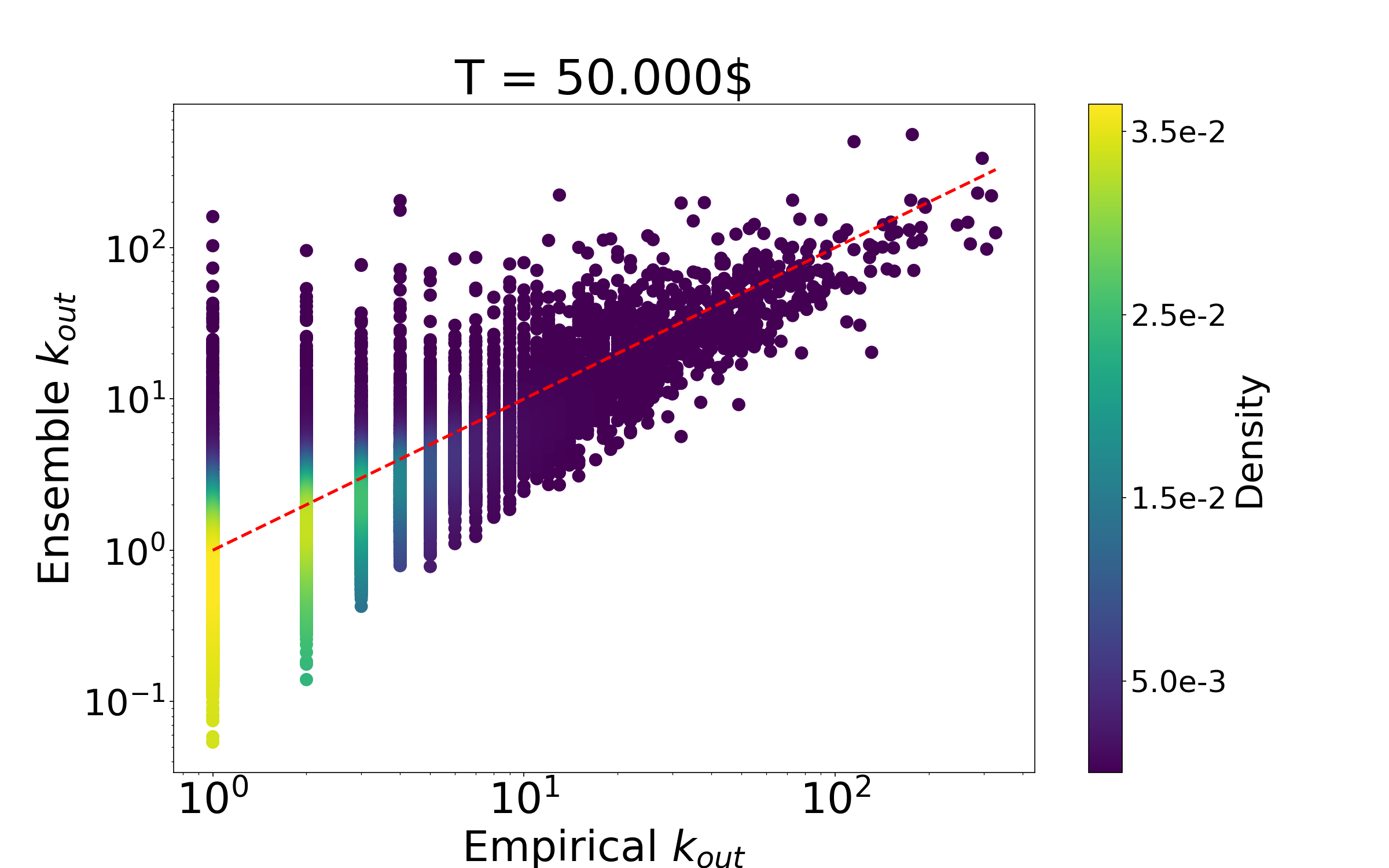}
\end{subfigure}
\caption{Empirical versus reconstructed out-degrees for the four, different thresholds.}
\label{fig13}
\end{figure*}

\begin{figure*}[t!]
\centering
\begin{subfigure}{0.02\textwidth}
    \textbf{a)}
\end{subfigure}
\begin{subfigure}[t]{0.47\textwidth}
\includegraphics[width=\textwidth,valign=t]{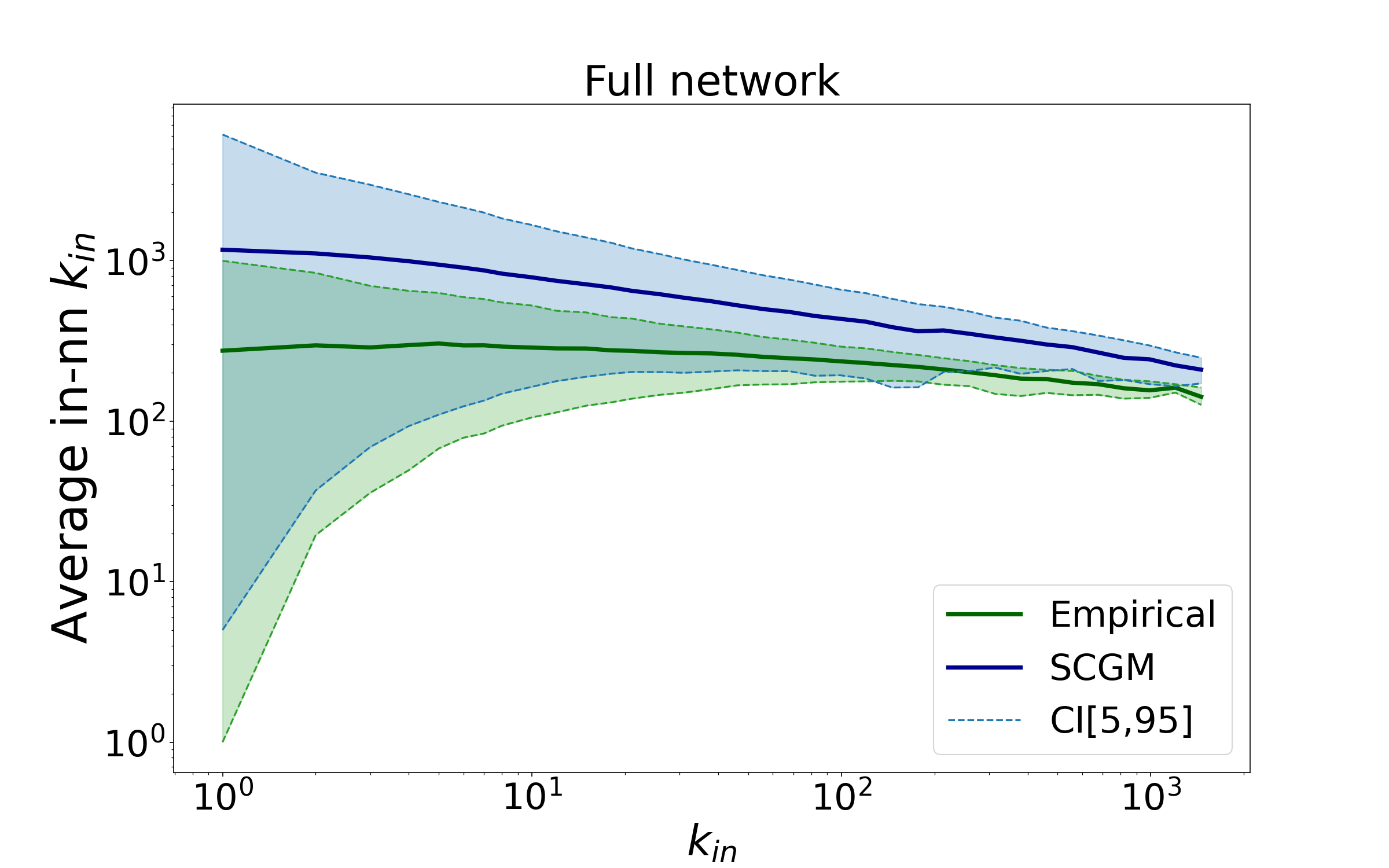}
\end{subfigure}
\begin{subfigure}{0.02\textwidth}
    \textbf{b)}
\end{subfigure}
\begin{subfigure}[t]{0.47\textwidth}
\includegraphics[width=\textwidth,valign=t]{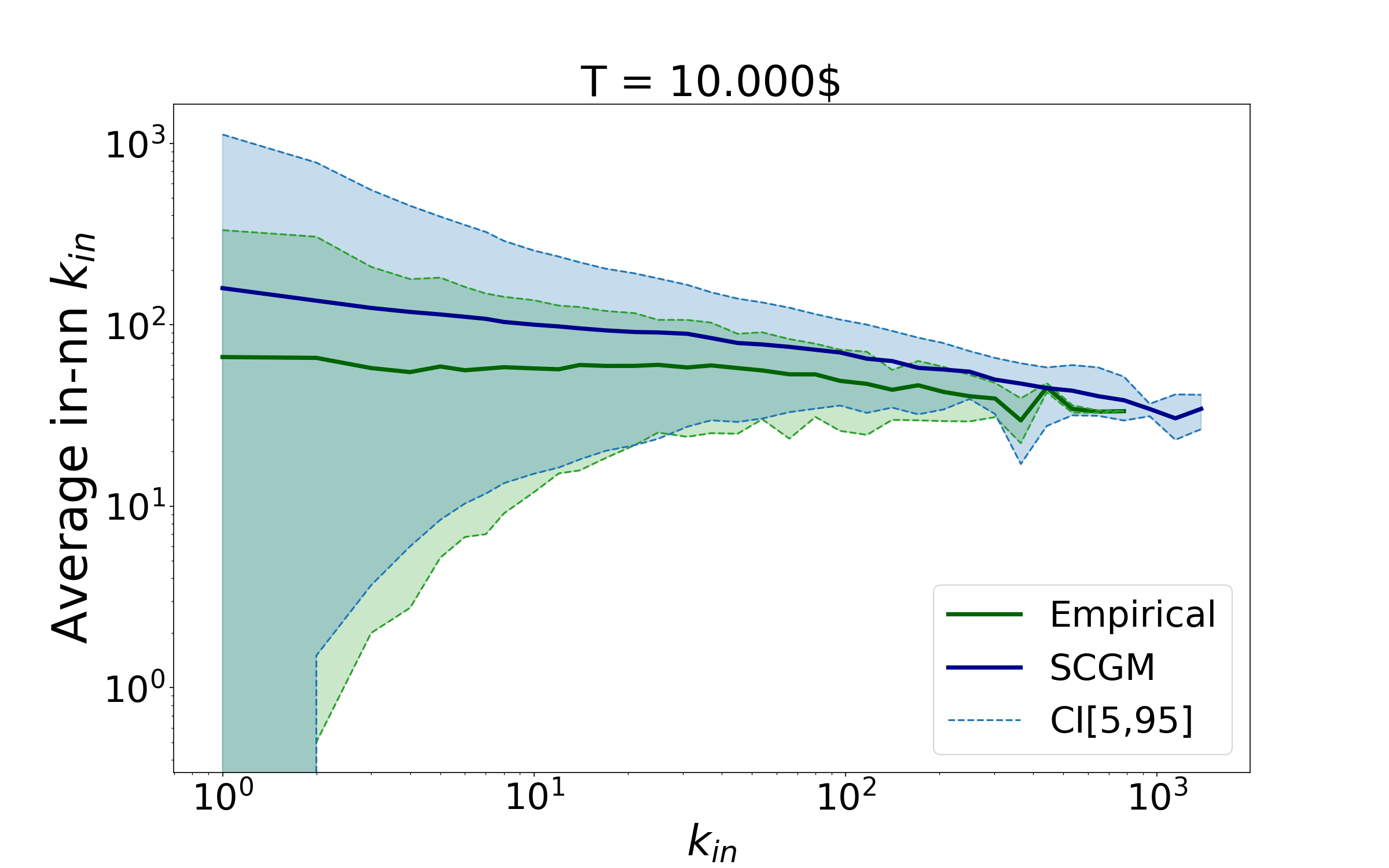}
\end{subfigure}
\begin{subfigure}{0.02\textwidth}
    \textbf{c)}
\end{subfigure}
\begin{subfigure}[t]{0.47\textwidth}
\includegraphics[width=\textwidth,valign=t]{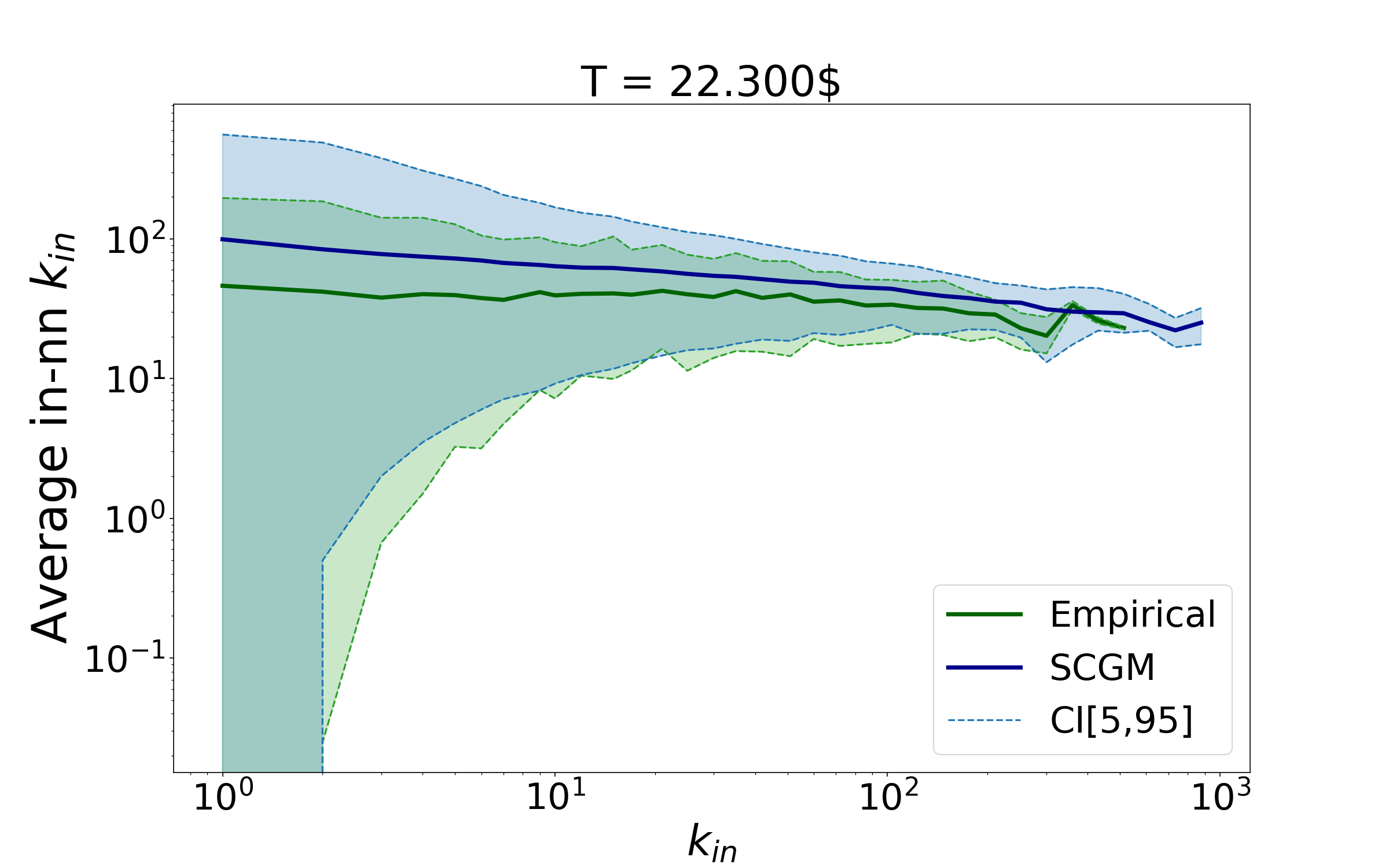}
\end{subfigure}
\begin{subfigure}{0.02\textwidth}
    \textbf{d)}
\end{subfigure}
\begin{subfigure}[t]{0.47\textwidth}
\includegraphics[width=\textwidth,valign=t]{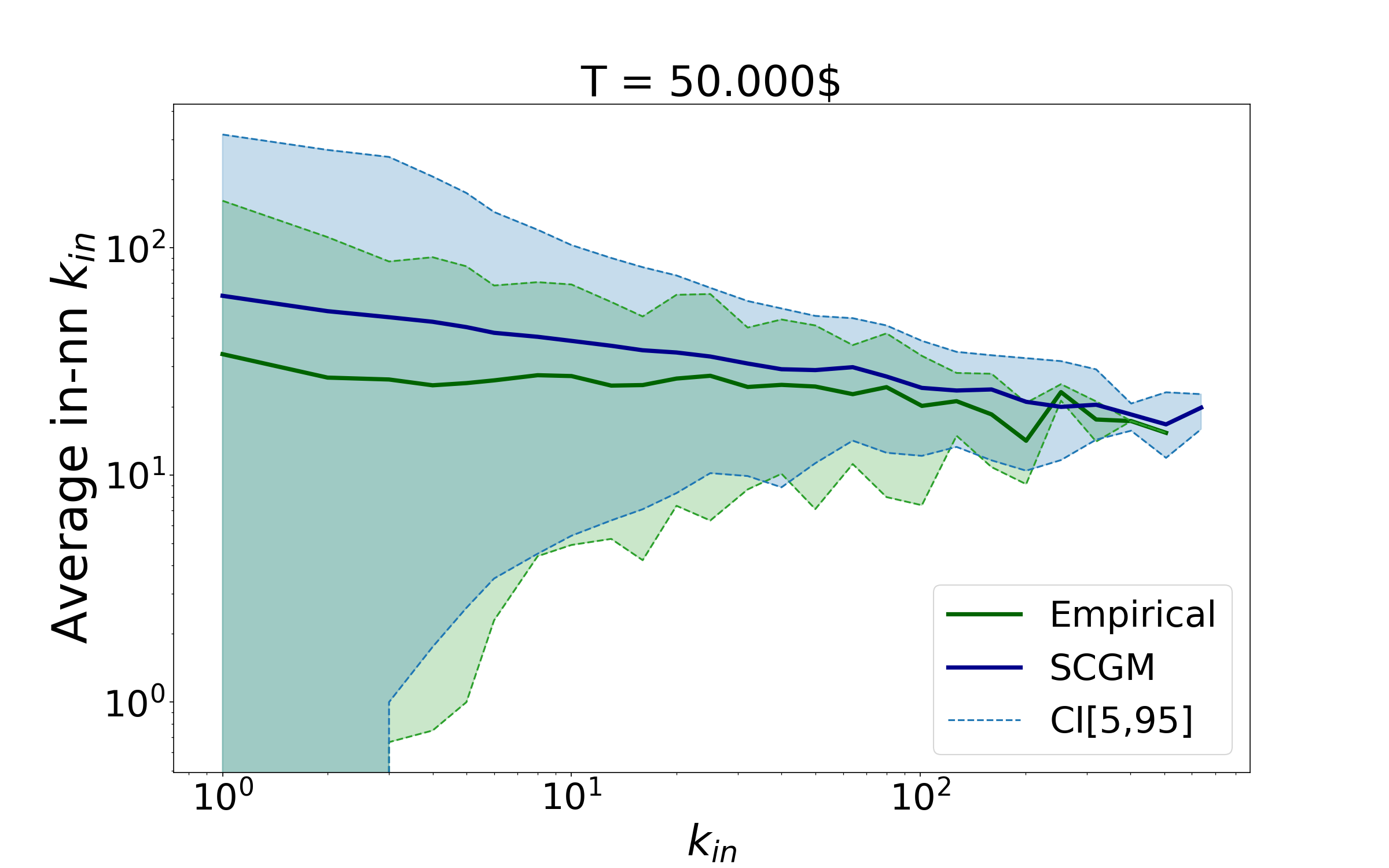}
\end{subfigure}
\caption{Scatter plots of $k_i^\text{in-in}$ versus $k_i^\text{in}$ for the four, different thresholds.}
\label{fig14}
\end{figure*}

\begin{figure*}[t!]
\centering
\begin{subfigure}{0.02\textwidth}
    \textbf{a)}
\end{subfigure}
\begin{subfigure}[t]{0.47\textwidth}
\includegraphics[width=\textwidth,valign=t]{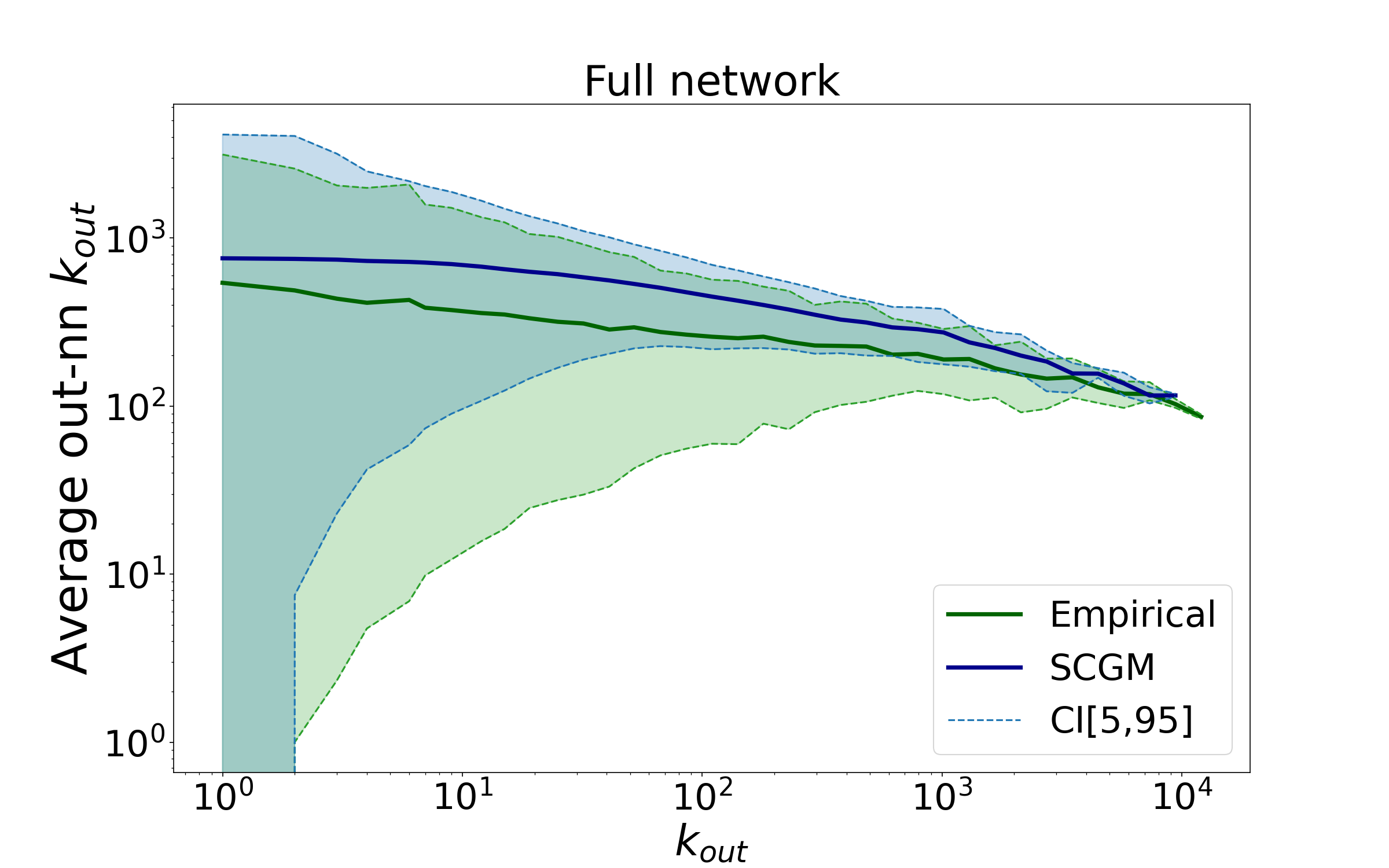}
\end{subfigure}
\begin{subfigure}{0.02\textwidth}
    \textbf{b)}
\end{subfigure}
\begin{subfigure}[t]{0.47\textwidth}
\includegraphics[width=\textwidth,valign=t]{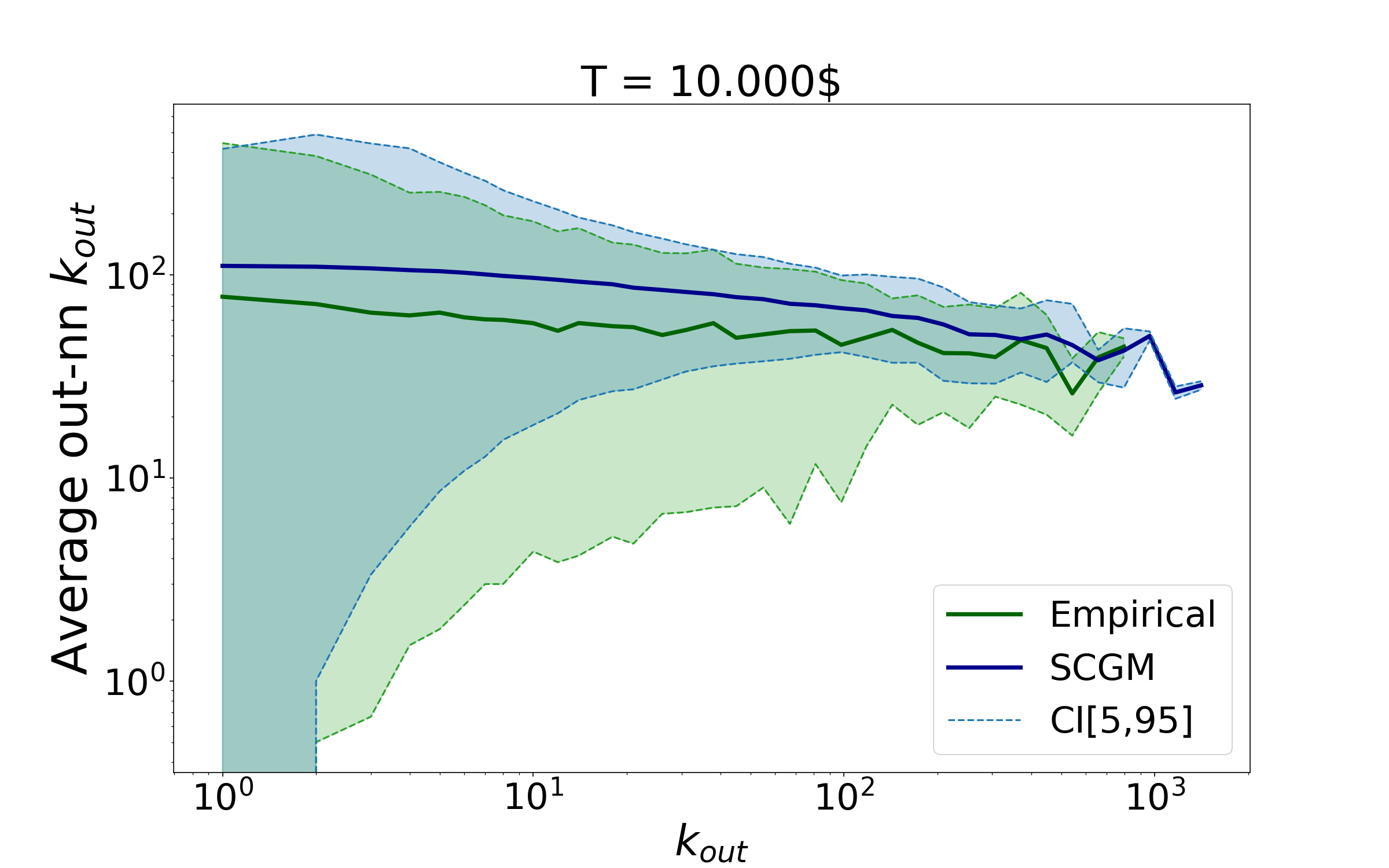}
\end{subfigure}
\begin{subfigure}{0.02\textwidth}
    \textbf{c)}
\end{subfigure}
\begin{subfigure}[t]{0.47\textwidth}
\includegraphics[width=\textwidth,valign=t]{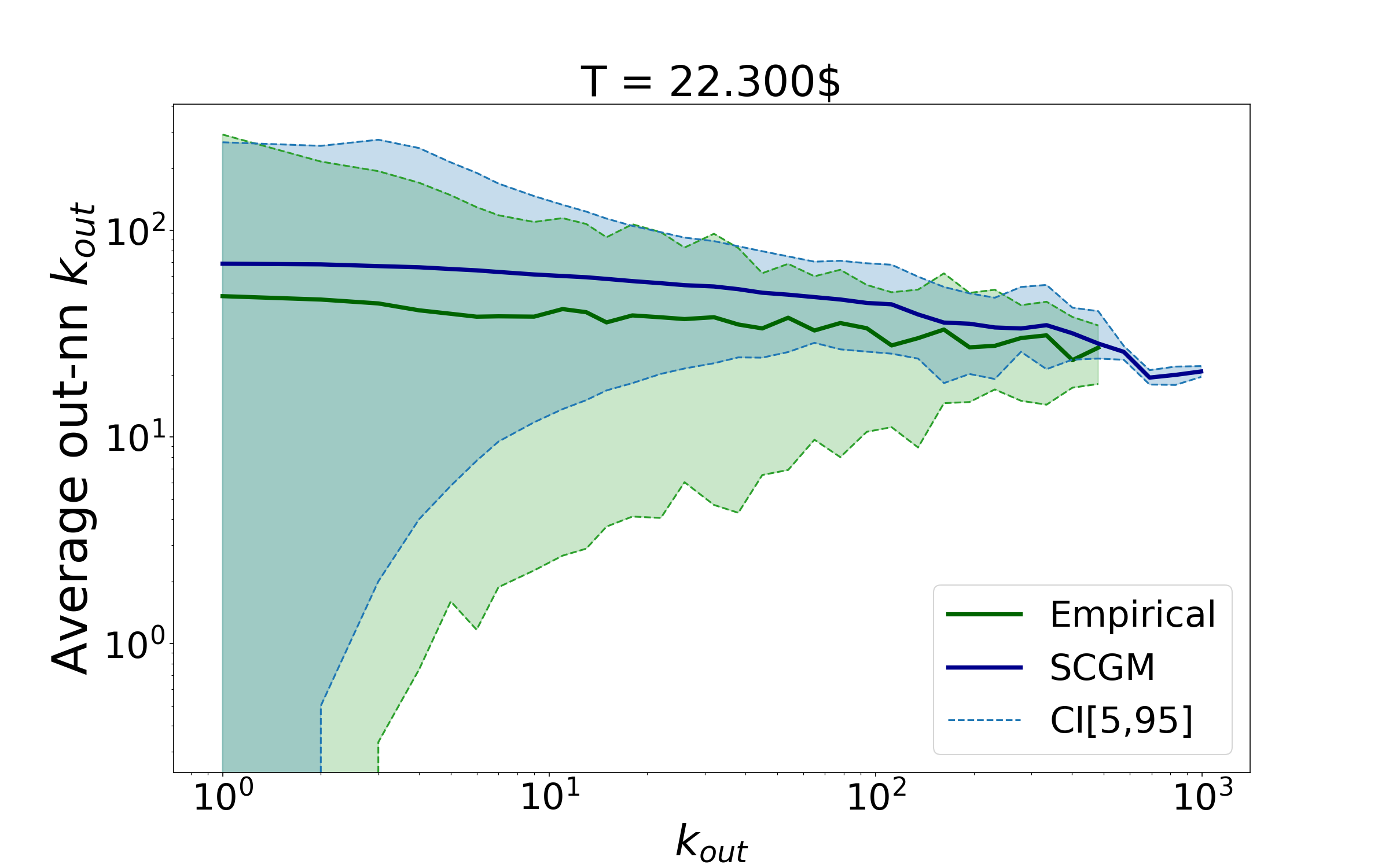}
\end{subfigure}
\begin{subfigure}{0.02\textwidth}
    \textbf{d)}
\end{subfigure}
\begin{subfigure}[t]{0.47\textwidth}
\includegraphics[width=\textwidth,valign=t]{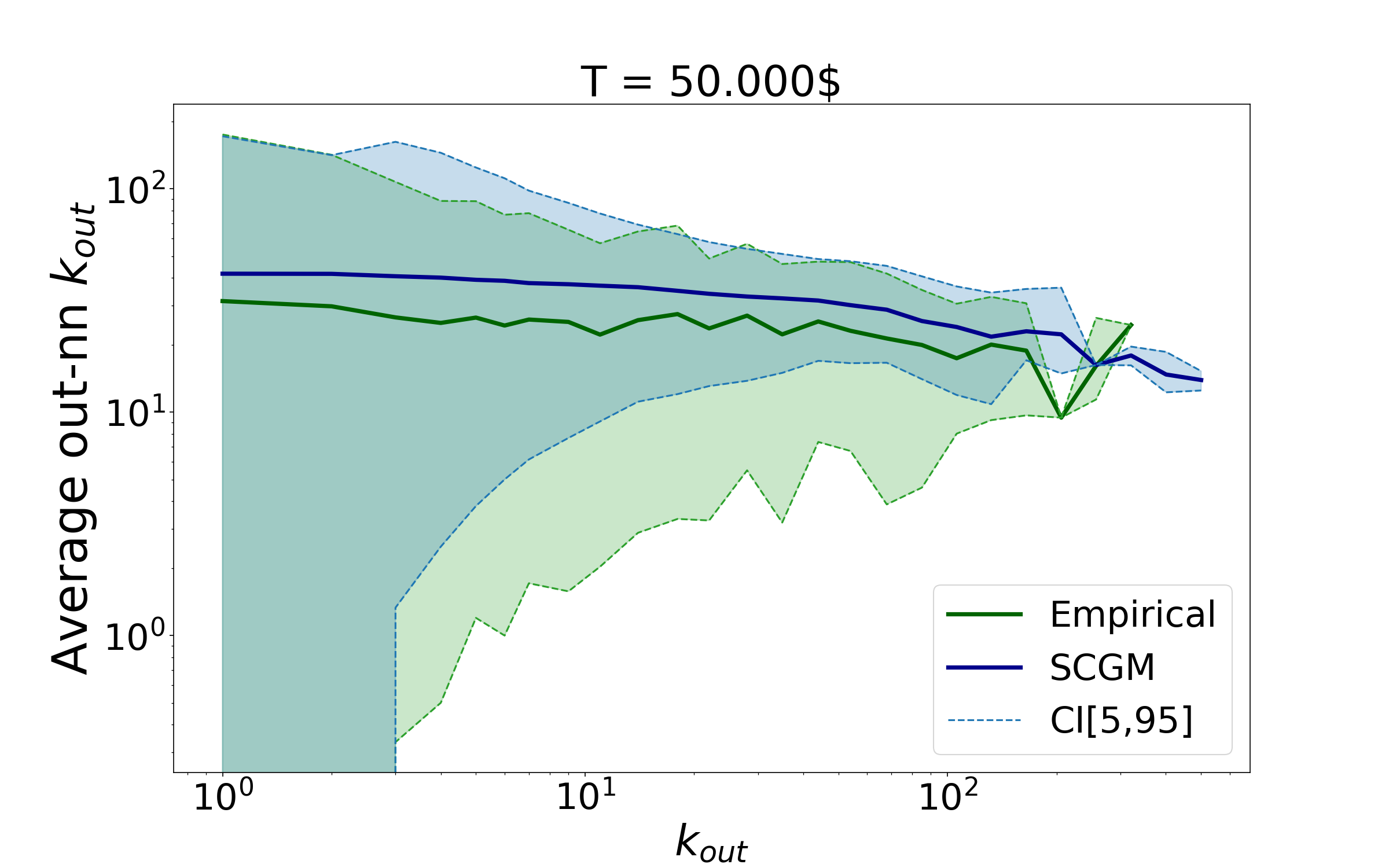}
\end{subfigure}
\caption{Scatter plots of $k_i^\text{out-out}$ versus $k_i^\text{out}$ for the four, different thresholds.}
\label{fig17}
\end{figure*}

For what concerns the ANND (see Figure \ref{fig14} and Figure \ref{fig17}), the SCGM achieves a better reconstruction on the filtered versions of the Ecuadorian production network. The difference is particularly evident in the case of $k_i^\text{in-in}$: this can be imputed to the fact that this variant of the ANND depends extensively on the in-degrees of the nodes and that, upon adopting a threshold, the reconstruction of the in-degrees experiences a bigger improvement than the reconstruction of the out-degrees (see Figure \ref{fig10} and Figure \ref{fig11}).

\begin{figure*}[t!]
\centering
\begin{subfigure}{0.02\textwidth}
    \textbf{a)}
\end{subfigure}
\begin{subfigure}[t]{0.47\textwidth}
\includegraphics[width=\textwidth,valign=t]{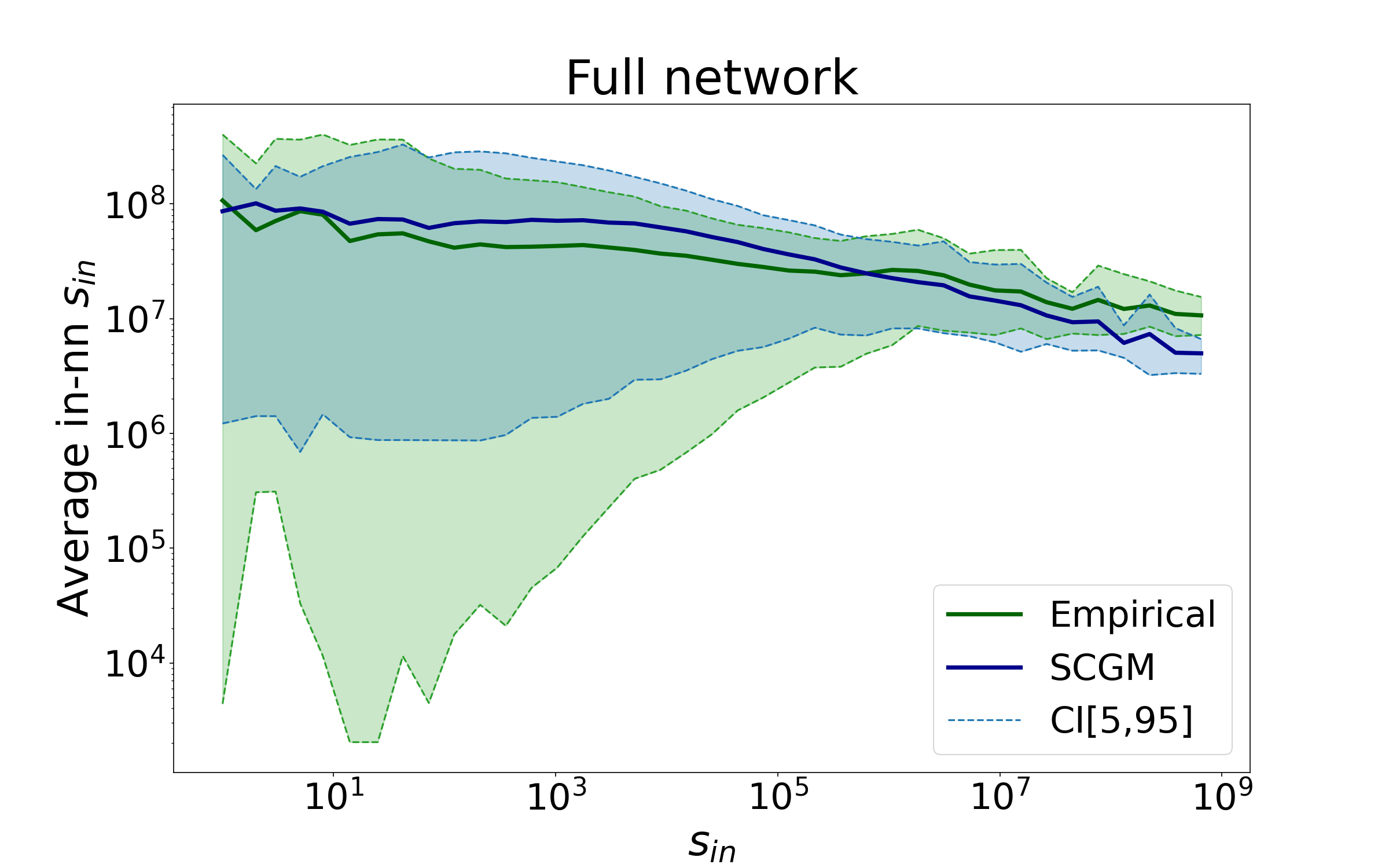}
\end{subfigure}
\begin{subfigure}{0.02\textwidth}
    \textbf{b)}
\end{subfigure}
\begin{subfigure}[t]{0.47\textwidth}
\includegraphics[width=\textwidth,valign=t]{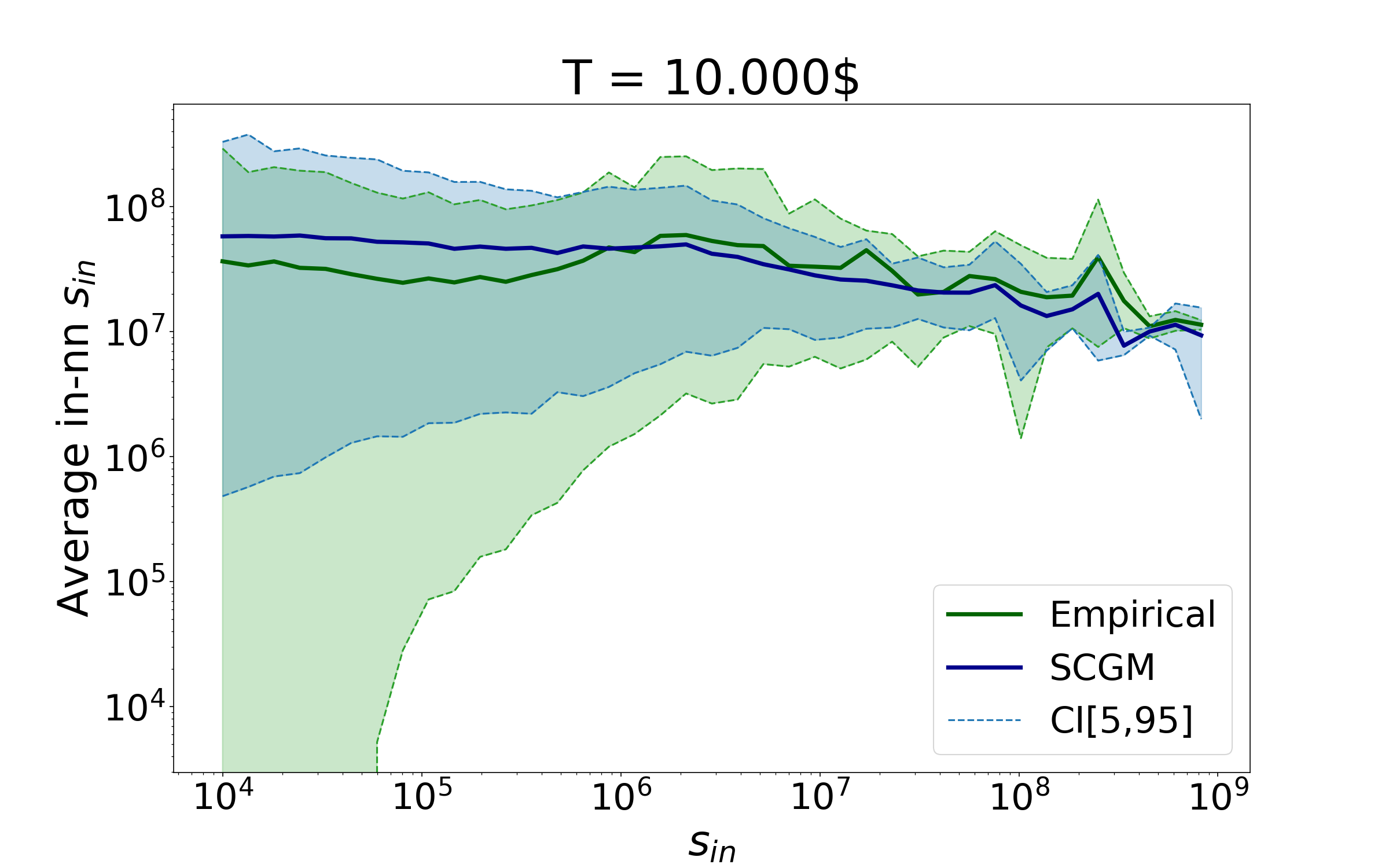}
\end{subfigure}
\begin{subfigure}{0.02\textwidth}
    \textbf{c)}
\end{subfigure}
\begin{subfigure}[t]{0.47\textwidth}
\includegraphics[width=\textwidth,valign=t]{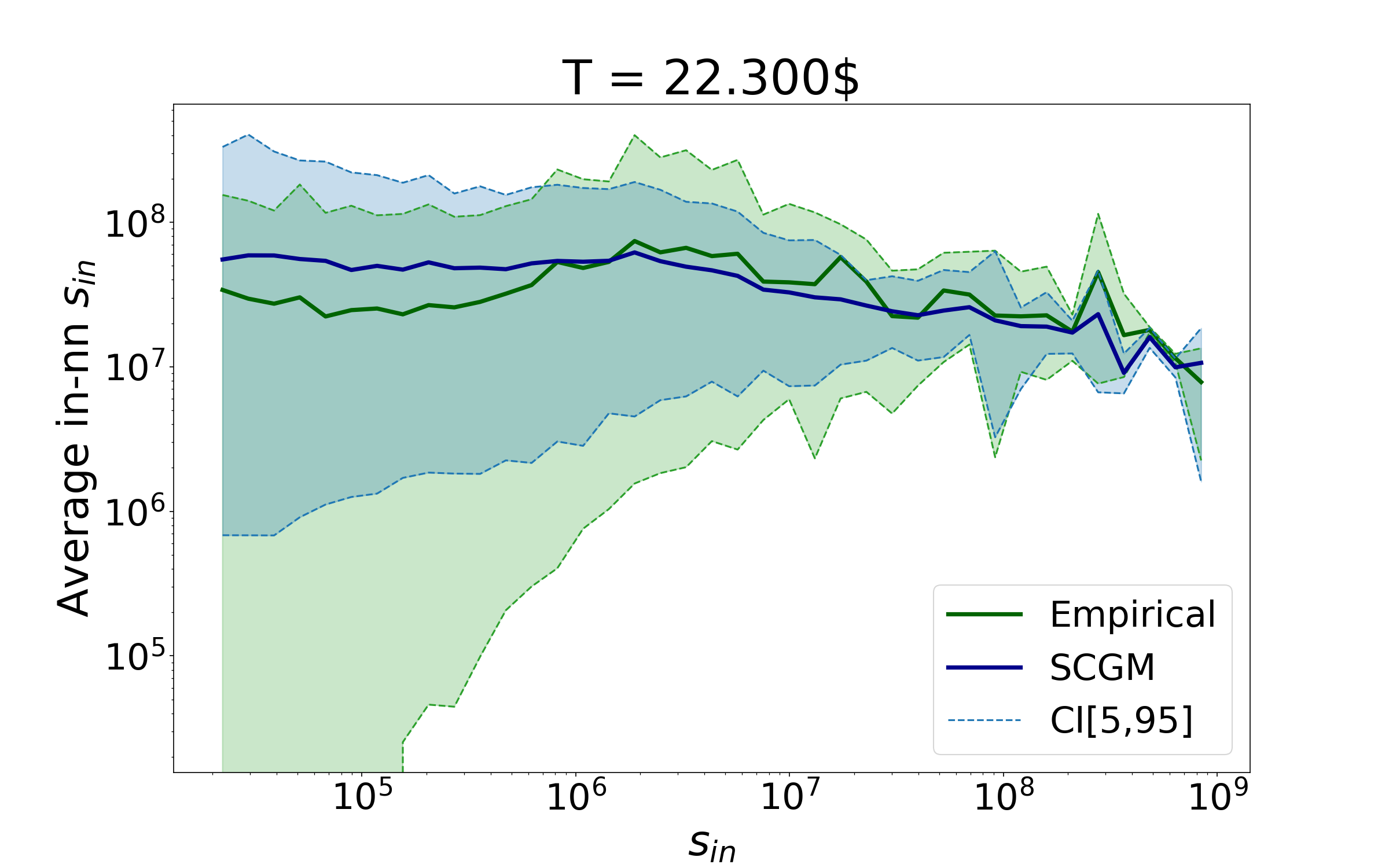}
\end{subfigure}
\begin{subfigure}{0.02\textwidth}
    \textbf{d)}
\end{subfigure}
\begin{subfigure}[t]{0.47\textwidth}
\includegraphics[width=\textwidth,valign=t]{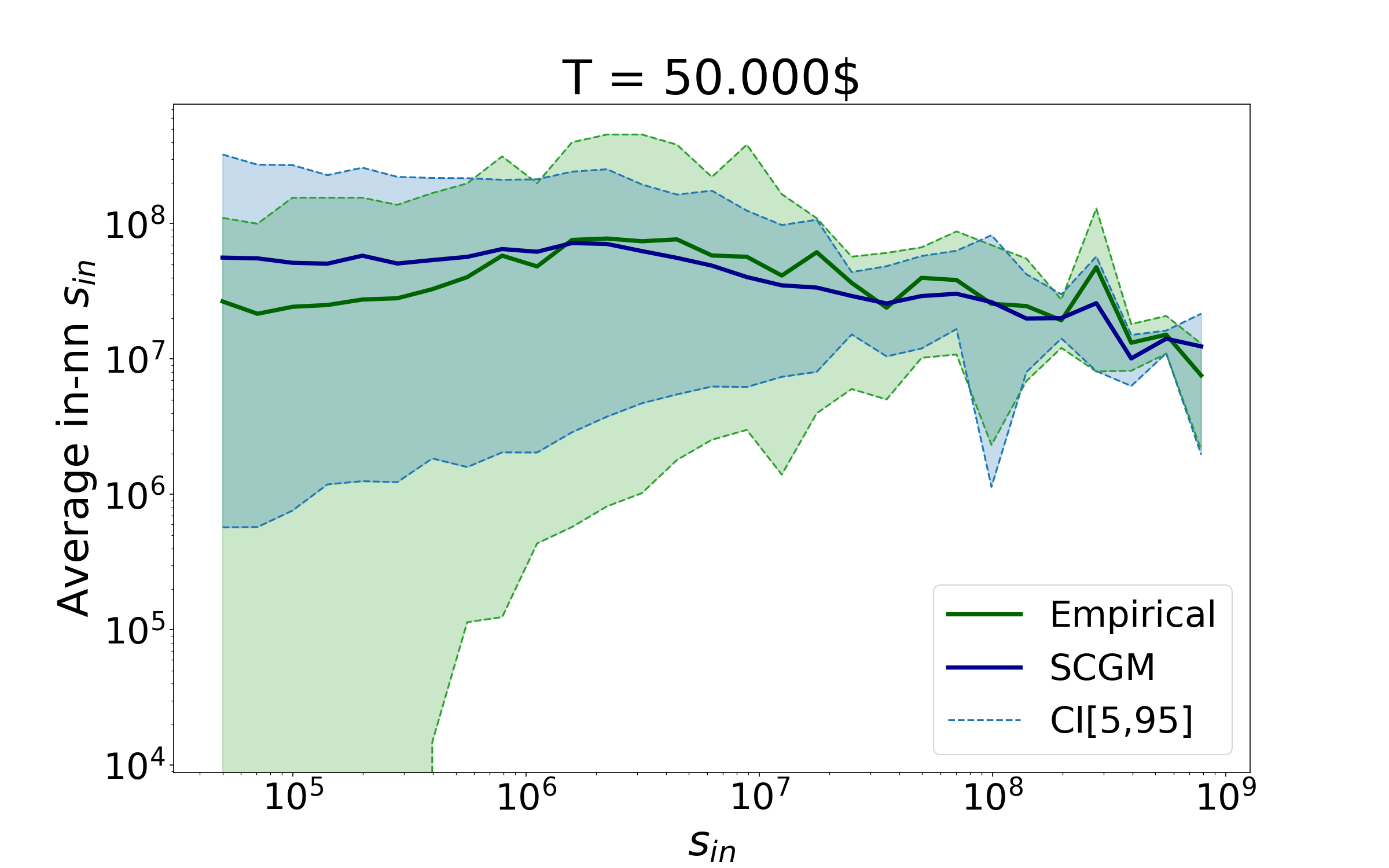}
\end{subfigure}
\caption{Scatter plots of $s_i^\text{in-in}$ versus $s_i^\text{in}$ for the four, different thresholds.}
\label{fig18}
\end{figure*}

\begin{figure*}[t!]
\centering
\begin{subfigure}{0.02\textwidth}
    \textbf{a)}
\end{subfigure}
\begin{subfigure}[t]{0.47\textwidth}
\includegraphics[width=\textwidth,valign=t]{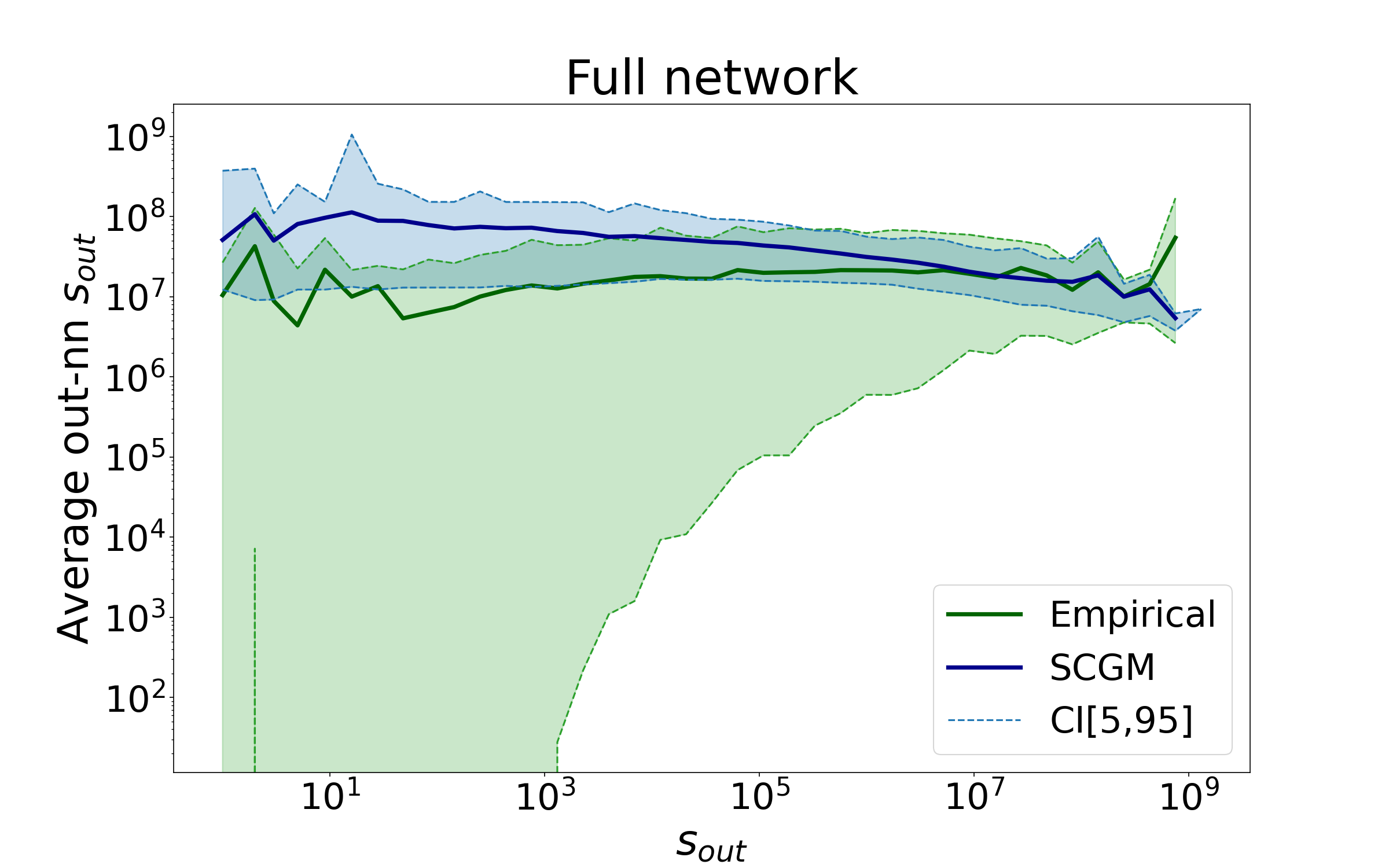}
\end{subfigure}
\begin{subfigure}{0.02\textwidth}
    \textbf{b)}
\end{subfigure}
\begin{subfigure}[t]{0.47\textwidth}
\includegraphics[width=\textwidth,valign=t]{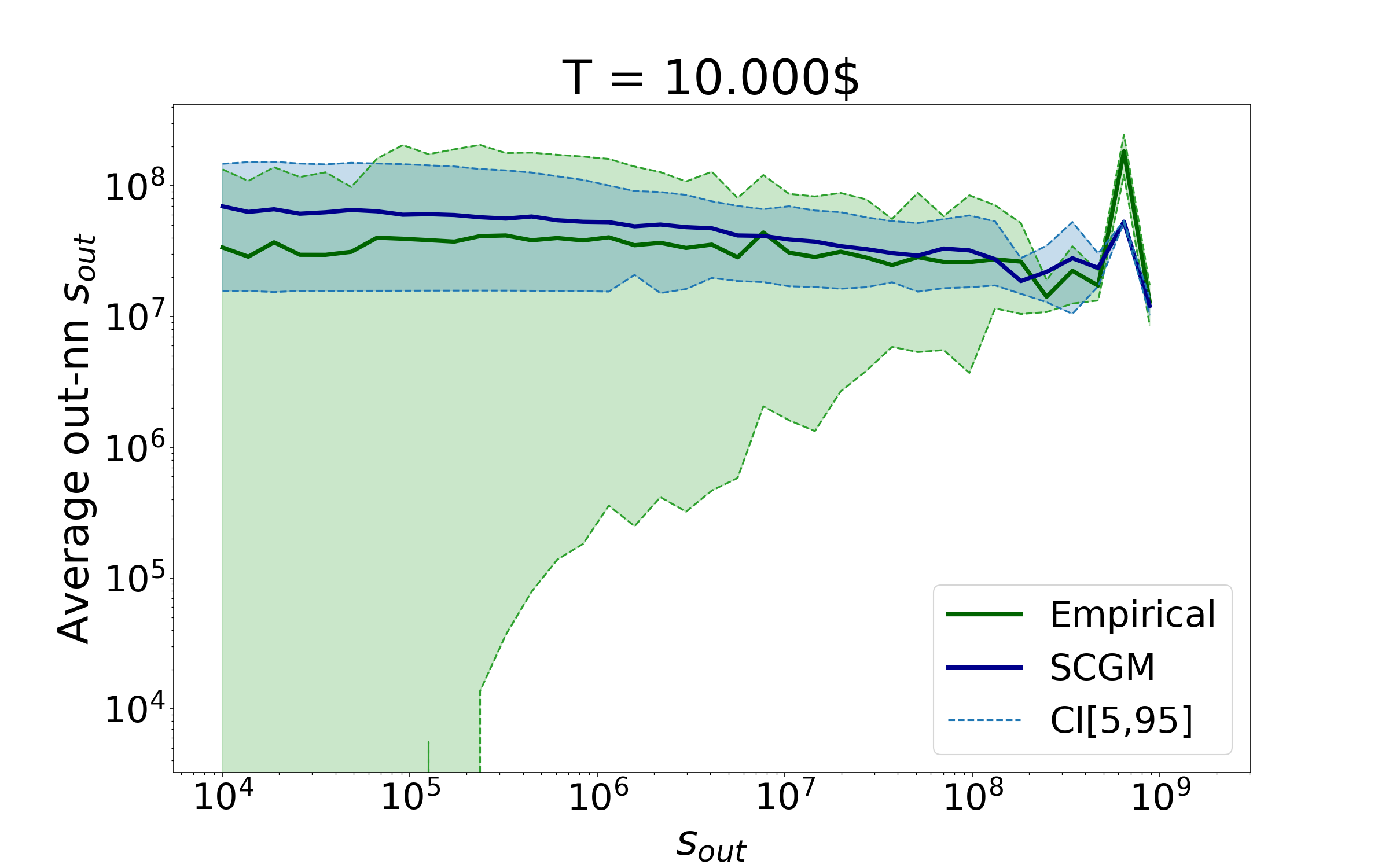}
\end{subfigure}
\begin{subfigure}{0.02\textwidth}
    \textbf{c)}
\end{subfigure}
\begin{subfigure}[t]{0.47\textwidth}
\includegraphics[width=\textwidth,valign=t]{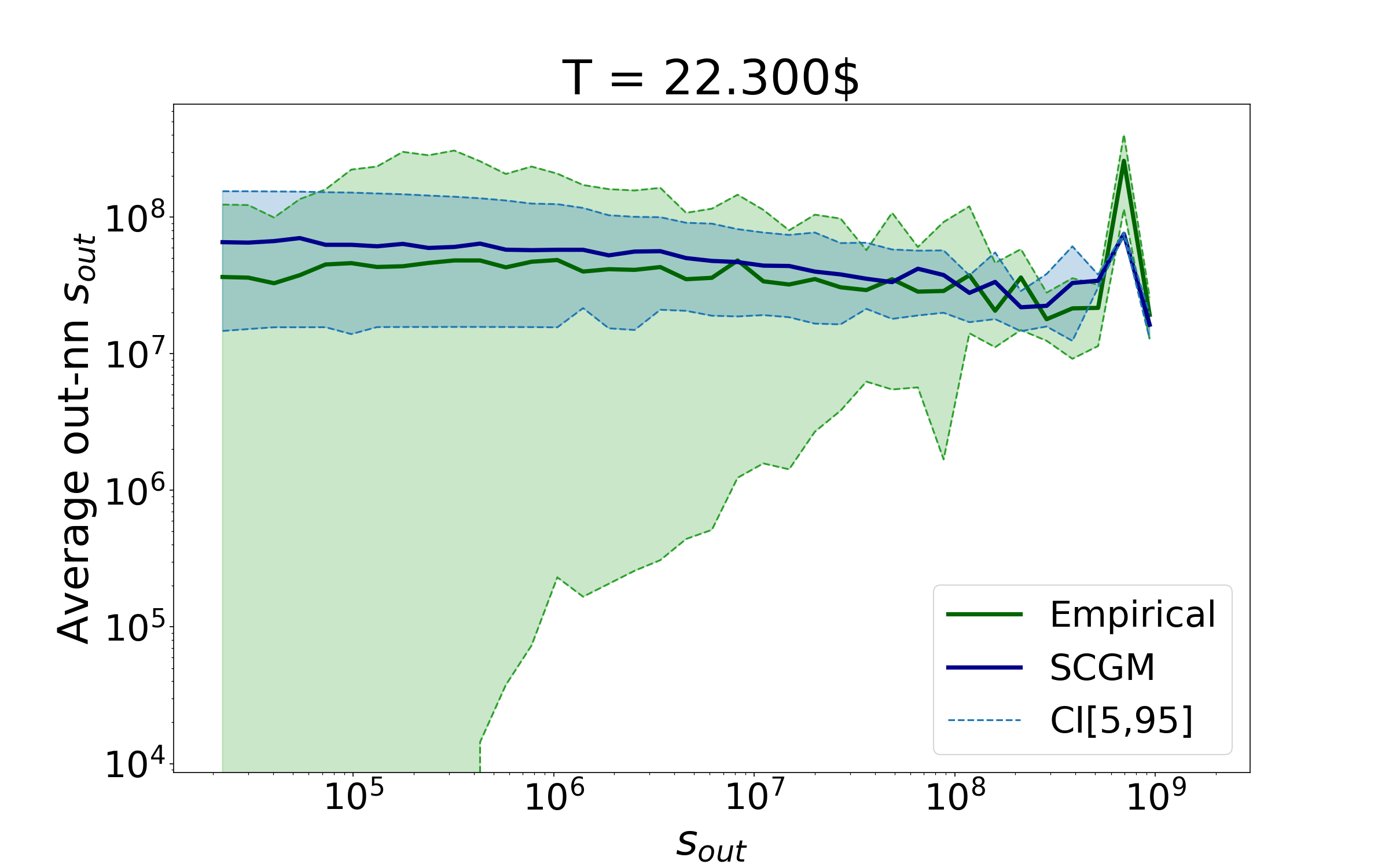}
\end{subfigure}
\begin{subfigure}{0.02\textwidth}
    \textbf{d)}
\end{subfigure}
\begin{subfigure}[t]{0.47\textwidth}
\includegraphics[width=\textwidth,valign=t]{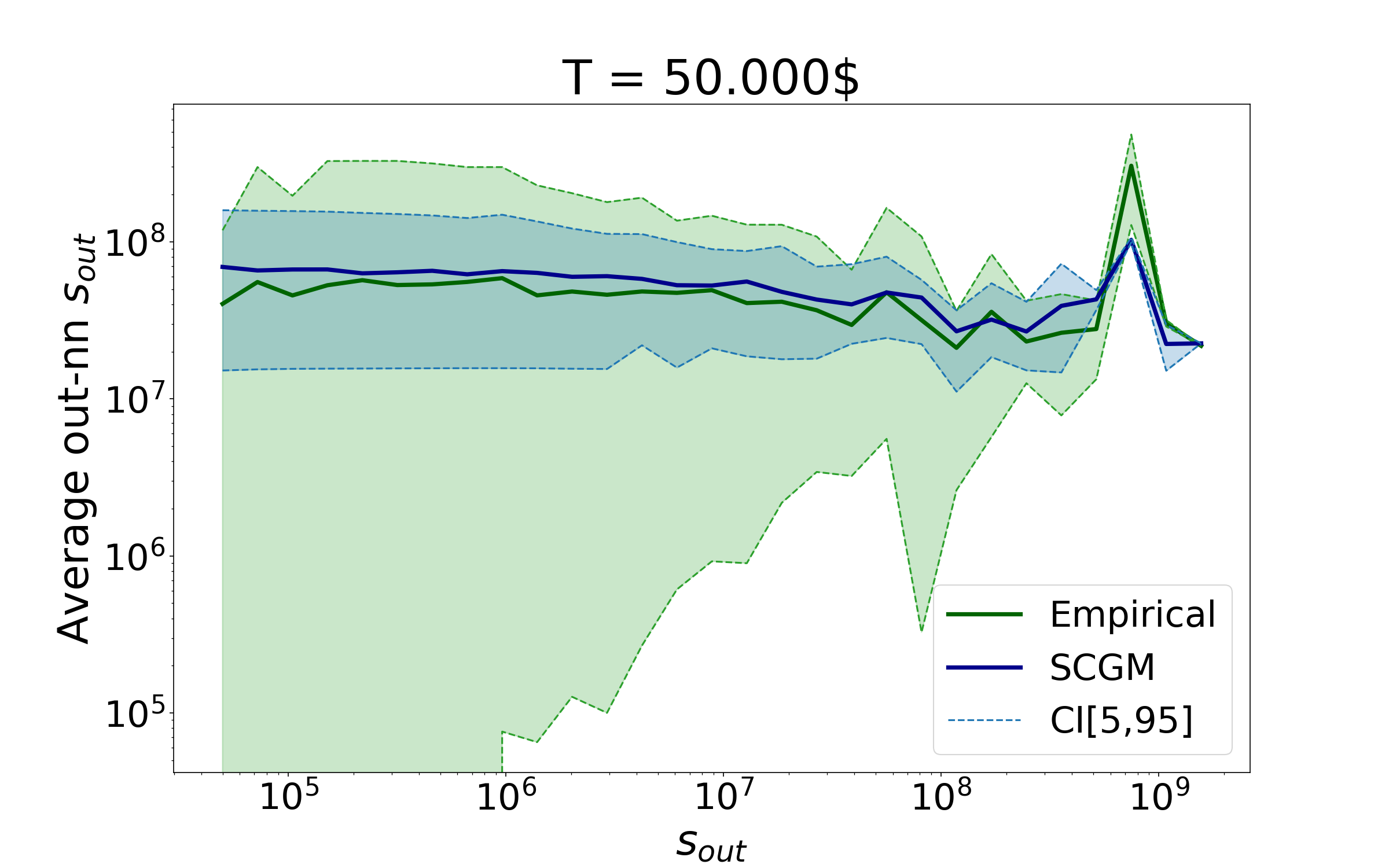}
\end{subfigure}
\caption{Scatter plots of $s_i^\text{out-out}$ versus $s_i^\text{out}$ for the four, different thresholds.}
\label{fig21}
\end{figure*}

For what concerns the ANNS (see Figure \ref{fig18} and Figure \ref{fig21}), the performance of the SCGM appears to be quite stable for all values of the threshold, the main difference being that its trend is characterised by a peak, for large values of the strengths, on the filtered versions of the Ecuadorian production network: this can be imputed to the removal of many, small firms linked to the bigger ones.

\clearpage

\clearpage

\clearpage

\clearpage

\clearpage

\section{Upstream and downstream components of ESRI}

Since the upstream and downstream components of an economic shock are computed independently, the upstream and downstream ESRI values of firm $i$ can be evaluated separately. Specifically,

\begin{align}
\text{ESRI}^u_i&=\sum_{j(\neq i)}\frac{s_j^\text{out}}{W_\text{tot}}(1-h_j^{u}),\\ 
\text{ESRI}^d_i&=\sum_{j(\neq i)}\frac{s_j^\text{out}}{W_\text{tot}}(1-h_j^d).
\end{align}

\begin{figure*}[t!]
\centering
\begin{subfigure}{0.02\textwidth}
    \textbf{a)}
\end{subfigure}
\begin{subfigure}[t]{0.47\textwidth}
\includegraphics[width=\textwidth,valign=t]{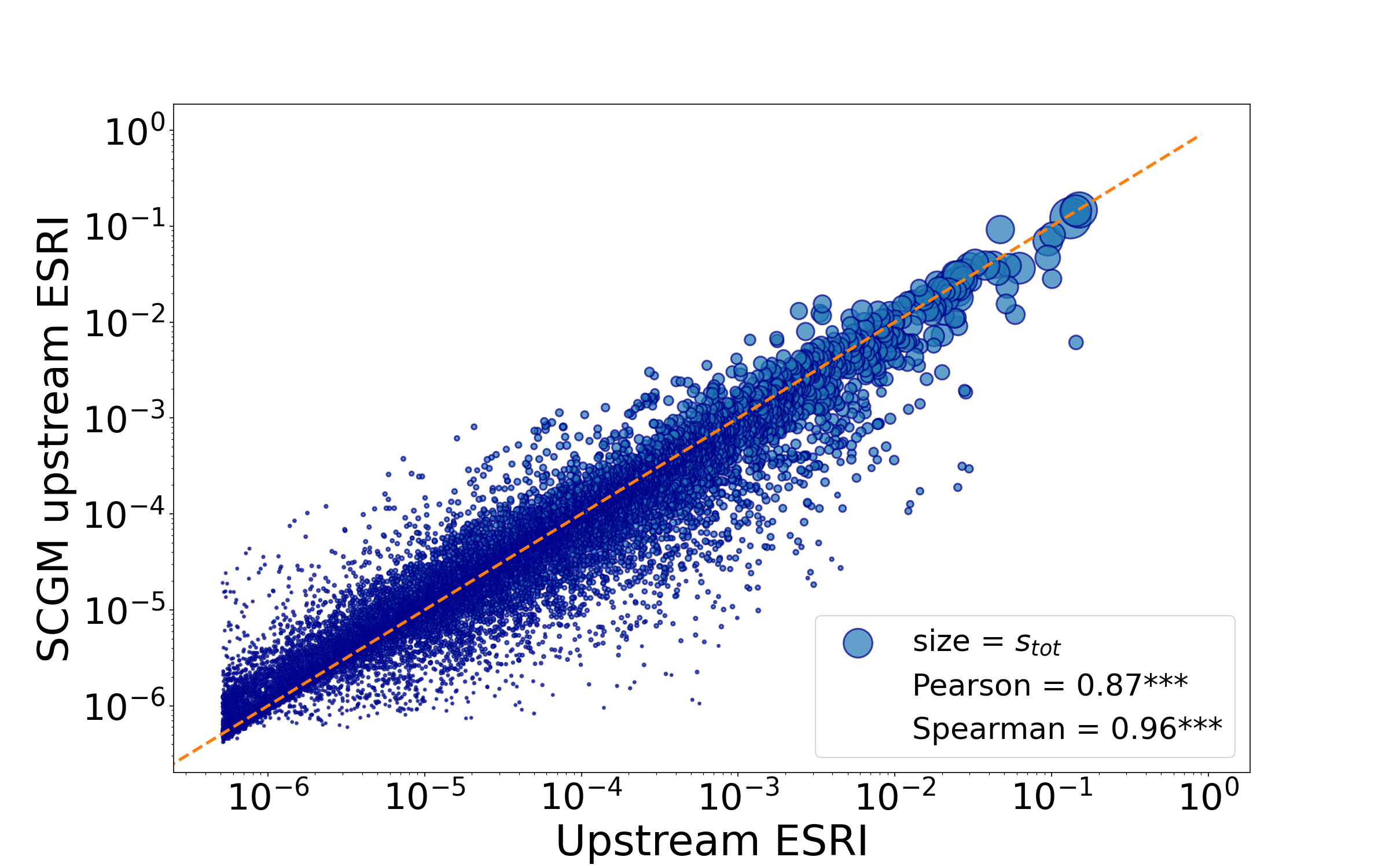}
\end{subfigure}
\begin{subfigure}{0.02\textwidth}
    \textbf{b)}
\end{subfigure}
\begin{subfigure}[t]{0.47\textwidth}
\includegraphics[width=\textwidth,valign=t]{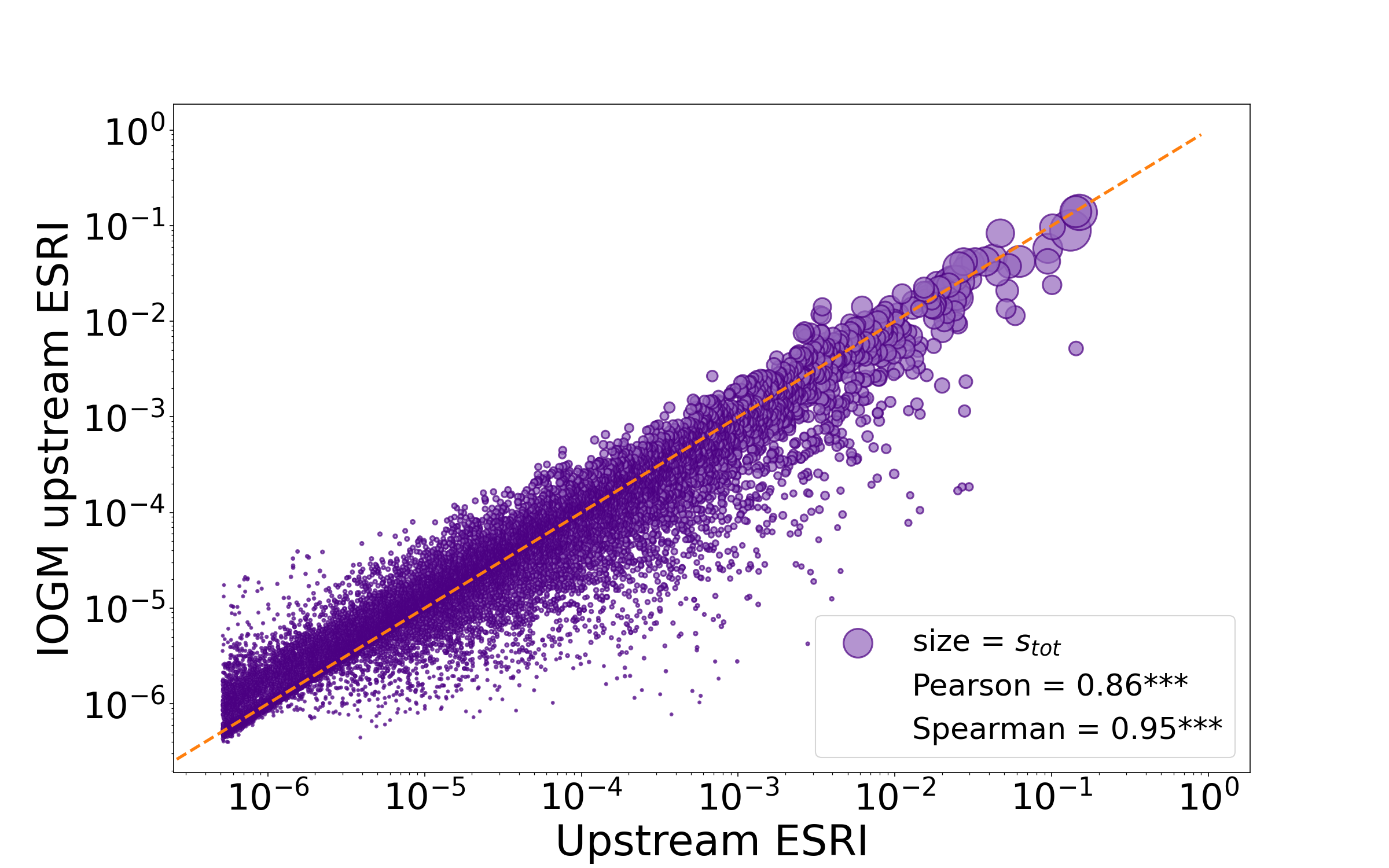}
\end{subfigure}
\begin{subfigure}{0.02\textwidth}
    \textbf{c)}
\end{subfigure}
\begin{subfigure}[t]{0.47\textwidth}
\includegraphics[width=\textwidth,valign=t]{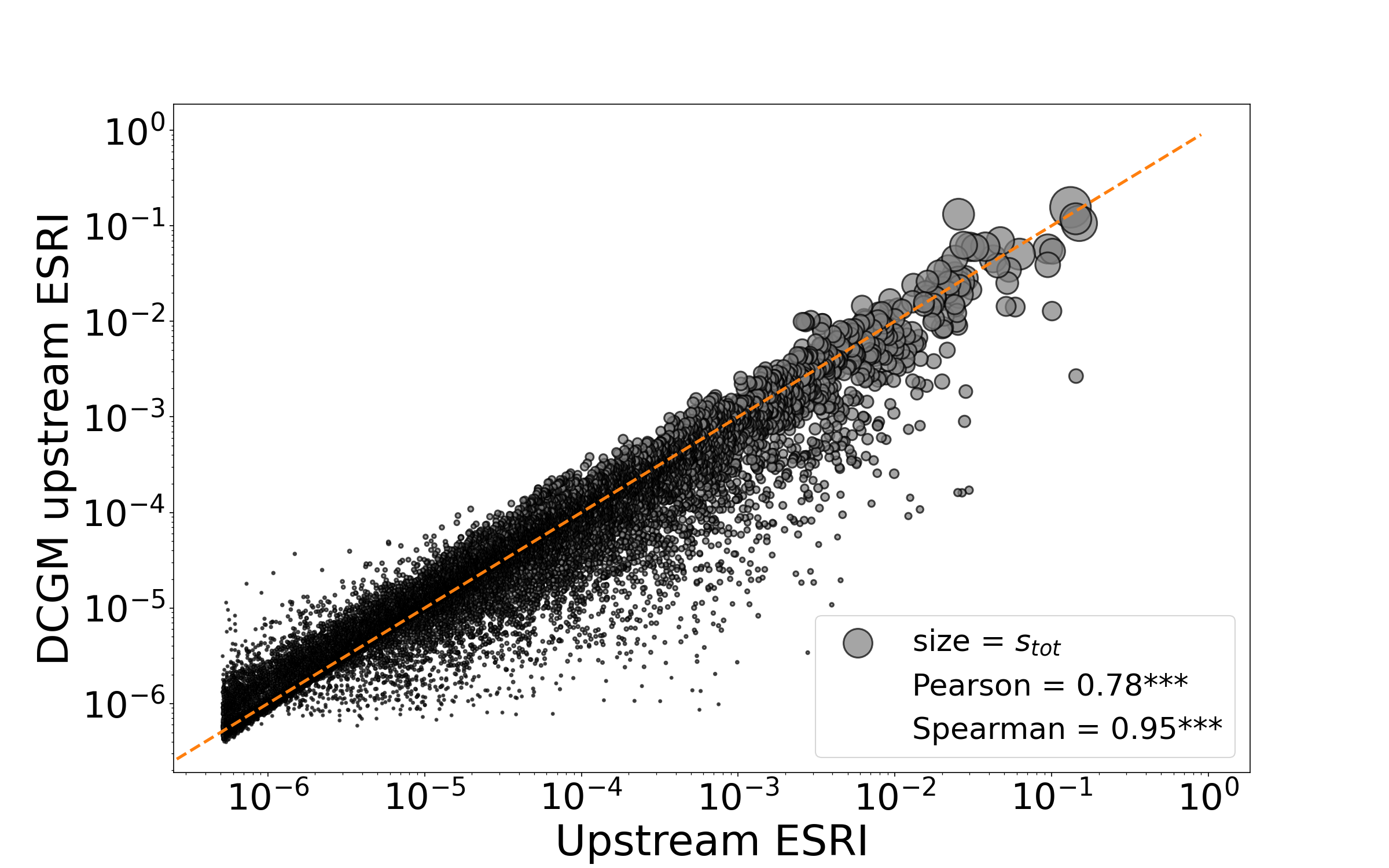}
\end{subfigure}
\begin{subfigure}{0.02\textwidth}
    \textbf{d)}
\end{subfigure}
\begin{subfigure}[t]{0.47\textwidth}
\includegraphics[width=\textwidth,valign=t]{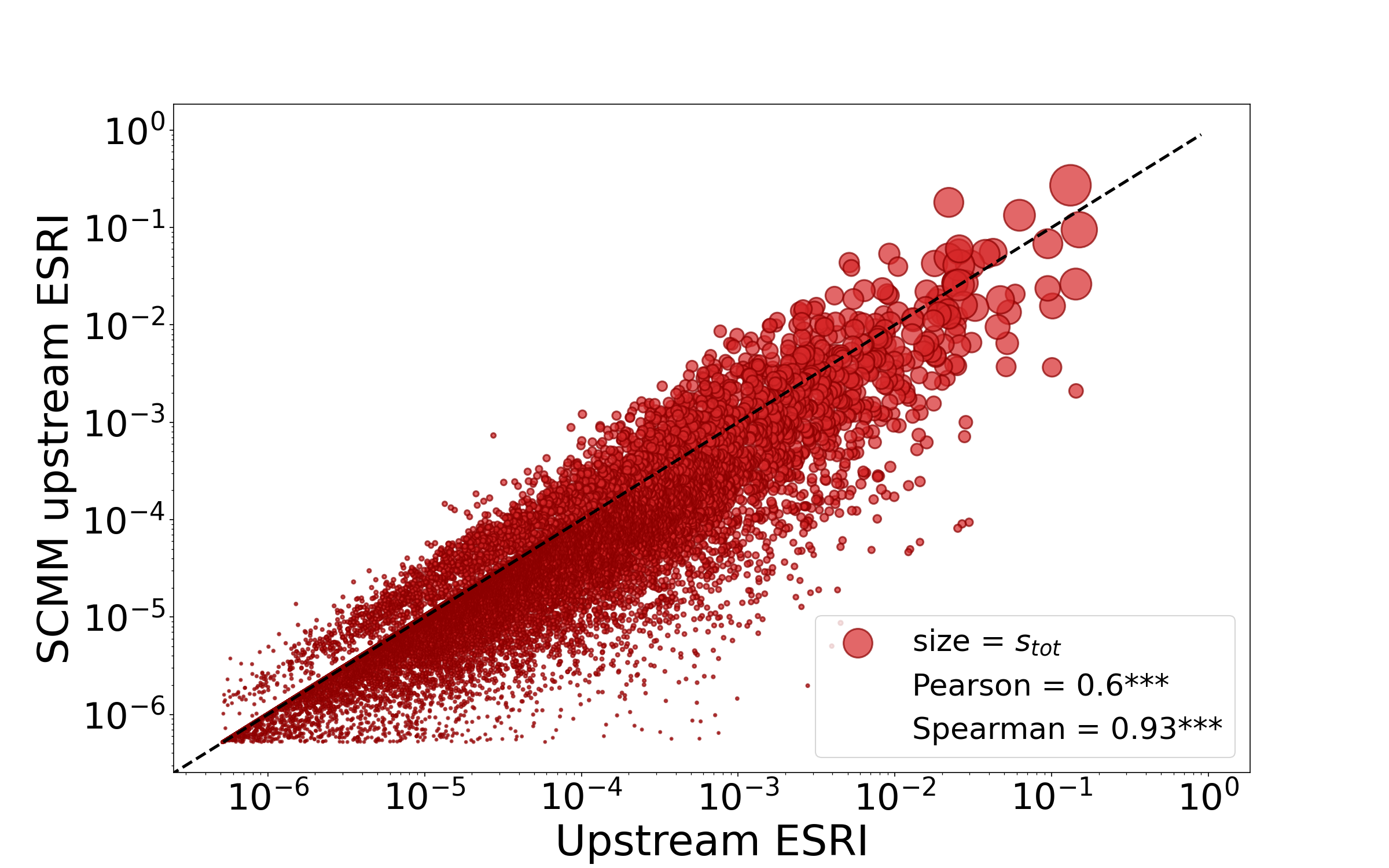}
\end{subfigure}
\caption{Empirical versus reconstructed upstream ESRI for the SCGM (a), IOGM (b), DCGM (c), SCMM (d).}
\label{fig22}
\end{figure*}

\begin{figure*}[t!]
\centering
\begin{subfigure}{0.02\textwidth}
    \textbf{a)}
\end{subfigure}
\begin{subfigure}[t]{0.47\textwidth}
\includegraphics[width=\textwidth,valign=t]{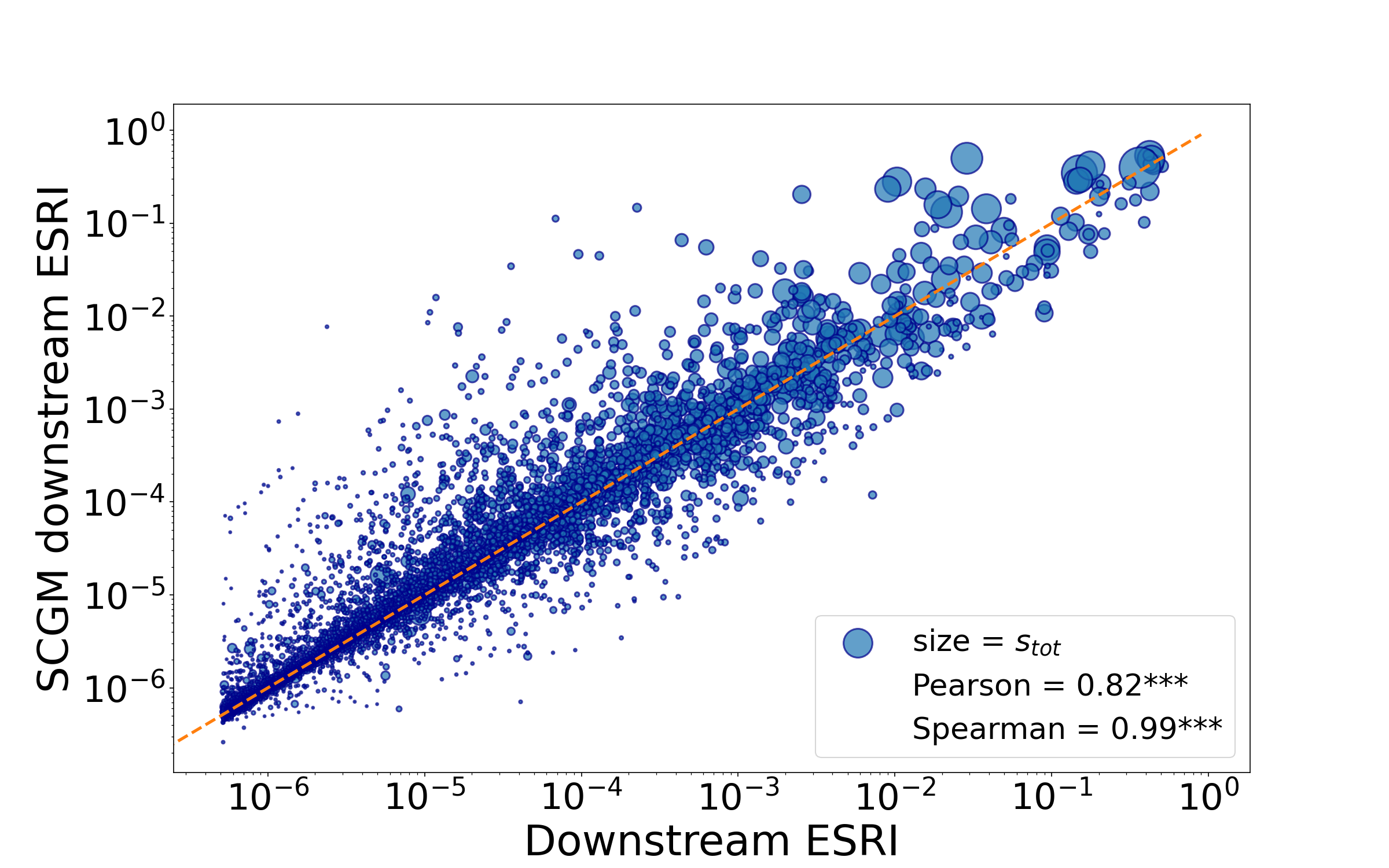}
\end{subfigure}
\begin{subfigure}{0.02\textwidth}
    \textbf{b)}
\end{subfigure}
\begin{subfigure}[t]{0.47\textwidth}
\includegraphics[width=\textwidth,valign=t]{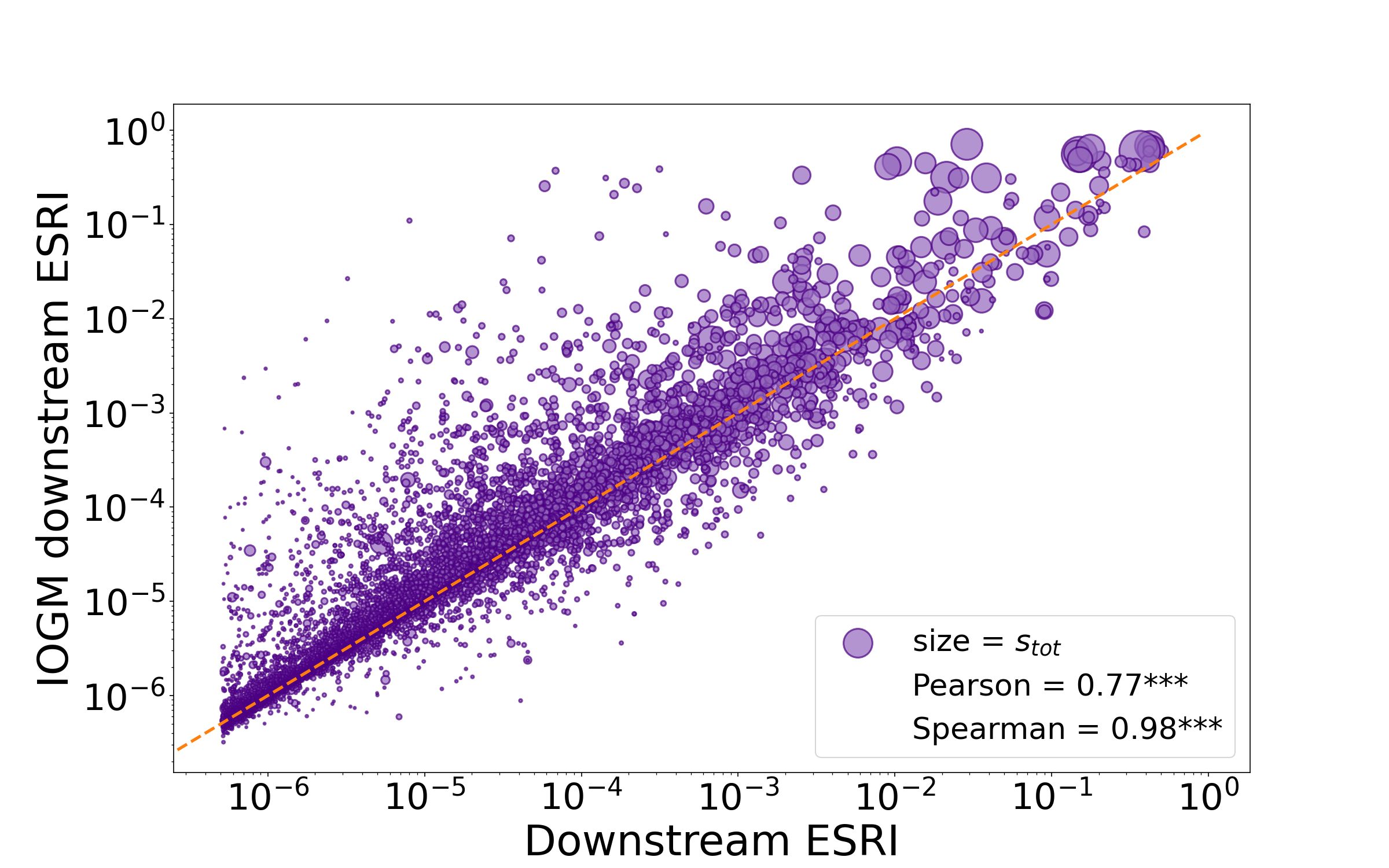}
\end{subfigure}
\begin{subfigure}{0.02\textwidth}
    \textbf{c)}
\end{subfigure}
\begin{subfigure}[t]{0.47\textwidth}
\includegraphics[width=\textwidth,valign=t]{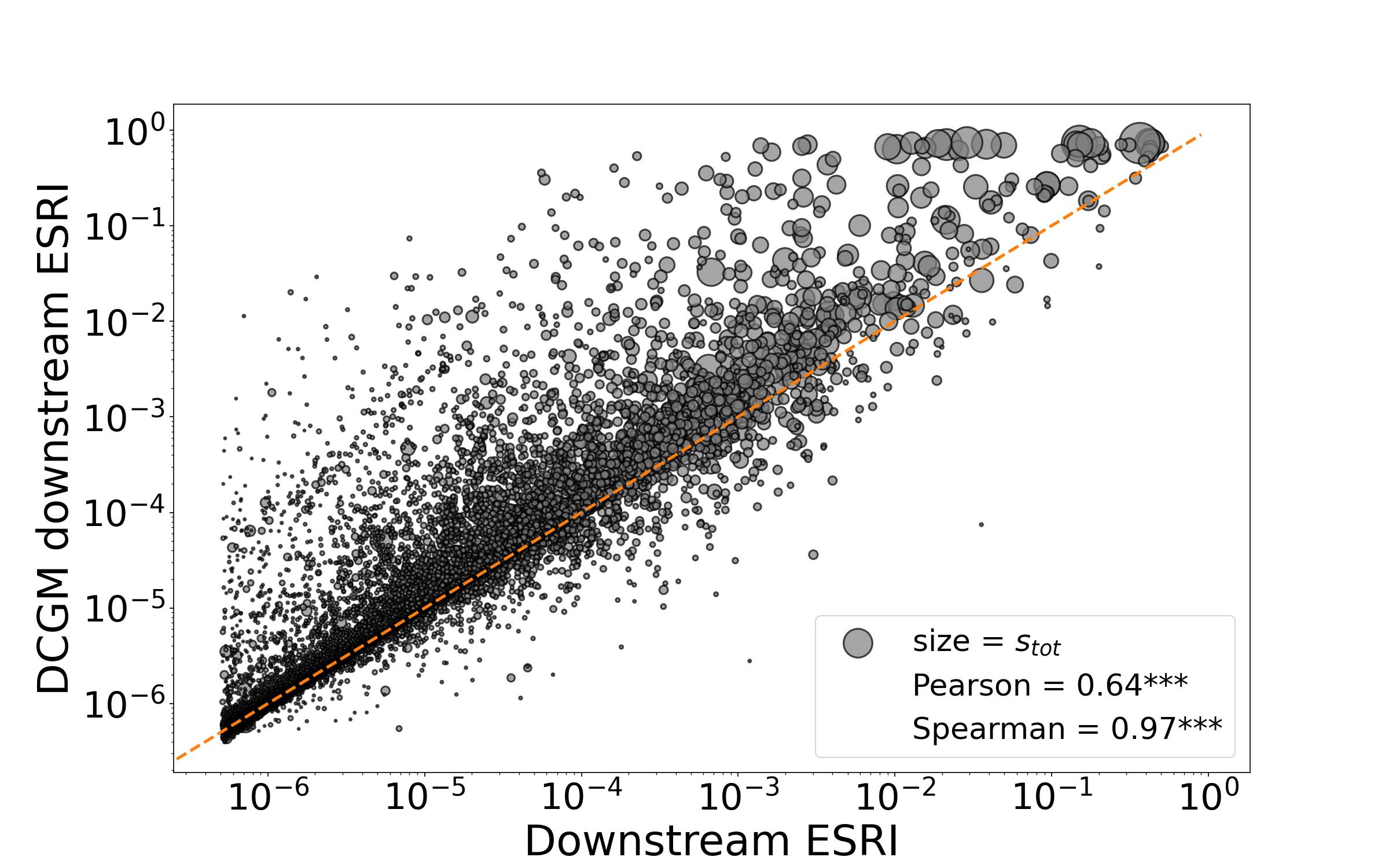}
\end{subfigure}
\begin{subfigure}{0.02\textwidth}
    \textbf{d)}
\end{subfigure}
\begin{subfigure}[t]{0.47\textwidth}
\includegraphics[width=\textwidth,valign=t]{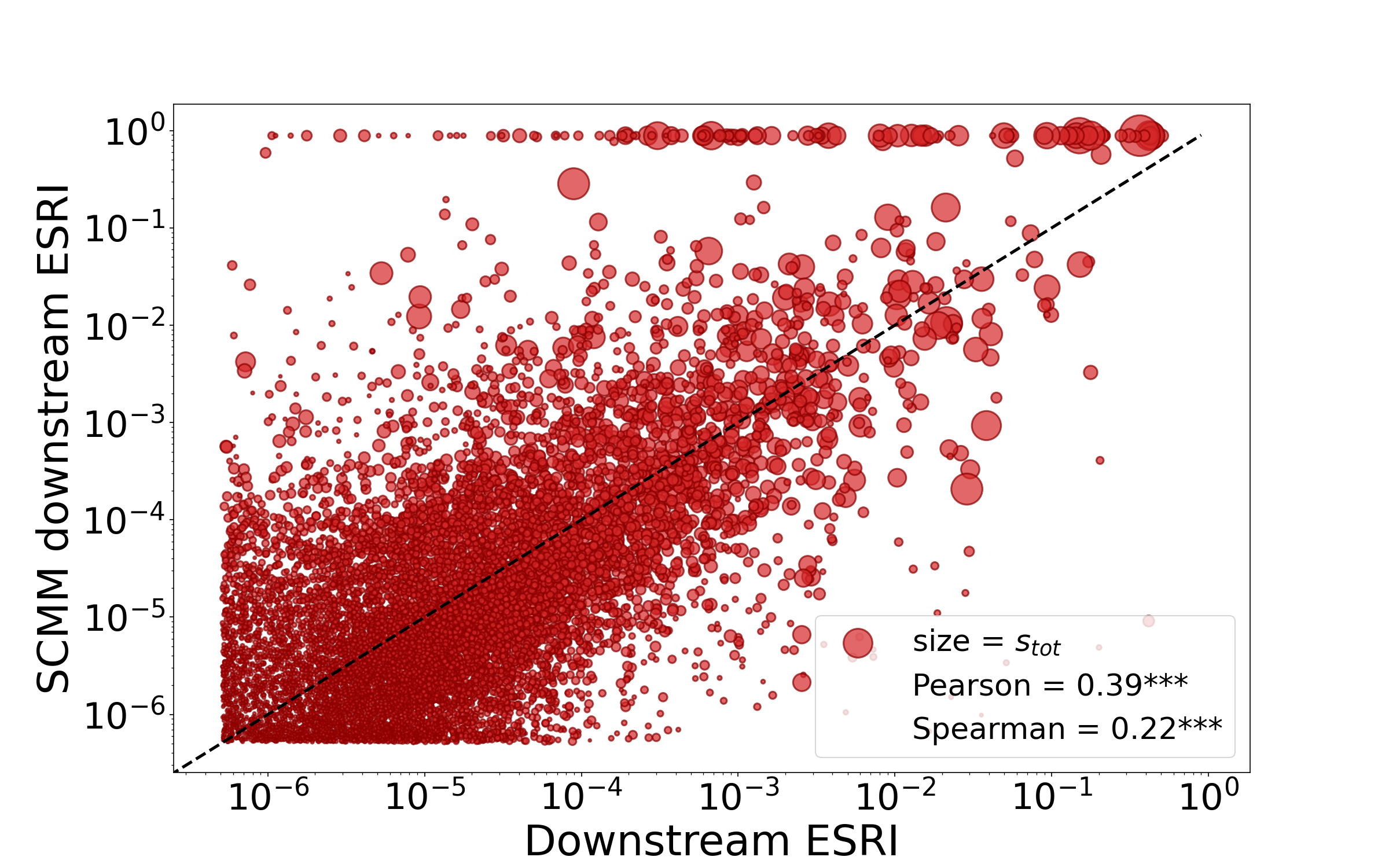}
\end{subfigure}
\caption{Empirical versus reconstructed downstream ESRI for the SCGM (a), IOGM (b), DCGM (c), SCMM (d).}
\label{fig23}
\end{figure*}

Figure \ref{fig22} and Figure \ref{fig23} show the empirical upstream and downstream ESRI values versus the reconstructed ones, for our, four models. The upstream ESRI values are obtained by solving a linear equation: hence, as confirmed by the large values of the correlation coefficients displayed by all models (the Pearson's one is larger than $0.60$; the Spearman's one is larger than $0.93$), they are easier to recover. For what concerns the downstream ESRI values, the performance of the SCMM experiences the biggest drop, a result signalling that the propagation of the downstream shock, obtained by solving a non-linear equation, is highly dependent on the network topology; all the other models, instead, perform better in recovering the ranking than the specific ESRI values.

A qualitative inspection of the plots suggests that, overall, our models underestimate the upstream ESRI values and overestimate the downstream ESRI values. This can be (at least, partially) imputed to the density of the sampled configurations: while distributing $s^\text{in}_i$ onto more incoming links decreases the upstream impact of firm $i$, in case of failure, distributing $s^\text{out}_i$ onto more outgoing links increases its downstream impact, since the number of firms whose output is hard-constrained by the essential input provided by firm $i$ increases as well.

\clearpage


\end{document}